\documentclass[sigconf]{acmart}

\copyrightyear{2024}
\acmYear{2024}
\setcopyright{acmlicensed}\acmConference[ASE '24]{39th IEEE/ACM International Conference on Automated Software Engineering }{October 27-November 1, 2024}{Sacramento, CA, USA}
\acmBooktitle{39th IEEE/ACM International Conference on Automated Software Engineering (ASE '24), October 27-November 1, 2024, Sacramento, CA, USA}
\acmDOI{10.1145/3691620.3695551}
\acmISBN{979-8-4007-1248-7/24/10}

\usepackage{xspace}
\usepackage{xcolor}
\usepackage{graphicx}
\usepackage[normalem]{ulem} %
\usepackage{amsmath}
\usepackage{amsfonts}
\usepackage{enumitem}
\usepackage{textcomp}

\usepackage{hyperref}
\hypersetup{linkcolor=black,citecolor=black,anchorcolor=black,filecolor=black,menucolor=black,runcolor=black,urlcolor=black,hidelinks}
\usepackage{breakurl}

\usepackage{booktabs}
\usepackage{multirow}
\usepackage{makecell}
\usepackage{ragged2e}

\usepackage{caption}
\usepackage{subcaption}

\usepackage{listings}

\usepackage{tikz}
\usetikzlibrary{calc}
\usetikzlibrary{shapes.geometric}
\usetikzlibrary{decorations.pathreplacing}
\usetikzlibrary{positioning}
\usetikzlibrary{shapes}

\usepackage{datetime} %
\usepackage{tcolorbox} %

\newcommand{\XSpace}[1]{}
\newcommand{\XComment}[1]{}

\makeatletter
\newcommand{\DefMacro}{\@ifstar\@DefMacroAllowRedefine\@DefMacro}
\newcommand{\@DefMacro}[2]{\expandafter\newcommand\csname rmk-#1\endcsname{#2}}
\newcommand{\@DefMacroAllowRedefine}[2]{\expandafter\providecommand\csname rmk-#1\endcsname{} \expandafter\renewcommand\csname rmk-#1\endcsname{#2}}
\makeatother
\newcommand{\UseMacro}[1]{\csname rmk-#1\endcsname}

\newcommand{\MyPara}[1]{\noindent\textbf{#1}.}

\newcommand{\InputWithSpace}[1]{\bgroup\def\arraystretch{1.1}\input{#1}\egroup}
\newcommand{\Code}[1]{{\ifmmode{\mathtt{#1}}\else$\mathtt{#1}$\fi}}
\newcommand{\CodeIn}[1]{{\ifmmode{\mathtt{#1}}\else$\mathtt{#1}$\fi}}

\newcolumntype{R}[1]{>{\RaggedLeft\arraybackslash}p{#1}}
\newcolumntype{L}[1]{>{\RaggedRight\arraybackslash}p{#1}}

\newcommand{\sunion}{\cup} %

\definecolor{gray}{RGB}{211,211,211}
\newcommand{\jbasicstyle}{\small\sffamily} %

\newcommand{\jnumberstyle}{\scriptsize}

\lstdefinelanguage{pseudo}
{
morekeywords={},
keywordstyle=\bfseries,
lineskip=-0.1em,
numbers=left, %
numberstyle=\jnumberstyle,
numbersep=4pt,
basicstyle=\jbasicstyle,
breaklines=true,
breakautoindent=true,
tabsize=2,
columns=fullflexible,
morecomment=*[l][\textsl]{//},
mathescape=true,
xleftmargin=10pt,
}

\lstdefinelanguage{todo-comment}
{
morekeywords={},
keywordstyle=\bfseries,
lineskip=-0.1em,
numbers=none,
basicstyle=\jbasicstyle,
breaklines=true,
breakautoindent=true,
tabsize=2,
columns=fullflexible,
morecomment=*[l][\textsl]{//},
mathescape=true,
xleftmargin=-10pt,
}

\lstdefinelanguage{java-pretty}
{
language=java,
numbers=left,
basicstyle=\scriptsize\ttfamily,
numberstyle=\scriptsize,
breaklines=true,
columns=fullflexible,
xleftmargin=16pt,
showstringspaces=false,
}

\lstdefinelanguage{java-pretty-no-number}
{
language=java,
numbers=none,
basicstyle=\scriptsize\ttfamily,
numberstyle=\scriptsize,
breaklines=true,
columns=fullflexible,
xleftmargin=16pt,
showstringspaces=false,
}

\newcommand\circled[1]{%
\tikz[baseline=(X.base)]
\node (X) [draw, shape=circle, inner sep=0] {\strut #1};}

\newcommand\Prepo[1]{\texttt{#1}\xspace}
\newcommand\Pversion[1]{\texttt{#1}\xspace}
\newcommand{\rqanswer}[2]{\begin{tcolorbox}[notitle,boxrule=0pt,boxsep=2pt,before skip=5pt,after skip=5pt,colback=gray!40,colframe=gray]\MyPara{Answer to #1} #2\end{tcolorbox}}

\newcommand{\ijacocoTool}{\textsc{iJaCoCo}\xspace}
\newcommand{\Title}{Efficient Incremental Code Coverage Analysis for \mbox{Regression Test Suites}}
\newcommand{\ijacocoURL}{https://github.com/uw-swag/ijacoco}

\newcommand{\jacoco}{JaCoCo\xspace}
\newcommand{\bjacoco}{\jacoco\xspace}
\newcommand{\ekstazi}{Ekstazi\xspace}
\newcommand{\retestall}{RetestAll\xspace}
\newcommand{\java}{Java\xspace}
\newcommand{\maven}{Maven\xspace}
\newcommand{\surefire}{Surefire\xspace}
\newcommand{\github}{GitHub\xspace}
\newcommand{\python}{Python\xspace}
\newcommand{\repo}{repository\xspace}
\newcommand{\repos}{repositories\xspace}
\newcommand{\Repo}{Repository\xspace}
\newcommand{\Repos}{Repositories\xspace}
\newcommand{\codecov}{code coverage\xspace}
\newcommand{\Codecov}{Code coverage\xspace}
\newcommand{\codecovanalysis}{code coverage analysis\xspace}
\newcommand{\Codecovanalysis}{Code coverage analysis\xspace}
\newcommand{\icodecovanalysis}{incremental code coverage analysis\xspace} %
\newcommand{\Icodecovanalysis}{Incremental code coverage analysis\xspace}

\newcommand{\revision}{version\xspace}
\newcommand{\revisions}{versions\xspace}
\newcommand{\analysis}{analysis\xspace}
\newcommand{\Analysis}{Analysis\xspace}
\newcommand{\analysisphase}{\analysis phase\xspace}
\newcommand{\execution}{execution\xspace}
\newcommand{\Execution}{Execution\xspace}
\newcommand{\executionphase}{\execution phase\xspace}
\newcommand{\collection}{collection\xspace}
\newcommand{\Collection}{Collection\xspace}
\newcommand{\collectionphase}{\collection phase\xspace}

\newcommand{\probe}{probe\xspace}
\newcommand{\probes}{probes\xspace}
\newcommand{\covdata}{coverage data\xspace} %
\newcommand{\codeelem}{code element\xspace}
\newcommand{\codeelems}{code elements\xspace}
\newcommand{\changeset}{changeset\xspace}
\newcommand{\nontest}{non-test\xspace}
\newcommand{\test}{test\xspace}
\newcommand{\tests}{tests\xspace}
\newcommand{\Tests}{Tests\xspace}
\newcommand{\testsuite}{test suite\xspace}
\newcommand{\testsuites}{test suites\xspace}
\newcommand{\testcases}{\tests{}\xspace} %
\newcommand{\codebase}{codebase\xspace}

\newcommand{\aTest}{\ensuremath{t}\xspace}

\newcommand{\aTestSet}{\ensuremath{\mathcal{T}}\xspace}
\newcommand{\aTestSetOld}{\ensuremath{\aTestSet}\xspace}
\newcommand{\aTestSetNew}{\ensuremath{\aTestSet'}\xspace}

\newcommand{\aTestSetNewSelectedRTS}{\ensuremath{\aTestSet'_{\mathtt{rts}}}\xspace}
\newcommand{\aTestSetNewSelectedI}{\ensuremath{\aTestSet'_{\mathtt{sel}}}\xspace}
\newcommand{\aProbe}{\ensuremath{p}\xspace}
\newcommand{\aProbeSet}{\ensuremath{\mathcal{P}}\xspace}

\newcommand{\aCoverageNew}{\ensuremath{cov'}\xspace}
\newcommand{\aDepGraph}{\ensuremath{\mathcal{G}}\xspace}
\newcommand{\aDepGraphOld}{\ensuremath{\aDepGraph}\xspace}
\newcommand{\aDepGraphNew}{\ensuremath{\aDepGraph'}\xspace}
\newcommand{\aDepGraphDiff}{\ensuremath{\aDepGraph_{\Delta}}\xspace}
\newcommand{\aClass}{\ensuremath{c}\xspace}
\newcommand{\aClassSet}{\ensuremath{\mathcal{C}}\xspace}
\newcommand{\aClassSetOld}{\ensuremath{\aClassSet}\xspace}
\newcommand{\aClassSetNew}{\ensuremath{\aClassSet'}\xspace}
\newcommand{\aClassSetChanged}{\ensuremath{\aClassSet_{\Delta}}\xspace}
\newcommand{\aClassSetCovInvalid}{\ensuremath{\aClassSet_{\mathtt{upd}}}\xspace}
\newcommand{\aCovData}{\ensuremath{\mathcal{D}}\xspace}
\newcommand{\aCovDataOld}{\ensuremath{\aCovData}\xspace}
\newcommand{\aCovDataNew}{\ensuremath{\aCovData'}\xspace}
\newcommand{\aCovDataDiff}{\ensuremath{\aCovData_{\Delta}}\xspace}
\newcommand{\aElem}{\ensuremath{e}\xspace}
\newcommand{\aElemSet}{\ensuremath{\mathcal{E}}\xspace}
\newcommand{\aElemSetCovered}{\ensuremath{\aElemSet_{\mathtt{cov}}}\xspace}
\newcommand{\aElemSetCoveredNew}{\ensuremath{\aElemSetCovered'}\xspace}
\newcommand{\aRTS}{\ensuremath{\mathtt{RTS}}\xspace}

\DefMacro{TH-sum}{$\Sigma$\xspace}
\DefMacro{TH-min}{min\xspace}
\DefMacro{TH-max}{max\xspace}
\DefMacro{TH-avg}{avg\xspace}
\DefMacro{TH-na}{-\xspace}

\DefMacro{TH-diff}{diff\xspace}
\DefMacro{TH-std}{std\xspace}
\DefMacro{TH-exact-same}{\makecell[c]{exact\\same}\xspace}
\DefMacro{THx-exact-same}{exact same\xspace}
\DefMacro{TH-no-stat-sign-diff}{\makecell[c]{no stat\\sign diff}\xspace}
\DefMacro{THx-no-stat-sign-diff}{no stat sign diff\xspace}
\DefMacro{TH-within-std}{\makecell[c]{within\\std}\xspace}
\DefMacro{THx-within-std}{within std\xspace}

\DefMacro{TH-project}{\Repo{}\xspace}
\DefMacro{TH-url}{URL\xspace}
\DefMacro{TH-head-revision}{\makecell[c]{First\\Ver.}\xspace}
\DefMacro{TH-num-revision}{\#Ver.\xspace}
\DefMacro{TH-project-size}{Code Base Size\xspace}
\DefMacro{TH-project-test}{Test Suite\xspace}
\DefMacro{TH-num-file}{\#File\xspace}
\DefMacro{TH-loc}{LOC\xspace}
\DefMacro{TH-num-test-class}{\#Class\xspace}
\DefMacro{TH-num-test-method}{\#Method\xspace}
\DefMacro{TH-test-time}{Time [s]\xspace}

\DefMacro{TH-selected-test-ratio}{Selection Rate [\%]\xspace}
\DefMacro{TH-speedup}{Speedup\xspace}
\DefMacro{TH-coverage}{Coverage [\%]\xspace}
\DefMacro{TH-correctness}{Correctness \xspace}

\DefMacro{TH-phase-time}{\ijacocoTool Phase Time [s] \xspace}
\DefMacro{TH-compile-time}{Compilation\xspace}
\DefMacro{TH-analysis-time}{\Analysis\xspace}
\DefMacro{TH-execution-collection-time}{ \makecell[c]{\Execution+\\\Collection} \xspace}
\DefMacro{TH-report-time}{Report\xspace}

\DefMacro{res-apache-commons-lang-retestall-num-files}{334}
\DefMacro{res-apache-commons-lang-retestall-loc}{77,789}
\DefMacro{res-apache-commons-lang-retestall-num-class}{148}
\DefMacro{res-apache-commons-lang-retestall-num-method}{4,108}
\DefMacro{res-apache-commons-lang-retestall-time}{27.15}
\DefMacro{res-apache-commons-lang-retestall-total-time}{1,384.52}
\DefMacro{res-apache-commons-io-retestall-num-files}{266}
\DefMacro{res-apache-commons-io-retestall-loc}{31,534}
\DefMacro{res-apache-commons-io-retestall-num-class}{112}
\DefMacro{res-apache-commons-io-retestall-num-method}{1,404}
\DefMacro{res-apache-commons-io-retestall-time}{47.42}
\DefMacro{res-apache-commons-io-retestall-total-time}{2,418.42}
\DefMacro{res-apache-commons-dbcp-retestall-num-files}{119}
\DefMacro{res-apache-commons-dbcp-retestall-loc}{29,895}
\DefMacro{res-apache-commons-dbcp-retestall-num-class}{42}
\DefMacro{res-apache-commons-dbcp-retestall-num-method}{1,309}
\DefMacro{res-apache-commons-dbcp-retestall-time}{107.80}
\DefMacro{res-apache-commons-dbcp-retestall-total-time}{5,498.03}
\DefMacro{res-apache-commons-net-retestall-num-files}{273}
\DefMacro{res-apache-commons-net-retestall-loc}{28,269}
\DefMacro{res-apache-commons-net-retestall-num-class}{44}
\DefMacro{res-apache-commons-net-retestall-num-method}{302}
\DefMacro{res-apache-commons-net-retestall-time}{76.90}
\DefMacro{res-apache-commons-net-retestall-total-time}{3,921.77}
\DefMacro{res-apache-commons-math-retestall-num-files}{1,453}
\DefMacro{res-apache-commons-math-retestall-loc}{189,786}
\DefMacro{res-apache-commons-math-retestall-num-class}{467}
\DefMacro{res-apache-commons-math-retestall-num-method}{6,004}
\DefMacro{res-apache-commons-math-retestall-time}{98.81}
\DefMacro{res-apache-commons-math-retestall-total-time}{5,039.51}
\DefMacro{res-davidmoten-rxjava-extras-retestall-num-files}{127}
\DefMacro{res-davidmoten-rxjava-extras-retestall-loc}{13,165}
\DefMacro{res-davidmoten-rxjava-extras-retestall-num-class}{48}
\DefMacro{res-davidmoten-rxjava-extras-retestall-num-method}{701}
\DefMacro{res-davidmoten-rxjava-extras-retestall-time}{93.91}
\DefMacro{res-davidmoten-rxjava-extras-retestall-total-time}{4,789.44}
\DefMacro{res-sonyxperiadev-gerrit-events-retestall-num-files}{107}
\DefMacro{res-sonyxperiadev-gerrit-events-retestall-loc}{7,389}
\DefMacro{res-sonyxperiadev-gerrit-events-retestall-num-class}{23}
\DefMacro{res-sonyxperiadev-gerrit-events-retestall-num-method}{121}
\DefMacro{res-sonyxperiadev-gerrit-events-retestall-time}{34.59}
\DefMacro{res-sonyxperiadev-gerrit-events-retestall-total-time}{1,764.25}
\DefMacro{res-apache-commons-collections-retestall-num-files}{541}
\DefMacro{res-apache-commons-collections-retestall-loc}{63,287}
\DefMacro{res-apache-commons-collections-retestall-num-class}{170}
\DefMacro{res-apache-commons-collections-retestall-num-method}{24,924}
\DefMacro{res-apache-commons-collections-retestall-time}{25.99}
\DefMacro{res-apache-commons-collections-retestall-total-time}{1,325.52}
\DefMacro{res-apache-commons-imaging-retestall-num-files}{504}
\DefMacro{res-apache-commons-imaging-retestall-loc}{38,797}
\DefMacro{res-apache-commons-imaging-retestall-num-class}{115}
\DefMacro{res-apache-commons-imaging-retestall-num-method}{591}
\DefMacro{res-apache-commons-imaging-retestall-time}{38.00}
\DefMacro{res-apache-commons-imaging-retestall-total-time}{1,937.97}
\DefMacro{res-apache-commons-codec-retestall-num-files}{139}
\DefMacro{res-apache-commons-codec-retestall-loc}{23,395}
\DefMacro{res-apache-commons-codec-retestall-num-class}{60}
\DefMacro{res-apache-commons-codec-retestall-num-method}{1,133}
\DefMacro{res-apache-commons-codec-retestall-time}{43.95}
\DefMacro{res-apache-commons-codec-retestall-total-time}{2,241.52}
\DefMacro{res-apache-commons-configuration-retestall-num-files}{467}
\DefMacro{res-apache-commons-configuration-retestall-loc}{68,019}
\DefMacro{res-apache-commons-configuration-retestall-num-class}{169}
\DefMacro{res-apache-commons-configuration-retestall-num-method}{2,832}
\DefMacro{res-apache-commons-configuration-retestall-time}{30.67}
\DefMacro{res-apache-commons-configuration-retestall-total-time}{1,564.07}
\DefMacro{res-asterisk-java-asterisk-java-retestall-num-files}{819}
\DefMacro{res-asterisk-java-asterisk-java-retestall-loc}{57,851}
\DefMacro{res-asterisk-java-asterisk-java-retestall-num-class}{46}
\DefMacro{res-asterisk-java-asterisk-java-retestall-num-method}{252}
\DefMacro{res-asterisk-java-asterisk-java-retestall-time}{32.56}
\DefMacro{res-asterisk-java-asterisk-java-retestall-total-time}{1,660.31}
\DefMacro{res-apache-commons-compress-retestall-num-files}{367}
\DefMacro{res-apache-commons-compress-retestall-loc}{48,300}
\DefMacro{res-apache-commons-compress-retestall-num-class}{142}
\DefMacro{res-apache-commons-compress-retestall-num-method}{1,180}
\DefMacro{res-apache-commons-compress-retestall-time}{51.53}
\DefMacro{res-apache-commons-compress-retestall-total-time}{2,628.12}
\DefMacro{res-alibaba-fastjson-retestall-num-files}{2,950}
\DefMacro{res-alibaba-fastjson-retestall-loc}{178,633}
\DefMacro{res-alibaba-fastjson-retestall-num-class}{2,292}
\DefMacro{res-alibaba-fastjson-retestall-num-method}{4,889}
\DefMacro{res-alibaba-fastjson-retestall-time}{82.76}
\DefMacro{res-alibaba-fastjson-retestall-total-time}{4,220.88}
\DefMacro{res-lmdbjava-lmdbjava-retestall-num-files}{46}
\DefMacro{res-lmdbjava-lmdbjava-retestall-loc}{5,195}
\DefMacro{res-lmdbjava-lmdbjava-retestall-num-class}{12}
\DefMacro{res-lmdbjava-lmdbjava-retestall-num-method}{147}
\DefMacro{res-lmdbjava-lmdbjava-retestall-time}{61.14}
\DefMacro{res-lmdbjava-lmdbjava-retestall-total-time}{3,117.97}
\DefMacro{res-brettwooldridge-HikariCP-retestall-num-files}{95}
\DefMacro{res-brettwooldridge-HikariCP-retestall-loc}{12,673}
\DefMacro{res-brettwooldridge-HikariCP-retestall-num-class}{38}
\DefMacro{res-brettwooldridge-HikariCP-retestall-num-method}{155}
\DefMacro{res-brettwooldridge-HikariCP-retestall-time}{127.57}
\DefMacro{res-brettwooldridge-HikariCP-retestall-total-time}{6,506.06}
\DefMacro{res-bullhorn-sdk-rest-retestall-num-files}{819}
\DefMacro{res-bullhorn-sdk-rest-retestall-loc}{58,586}
\DefMacro{res-bullhorn-sdk-rest-retestall-num-class}{24}
\DefMacro{res-bullhorn-sdk-rest-retestall-num-method}{366}
\DefMacro{res-bullhorn-sdk-rest-retestall-time}{307.53}
\DefMacro{res-bullhorn-sdk-rest-retestall-total-time}{15,684.03}
\DefMacro{res-tabulapdf-tabula-java-retestall-num-files}{52}
\DefMacro{res-tabulapdf-tabula-java-retestall-loc}{6,641}
\DefMacro{res-tabulapdf-tabula-java-retestall-num-class}{15}
\DefMacro{res-tabulapdf-tabula-java-retestall-num-method}{204}
\DefMacro{res-tabulapdf-tabula-java-retestall-time}{86.64}
\DefMacro{res-tabulapdf-tabula-java-retestall-total-time}{4,418.67}
\DefMacro{res-apache-commons-pool-retestall-num-files}{92}
\DefMacro{res-apache-commons-pool-retestall-loc}{14,473}
\DefMacro{res-apache-commons-pool-retestall-num-class}{22}
\DefMacro{res-apache-commons-pool-retestall-num-method}{293}
\DefMacro{res-apache-commons-pool-retestall-time}{336.66}
\DefMacro{res-apache-commons-pool-retestall-total-time}{17,169.61}
\DefMacro{res-apache-commons-beanutils-retestall-num-files}{255}
\DefMacro{res-apache-commons-beanutils-retestall-loc}{32,297}
\DefMacro{res-apache-commons-beanutils-retestall-num-class}{100}
\DefMacro{res-apache-commons-beanutils-retestall-num-method}{1,302}
\DefMacro{res-apache-commons-beanutils-retestall-time}{9.95}
\DefMacro{res-apache-commons-beanutils-retestall-total-time}{507.50}
\DefMacro{res-finmath-finmath-lib-retestall-num-files}{1,410}
\DefMacro{res-finmath-finmath-lib-retestall-loc}{85,558}
\DefMacro{res-finmath-finmath-lib-retestall-num-class}{97}
\DefMacro{res-finmath-finmath-lib-retestall-num-method}{517}
\DefMacro{res-finmath-finmath-lib-retestall-time}{1,550.91}
\DefMacro{res-finmath-finmath-lib-retestall-total-time}{79,096.45}
\DefMacro{res-logic-ng-LogicNG-retestall-num-files}{350}
\DefMacro{res-logic-ng-LogicNG-retestall-loc}{40,066}
\DefMacro{res-logic-ng-LogicNG-retestall-num-class}{119}
\DefMacro{res-logic-ng-LogicNG-retestall-num-method}{976}
\DefMacro{res-logic-ng-LogicNG-retestall-time}{522.37}
\DefMacro{res-logic-ng-LogicNG-retestall-total-time}{26,640.95}
\DefMacro{res-apache-commons-lang-ekstazi-num-files}{334}
\DefMacro{res-apache-commons-lang-ekstazi-loc}{77,789}
\DefMacro{res-apache-commons-lang-ekstazi-num-class}{17}
\DefMacro{res-apache-commons-lang-ekstazi-num-method}{581}
\DefMacro{res-apache-commons-lang-ekstazi-time}{12.34}
\DefMacro{res-apache-commons-lang-ekstazi-total-time}{629.09}
\DefMacro{res-apache-commons-lang-ekstazi-selected-rate}{11.58}
\DefMacro{res-apache-commons-io-ekstazi-num-files}{266}
\DefMacro{res-apache-commons-io-ekstazi-loc}{31,534}
\DefMacro{res-apache-commons-io-ekstazi-num-class}{9}
\DefMacro{res-apache-commons-io-ekstazi-num-method}{150}
\DefMacro{res-apache-commons-io-ekstazi-time}{78.67}
\DefMacro{res-apache-commons-io-ekstazi-total-time}{4,012.30}
\DefMacro{res-apache-commons-io-ekstazi-selected-rate}{8.01}
\DefMacro{res-apache-commons-dbcp-ekstazi-num-files}{119}
\DefMacro{res-apache-commons-dbcp-ekstazi-loc}{29,895}
\DefMacro{res-apache-commons-dbcp-ekstazi-num-class}{7}
\DefMacro{res-apache-commons-dbcp-ekstazi-num-method}{257}
\DefMacro{res-apache-commons-dbcp-ekstazi-time}{39.52}
\DefMacro{res-apache-commons-dbcp-ekstazi-total-time}{2,015.29}
\DefMacro{res-apache-commons-dbcp-ekstazi-selected-rate}{18.59}
\DefMacro{res-apache-commons-net-ekstazi-num-files}{273}
\DefMacro{res-apache-commons-net-ekstazi-loc}{28,269}
\DefMacro{res-apache-commons-net-ekstazi-num-class}{4}
\DefMacro{res-apache-commons-net-ekstazi-num-method}{30}
\DefMacro{res-apache-commons-net-ekstazi-time}{17.29}
\DefMacro{res-apache-commons-net-ekstazi-total-time}{881.79}
\DefMacro{res-apache-commons-net-ekstazi-selected-rate}{9.40}
\DefMacro{res-apache-commons-math-ekstazi-num-files}{1,453}
\DefMacro{res-apache-commons-math-ekstazi-loc}{189,786}
\DefMacro{res-apache-commons-math-ekstazi-num-class}{45}
\DefMacro{res-apache-commons-math-ekstazi-num-method}{737}
\DefMacro{res-apache-commons-math-ekstazi-time}{32.78}
\DefMacro{res-apache-commons-math-ekstazi-total-time}{1,671.68}
\DefMacro{res-apache-commons-math-ekstazi-selected-rate}{9.76}
\DefMacro{res-davidmoten-rxjava-extras-ekstazi-num-files}{127}
\DefMacro{res-davidmoten-rxjava-extras-ekstazi-loc}{13,165}
\DefMacro{res-davidmoten-rxjava-extras-ekstazi-num-class}{4}
\DefMacro{res-davidmoten-rxjava-extras-ekstazi-num-method}{67}
\DefMacro{res-davidmoten-rxjava-extras-ekstazi-time}{19.22}
\DefMacro{res-davidmoten-rxjava-extras-ekstazi-total-time}{980.02}
\DefMacro{res-davidmoten-rxjava-extras-ekstazi-selected-rate}{8.77}
\DefMacro{res-sonyxperiadev-gerrit-events-ekstazi-num-files}{107}
\DefMacro{res-sonyxperiadev-gerrit-events-ekstazi-loc}{7,389}
\DefMacro{res-sonyxperiadev-gerrit-events-ekstazi-num-class}{4}
\DefMacro{res-sonyxperiadev-gerrit-events-ekstazi-num-method}{30}
\DefMacro{res-sonyxperiadev-gerrit-events-ekstazi-time}{34.44}
\DefMacro{res-sonyxperiadev-gerrit-events-ekstazi-total-time}{1,756.44}
\DefMacro{res-sonyxperiadev-gerrit-events-ekstazi-selected-rate}{19.51}
\DefMacro{res-apache-commons-collections-ekstazi-num-files}{541}
\DefMacro{res-apache-commons-collections-ekstazi-loc}{63,287}
\DefMacro{res-apache-commons-collections-ekstazi-num-class}{12}
\DefMacro{res-apache-commons-collections-ekstazi-num-method}{3,231}
\DefMacro{res-apache-commons-collections-ekstazi-time}{12.80}
\DefMacro{res-apache-commons-collections-ekstazi-total-time}{653.05}
\DefMacro{res-apache-commons-collections-ekstazi-selected-rate}{7.53}
\DefMacro{res-apache-commons-imaging-ekstazi-num-files}{504}
\DefMacro{res-apache-commons-imaging-ekstazi-loc}{38,797}
\DefMacro{res-apache-commons-imaging-ekstazi-num-class}{25}
\DefMacro{res-apache-commons-imaging-ekstazi-num-method}{199}
\DefMacro{res-apache-commons-imaging-ekstazi-time}{21.72}
\DefMacro{res-apache-commons-imaging-ekstazi-total-time}{1,107.63}
\DefMacro{res-apache-commons-imaging-ekstazi-selected-rate}{21.97}
\DefMacro{res-apache-commons-codec-ekstazi-num-files}{139}
\DefMacro{res-apache-commons-codec-ekstazi-loc}{23,395}
\DefMacro{res-apache-commons-codec-ekstazi-num-class}{3}
\DefMacro{res-apache-commons-codec-ekstazi-num-method}{97}
\DefMacro{res-apache-commons-codec-ekstazi-time}{12.38}
\DefMacro{res-apache-commons-codec-ekstazi-total-time}{631.21}
\DefMacro{res-apache-commons-codec-ekstazi-selected-rate}{6.34}
\DefMacro{res-apache-commons-configuration-ekstazi-num-files}{467}
\DefMacro{res-apache-commons-configuration-ekstazi-loc}{68,019}
\DefMacro{res-apache-commons-configuration-ekstazi-num-class}{30}
\DefMacro{res-apache-commons-configuration-ekstazi-num-method}{633}
\DefMacro{res-apache-commons-configuration-ekstazi-time}{21.03}
\DefMacro{res-apache-commons-configuration-ekstazi-total-time}{1,072.38}
\DefMacro{res-apache-commons-configuration-ekstazi-selected-rate}{17.89}
\DefMacro{res-asterisk-java-asterisk-java-ekstazi-num-files}{819}
\DefMacro{res-asterisk-java-asterisk-java-ekstazi-loc}{57,851}
\DefMacro{res-asterisk-java-asterisk-java-ekstazi-num-class}{2}
\DefMacro{res-asterisk-java-asterisk-java-ekstazi-num-method}{26}
\DefMacro{res-asterisk-java-asterisk-java-ekstazi-time}{11.04}
\DefMacro{res-asterisk-java-asterisk-java-ekstazi-total-time}{563.02}
\DefMacro{res-asterisk-java-asterisk-java-ekstazi-selected-rate}{5.97}
\DefMacro{res-apache-commons-compress-ekstazi-num-files}{367}
\DefMacro{res-apache-commons-compress-ekstazi-loc}{48,300}
\DefMacro{res-apache-commons-compress-ekstazi-num-class}{28}
\DefMacro{res-apache-commons-compress-ekstazi-num-method}{294}
\DefMacro{res-apache-commons-compress-ekstazi-time}{20.88}
\DefMacro{res-apache-commons-compress-ekstazi-total-time}{1,065.04}
\DefMacro{res-apache-commons-compress-ekstazi-selected-rate}{20.31}
\DefMacro{res-alibaba-fastjson-ekstazi-num-files}{2,950}
\DefMacro{res-alibaba-fastjson-ekstazi-loc}{178,633}
\DefMacro{res-alibaba-fastjson-ekstazi-num-class}{655}
\DefMacro{res-alibaba-fastjson-ekstazi-num-method}{1,392}
\DefMacro{res-alibaba-fastjson-ekstazi-time}{57.59}
\DefMacro{res-alibaba-fastjson-ekstazi-total-time}{2,937.18}
\DefMacro{res-alibaba-fastjson-ekstazi-selected-rate}{28.59}
\DefMacro{res-lmdbjava-lmdbjava-ekstazi-num-files}{46}
\DefMacro{res-lmdbjava-lmdbjava-ekstazi-loc}{5,195}
\DefMacro{res-lmdbjava-lmdbjava-ekstazi-num-class}{3}
\DefMacro{res-lmdbjava-lmdbjava-ekstazi-num-method}{49}
\DefMacro{res-lmdbjava-lmdbjava-ekstazi-time}{41.63}
\DefMacro{res-lmdbjava-lmdbjava-ekstazi-total-time}{2,122.99}
\DefMacro{res-lmdbjava-lmdbjava-ekstazi-selected-rate}{28.10}
\DefMacro{res-brettwooldridge-HikariCP-ekstazi-num-files}{94}
\DefMacro{res-brettwooldridge-HikariCP-ekstazi-loc}{12,606}
\DefMacro{res-brettwooldridge-HikariCP-ekstazi-num-class}{18}
\DefMacro{res-brettwooldridge-HikariCP-ekstazi-num-method}{80}
\DefMacro{res-brettwooldridge-HikariCP-ekstazi-time}{125.66}
\DefMacro{res-brettwooldridge-HikariCP-ekstazi-total-time}{6,408.81}
\DefMacro{res-brettwooldridge-HikariCP-ekstazi-selected-rate}{47.99}
\DefMacro{res-bullhorn-sdk-rest-ekstazi-num-files}{819}
\DefMacro{res-bullhorn-sdk-rest-ekstazi-loc}{58,586}
\DefMacro{res-bullhorn-sdk-rest-ekstazi-num-class}{20}
\DefMacro{res-bullhorn-sdk-rest-ekstazi-num-method}{337}
\DefMacro{res-bullhorn-sdk-rest-ekstazi-time}{317.13}
\DefMacro{res-bullhorn-sdk-rest-ekstazi-total-time}{16,173.88}
\DefMacro{res-bullhorn-sdk-rest-ekstazi-selected-rate}{85.13}
\DefMacro{res-tabulapdf-tabula-java-ekstazi-num-files}{52}
\DefMacro{res-tabulapdf-tabula-java-ekstazi-loc}{6,641}
\DefMacro{res-tabulapdf-tabula-java-ekstazi-num-class}{5}
\DefMacro{res-tabulapdf-tabula-java-ekstazi-num-method}{92}
\DefMacro{res-tabulapdf-tabula-java-ekstazi-time}{55.48}
\DefMacro{res-tabulapdf-tabula-java-ekstazi-total-time}{2,829.27}
\DefMacro{res-tabulapdf-tabula-java-ekstazi-selected-rate}{33.72}
\DefMacro{res-apache-commons-pool-ekstazi-num-files}{92}
\DefMacro{res-apache-commons-pool-ekstazi-loc}{14,473}
\DefMacro{res-apache-commons-pool-ekstazi-num-class}{2}
\DefMacro{res-apache-commons-pool-ekstazi-num-method}{58}
\DefMacro{res-apache-commons-pool-ekstazi-time}{96.13}
\DefMacro{res-apache-commons-pool-ekstazi-total-time}{4,902.83}
\DefMacro{res-apache-commons-pool-ekstazi-selected-rate}{13.55}
\DefMacro{res-apache-commons-beanutils-ekstazi-num-files}{255}
\DefMacro{res-apache-commons-beanutils-ekstazi-loc}{32,297}
\DefMacro{res-apache-commons-beanutils-ekstazi-num-class}{12}
\DefMacro{res-apache-commons-beanutils-ekstazi-num-method}{202}
\DefMacro{res-apache-commons-beanutils-ekstazi-time}{75.65}
\DefMacro{res-apache-commons-beanutils-ekstazi-total-time}{3,858.11}
\DefMacro{res-apache-commons-beanutils-ekstazi-selected-rate}{12.47}
\DefMacro{res-finmath-finmath-lib-ekstazi-num-files}{1,410}
\DefMacro{res-finmath-finmath-lib-ekstazi-loc}{85,558}
\DefMacro{res-finmath-finmath-lib-ekstazi-num-class}{7}
\DefMacro{res-finmath-finmath-lib-ekstazi-num-method}{46}
\DefMacro{res-finmath-finmath-lib-ekstazi-time}{261.91}
\DefMacro{res-finmath-finmath-lib-ekstazi-total-time}{13,357.26}
\DefMacro{res-finmath-finmath-lib-ekstazi-selected-rate}{8.13}
\DefMacro{res-logic-ng-LogicNG-ekstazi-num-files}{350}
\DefMacro{res-logic-ng-LogicNG-ekstazi-loc}{40,066}
\DefMacro{res-logic-ng-LogicNG-ekstazi-num-class}{36}
\DefMacro{res-logic-ng-LogicNG-ekstazi-num-method}{352}
\DefMacro{res-logic-ng-LogicNG-ekstazi-time}{344.20}
\DefMacro{res-logic-ng-LogicNG-ekstazi-total-time}{17,554.41}
\DefMacro{res-logic-ng-LogicNG-ekstazi-selected-rate}{30.46}
\DefMacro{res-alibaba-fastjson-ijacoco-time}{87.46}
\DefMacro{res-alibaba-fastjson-bjacoco-time}{96.98}
\DefMacro{res-alibaba-fastjson-ibjacoco-speedup-time}{1.11}
\DefMacro{res-alibaba-fastjson-max-ijacoco-time}{162.08}
\DefMacro{res-alibaba-fastjson-min-ijacoco-time}{31.75}
\DefMacro{res-alibaba-fastjson-max-bjacoco-time}{100.19}
\DefMacro{res-alibaba-fastjson-min-bjacoco-time}{95.77}
\DefMacro{res-alibaba-fastjson-total-ijacoco-time}{4,460.33}
\DefMacro{res-alibaba-fastjson-total-bjacoco-time}{4,945.99}
\DefMacro{res-alibaba-fastjson-ratio-time}{0.90}
\DefMacro{res-alibaba-fastjson-ijacoco-max-coverage}{88.1}
\DefMacro{res-alibaba-fastjson-bjacoco-max-coverage}{88.0}
\DefMacro{res-alibaba-fastjson-ijacoco-min-coverage}{87.6}
\DefMacro{res-alibaba-fastjson-bjacoco-min-coverage}{87.8}
\DefMacro{res-alibaba-fastjson-ijacoco-selected-rate}{41.41}
\DefMacro{res-apache-commons-beanutils-ijacoco-time}{12.37}
\DefMacro{res-apache-commons-beanutils-bjacoco-time}{11.03}
\DefMacro{res-apache-commons-beanutils-ibjacoco-speedup-time}{0.89}
\DefMacro{res-apache-commons-beanutils-max-ijacoco-time}{13.20}
\DefMacro{res-apache-commons-beanutils-min-ijacoco-time}{8.47}
\DefMacro{res-apache-commons-beanutils-max-bjacoco-time}{11.49}
\DefMacro{res-apache-commons-beanutils-min-bjacoco-time}{10.66}
\DefMacro{res-apache-commons-beanutils-total-ijacoco-time}{630.75}
\DefMacro{res-apache-commons-beanutils-total-bjacoco-time}{562.28}
\DefMacro{res-apache-commons-beanutils-ratio-time}{1.12}
\DefMacro{res-apache-commons-beanutils-ijacoco-max-coverage}{72.5}
\DefMacro{res-apache-commons-beanutils-bjacoco-max-coverage}{72.6}
\DefMacro{res-apache-commons-beanutils-ijacoco-min-coverage}{71.3}
\DefMacro{res-apache-commons-beanutils-bjacoco-min-coverage}{71.3}
\DefMacro{res-apache-commons-beanutils-ijacoco-selected-rate}{85.09}
\DefMacro{res-apache-commons-codec-ijacoco-time}{18.96}
\DefMacro{res-apache-commons-codec-bjacoco-time}{45.17}
\DefMacro{res-apache-commons-codec-ibjacoco-speedup-time}{2.38}
\DefMacro{res-apache-commons-codec-max-ijacoco-time}{50.65}
\DefMacro{res-apache-commons-codec-min-ijacoco-time}{7.31}
\DefMacro{res-apache-commons-codec-max-bjacoco-time}{46.30}
\DefMacro{res-apache-commons-codec-min-bjacoco-time}{44.34}
\DefMacro{res-apache-commons-codec-total-ijacoco-time}{967.08}
\DefMacro{res-apache-commons-codec-total-bjacoco-time}{2,303.74}
\DefMacro{res-apache-commons-codec-ratio-time}{0.42}
\DefMacro{res-apache-commons-codec-ijacoco-max-coverage}{94.5}
\DefMacro{res-apache-commons-codec-bjacoco-max-coverage}{94.5}
\DefMacro{res-apache-commons-codec-ijacoco-min-coverage}{93.9}
\DefMacro{res-apache-commons-codec-bjacoco-min-coverage}{93.9}
\DefMacro{res-apache-commons-codec-ijacoco-selected-rate}{17.28}
\DefMacro{res-apache-commons-collections-ijacoco-time}{20.28}
\DefMacro{res-apache-commons-collections-bjacoco-time}{28.26}
\DefMacro{res-apache-commons-collections-ibjacoco-speedup-time}{1.39}
\DefMacro{res-apache-commons-collections-max-ijacoco-time}{31.88}
\DefMacro{res-apache-commons-collections-min-ijacoco-time}{11.38}
\DefMacro{res-apache-commons-collections-max-bjacoco-time}{28.74}
\DefMacro{res-apache-commons-collections-min-bjacoco-time}{27.68}
\DefMacro{res-apache-commons-collections-total-ijacoco-time}{1,034.04}
\DefMacro{res-apache-commons-collections-total-bjacoco-time}{1,441.07}
\DefMacro{res-apache-commons-collections-ratio-time}{0.72}
\DefMacro{res-apache-commons-collections-ijacoco-max-coverage}{88.7}
\DefMacro{res-apache-commons-collections-bjacoco-max-coverage}{88.7}
\DefMacro{res-apache-commons-collections-ijacoco-min-coverage}{87.8}
\DefMacro{res-apache-commons-collections-bjacoco-min-coverage}{87.8}
\DefMacro{res-apache-commons-collections-ijacoco-selected-rate}{34.60}
\DefMacro{res-apache-commons-compress-ijacoco-time}{46.39}
\DefMacro{res-apache-commons-compress-bjacoco-time}{65.93}
\DefMacro{res-apache-commons-compress-ibjacoco-speedup-time}{1.42}
\DefMacro{res-apache-commons-compress-max-ijacoco-time}{91.66}
\DefMacro{res-apache-commons-compress-min-ijacoco-time}{10.64}
\DefMacro{res-apache-commons-compress-max-bjacoco-time}{87.39}
\DefMacro{res-apache-commons-compress-min-bjacoco-time}{25.94}
\DefMacro{res-apache-commons-compress-total-ijacoco-time}{2,366.01}
\DefMacro{res-apache-commons-compress-total-bjacoco-time}{3,362.41}
\DefMacro{res-apache-commons-compress-ratio-time}{0.70}
\DefMacro{res-apache-commons-compress-ijacoco-max-coverage}{87.3}
\DefMacro{res-apache-commons-compress-bjacoco-max-coverage}{87.3}
\DefMacro{res-apache-commons-compress-ijacoco-min-coverage}{85.9}
\DefMacro{res-apache-commons-compress-bjacoco-min-coverage}{85.9}
\DefMacro{res-apache-commons-compress-ijacoco-selected-rate}{48.73}
\DefMacro{res-apache-commons-configuration-ijacoco-time}{31.16}
\DefMacro{res-apache-commons-configuration-bjacoco-time}{34.38}
\DefMacro{res-apache-commons-configuration-ibjacoco-speedup-time}{1.10}
\DefMacro{res-apache-commons-configuration-max-ijacoco-time}{43.51}
\DefMacro{res-apache-commons-configuration-min-ijacoco-time}{13.04}
\DefMacro{res-apache-commons-configuration-max-bjacoco-time}{34.96}
\DefMacro{res-apache-commons-configuration-min-bjacoco-time}{34.05}
\DefMacro{res-apache-commons-configuration-total-ijacoco-time}{1,588.92}
\DefMacro{res-apache-commons-configuration-total-bjacoco-time}{1,753.52}
\DefMacro{res-apache-commons-configuration-ratio-time}{0.91}
\DefMacro{res-apache-commons-configuration-ijacoco-max-coverage}{89.5}
\DefMacro{res-apache-commons-configuration-bjacoco-max-coverage}{89.5}
\DefMacro{res-apache-commons-configuration-ijacoco-min-coverage}{88.9}
\DefMacro{res-apache-commons-configuration-bjacoco-min-coverage}{88.9}
\DefMacro{res-apache-commons-configuration-ijacoco-selected-rate}{48.01}
\DefMacro{res-apache-commons-dbcp-ijacoco-time}{63.65}
\DefMacro{res-apache-commons-dbcp-bjacoco-time}{101.08}
\DefMacro{res-apache-commons-dbcp-ibjacoco-speedup-time}{1.59}
\DefMacro{res-apache-commons-dbcp-max-ijacoco-time}{319.47}
\DefMacro{res-apache-commons-dbcp-min-ijacoco-time}{8.78}
\DefMacro{res-apache-commons-dbcp-max-bjacoco-time}{104.91}
\DefMacro{res-apache-commons-dbcp-min-bjacoco-time}{97.96}
\DefMacro{res-apache-commons-dbcp-total-ijacoco-time}{3,246.08}
\DefMacro{res-apache-commons-dbcp-total-bjacoco-time}{5,155.07}
\DefMacro{res-apache-commons-dbcp-ratio-time}{0.63}
\DefMacro{res-apache-commons-dbcp-ijacoco-max-coverage}{64.1}
\DefMacro{res-apache-commons-dbcp-bjacoco-max-coverage}{64.1}
\DefMacro{res-apache-commons-dbcp-ijacoco-min-coverage}{40.7}
\DefMacro{res-apache-commons-dbcp-bjacoco-min-coverage}{40.6}
\DefMacro{res-apache-commons-dbcp-ijacoco-selected-rate}{31.62}
\DefMacro{res-apache-commons-imaging-ijacoco-time}{29.38}
\DefMacro{res-apache-commons-imaging-bjacoco-time}{41.88}
\DefMacro{res-apache-commons-imaging-ibjacoco-speedup-time}{1.43}
\DefMacro{res-apache-commons-imaging-max-ijacoco-time}{46.94}
\DefMacro{res-apache-commons-imaging-min-ijacoco-time}{9.83}
\DefMacro{res-apache-commons-imaging-max-bjacoco-time}{45.75}
\DefMacro{res-apache-commons-imaging-min-bjacoco-time}{40.66}
\DefMacro{res-apache-commons-imaging-total-ijacoco-time}{1,498.40}
\DefMacro{res-apache-commons-imaging-total-bjacoco-time}{2,135.98}
\DefMacro{res-apache-commons-imaging-ratio-time}{0.70}
\DefMacro{res-apache-commons-imaging-ijacoco-max-coverage}{74.8}
\DefMacro{res-apache-commons-imaging-bjacoco-max-coverage}{74.8}
\DefMacro{res-apache-commons-imaging-ijacoco-min-coverage}{73.5}
\DefMacro{res-apache-commons-imaging-bjacoco-min-coverage}{73.5}
\DefMacro{res-apache-commons-imaging-ijacoco-selected-rate}{32.70}
\DefMacro{res-apache-commons-io-ijacoco-time}{17.23}
\DefMacro{res-apache-commons-io-bjacoco-time}{49.19}
\DefMacro{res-apache-commons-io-ibjacoco-speedup-time}{2.85}
\DefMacro{res-apache-commons-io-max-ijacoco-time}{52.79}
\DefMacro{res-apache-commons-io-min-ijacoco-time}{9.00}
\DefMacro{res-apache-commons-io-max-bjacoco-time}{49.87}
\DefMacro{res-apache-commons-io-min-bjacoco-time}{47.89}
\DefMacro{res-apache-commons-io-total-ijacoco-time}{878.83}
\DefMacro{res-apache-commons-io-total-bjacoco-time}{2,508.59}
\DefMacro{res-apache-commons-io-ratio-time}{0.35}
\DefMacro{res-apache-commons-io-ijacoco-max-coverage}{88.9}
\DefMacro{res-apache-commons-io-bjacoco-max-coverage}{88.9}
\DefMacro{res-apache-commons-io-ijacoco-min-coverage}{88.1}
\DefMacro{res-apache-commons-io-bjacoco-min-coverage}{88.1}
\DefMacro{res-apache-commons-io-ijacoco-selected-rate}{11.56}
\DefMacro{res-apache-commons-lang-ijacoco-time}{17.08}
\DefMacro{res-apache-commons-lang-bjacoco-time}{28.72}
\DefMacro{res-apache-commons-lang-ibjacoco-speedup-time}{1.68}
\DefMacro{res-apache-commons-lang-max-ijacoco-time}{33.63}
\DefMacro{res-apache-commons-lang-min-ijacoco-time}{11.63}
\DefMacro{res-apache-commons-lang-max-bjacoco-time}{29.77}
\DefMacro{res-apache-commons-lang-min-bjacoco-time}{28.24}
\DefMacro{res-apache-commons-lang-total-ijacoco-time}{871.31}
\DefMacro{res-apache-commons-lang-total-bjacoco-time}{1,464.68}
\DefMacro{res-apache-commons-lang-ratio-time}{0.59}
\DefMacro{res-apache-commons-lang-ijacoco-max-coverage}{95.3}
\DefMacro{res-apache-commons-lang-bjacoco-max-coverage}{95.3}
\DefMacro{res-apache-commons-lang-ijacoco-min-coverage}{95.2}
\DefMacro{res-apache-commons-lang-bjacoco-min-coverage}{95.2}
\DefMacro{res-apache-commons-lang-ijacoco-selected-rate}{22.47}
\DefMacro{res-apache-commons-math-ijacoco-time}{175.71}
\DefMacro{res-apache-commons-math-bjacoco-time}{131.67}
\DefMacro{res-apache-commons-math-ibjacoco-speedup-time}{0.75}
\DefMacro{res-apache-commons-math-max-ijacoco-time}{203.33}
\DefMacro{res-apache-commons-math-min-ijacoco-time}{148.02}
\DefMacro{res-apache-commons-math-max-bjacoco-time}{153.07}
\DefMacro{res-apache-commons-math-min-bjacoco-time}{110.31}
\DefMacro{res-apache-commons-math-total-ijacoco-time}{8,961.26}
\DefMacro{res-apache-commons-math-total-bjacoco-time}{6,715.13}
\DefMacro{res-apache-commons-math-ratio-time}{1.33}
\DefMacro{res-apache-commons-math-ijacoco-max-coverage}{90.9}
\DefMacro{res-apache-commons-math-bjacoco-max-coverage}{90.9}
\DefMacro{res-apache-commons-math-ijacoco-min-coverage}{90.3}
\DefMacro{res-apache-commons-math-bjacoco-min-coverage}{90.3}
\DefMacro{res-apache-commons-math-ijacoco-selected-rate}{96.39}
\DefMacro{res-apache-commons-net-ijacoco-time}{26.26}
\DefMacro{res-apache-commons-net-bjacoco-time}{78.57}
\DefMacro{res-apache-commons-net-ibjacoco-speedup-time}{2.99}
\DefMacro{res-apache-commons-net-max-ijacoco-time}{80.86}
\DefMacro{res-apache-commons-net-min-ijacoco-time}{7.83}
\DefMacro{res-apache-commons-net-max-bjacoco-time}{80.56}
\DefMacro{res-apache-commons-net-min-bjacoco-time}{77.50}
\DefMacro{res-apache-commons-net-total-ijacoco-time}{1,339.08}
\DefMacro{res-apache-commons-net-total-bjacoco-time}{4,007.25}
\DefMacro{res-apache-commons-net-ratio-time}{0.33}
\DefMacro{res-apache-commons-net-ijacoco-max-coverage}{35.1}
\DefMacro{res-apache-commons-net-bjacoco-max-coverage}{35.1}
\DefMacro{res-apache-commons-net-ijacoco-min-coverage}{34.4}
\DefMacro{res-apache-commons-net-bjacoco-min-coverage}{34.4}
\DefMacro{res-apache-commons-net-ijacoco-selected-rate}{18.00}
\DefMacro{res-apache-commons-pool-ijacoco-time}{153.39}
\DefMacro{res-apache-commons-pool-bjacoco-time}{349.86}
\DefMacro{res-apache-commons-pool-ibjacoco-speedup-time}{2.28}
\DefMacro{res-apache-commons-pool-max-ijacoco-time}{357.61}
\DefMacro{res-apache-commons-pool-min-ijacoco-time}{8.34}
\DefMacro{res-apache-commons-pool-max-bjacoco-time}{362.42}
\DefMacro{res-apache-commons-pool-min-bjacoco-time}{342.53}
\DefMacro{res-apache-commons-pool-total-ijacoco-time}{7,822.66}
\DefMacro{res-apache-commons-pool-total-bjacoco-time}{17,842.92}
\DefMacro{res-apache-commons-pool-ratio-time}{0.44}
\DefMacro{res-apache-commons-pool-ijacoco-max-coverage}{85.1}
\DefMacro{res-apache-commons-pool-bjacoco-max-coverage}{85.1}
\DefMacro{res-apache-commons-pool-ijacoco-min-coverage}{84.8}
\DefMacro{res-apache-commons-pool-bjacoco-min-coverage}{84.8}
\DefMacro{res-apache-commons-pool-ijacoco-selected-rate}{34.49}
\DefMacro{res-asterisk-java-asterisk-java-ijacoco-time}{12.80}
\DefMacro{res-asterisk-java-asterisk-java-bjacoco-time}{34.57}
\DefMacro{res-asterisk-java-asterisk-java-ibjacoco-speedup-time}{2.70}
\DefMacro{res-asterisk-java-asterisk-java-max-ijacoco-time}{37.51}
\DefMacro{res-asterisk-java-asterisk-java-min-ijacoco-time}{9.81}
\DefMacro{res-asterisk-java-asterisk-java-max-bjacoco-time}{35.07}
\DefMacro{res-asterisk-java-asterisk-java-min-bjacoco-time}{34.01}
\DefMacro{res-asterisk-java-asterisk-java-total-ijacoco-time}{652.89}
\DefMacro{res-asterisk-java-asterisk-java-total-bjacoco-time}{1,763.25}
\DefMacro{res-asterisk-java-asterisk-java-ratio-time}{0.37}
\DefMacro{res-asterisk-java-asterisk-java-ijacoco-max-coverage}{14.1}
\DefMacro{res-asterisk-java-asterisk-java-bjacoco-max-coverage}{14.1}
\DefMacro{res-asterisk-java-asterisk-java-ijacoco-min-coverage}{13.5}
\DefMacro{res-asterisk-java-asterisk-java-bjacoco-min-coverage}{13.5}
\DefMacro{res-asterisk-java-asterisk-java-ijacoco-selected-rate}{5.97}
\DefMacro{res-brettwooldridge-HikariCP-ijacoco-time}{100.06}
\DefMacro{res-brettwooldridge-HikariCP-bjacoco-time}{129.56}
\DefMacro{res-brettwooldridge-HikariCP-ibjacoco-speedup-time}{1.29}
\DefMacro{res-brettwooldridge-HikariCP-max-ijacoco-time}{281.75}
\DefMacro{res-brettwooldridge-HikariCP-min-ijacoco-time}{9.13}
\DefMacro{res-brettwooldridge-HikariCP-max-bjacoco-time}{282.57}
\DefMacro{res-brettwooldridge-HikariCP-min-bjacoco-time}{114.58}
\DefMacro{res-brettwooldridge-HikariCP-total-ijacoco-time}{5,103.12}
\DefMacro{res-brettwooldridge-HikariCP-total-bjacoco-time}{6,607.77}
\DefMacro{res-brettwooldridge-HikariCP-ratio-time}{0.77}
\DefMacro{res-brettwooldridge-HikariCP-ijacoco-max-coverage}{83.4}
\DefMacro{res-brettwooldridge-HikariCP-bjacoco-max-coverage}{83.3}
\DefMacro{res-brettwooldridge-HikariCP-ijacoco-min-coverage}{78.4}
\DefMacro{res-brettwooldridge-HikariCP-bjacoco-min-coverage}{78.4}
\DefMacro{res-brettwooldridge-HikariCP-ijacoco-selected-rate}{59.27}
\DefMacro{res-bullhorn-sdk-rest-ijacoco-time}{432.54}
\DefMacro{res-bullhorn-sdk-rest-bjacoco-time}{405.39}
\DefMacro{res-bullhorn-sdk-rest-ibjacoco-speedup-time}{0.94}
\DefMacro{res-bullhorn-sdk-rest-max-ijacoco-time}{540.06}
\DefMacro{res-bullhorn-sdk-rest-min-ijacoco-time}{15.75}
\DefMacro{res-bullhorn-sdk-rest-max-bjacoco-time}{473.32}
\DefMacro{res-bullhorn-sdk-rest-min-bjacoco-time}{325.31}
\DefMacro{res-bullhorn-sdk-rest-total-ijacoco-time}{22,059.67}
\DefMacro{res-bullhorn-sdk-rest-total-bjacoco-time}{20,675.11}
\DefMacro{res-bullhorn-sdk-rest-ratio-time}{1.07}
\DefMacro{res-bullhorn-sdk-rest-ijacoco-max-coverage}{66.3}
\DefMacro{res-bullhorn-sdk-rest-bjacoco-max-coverage}{66.3}
\DefMacro{res-bullhorn-sdk-rest-ijacoco-min-coverage}{62.8}
\DefMacro{res-bullhorn-sdk-rest-bjacoco-min-coverage}{62.8}
\DefMacro{res-bullhorn-sdk-rest-ijacoco-selected-rate}{87.17}
\DefMacro{res-davidmoten-rxjava-extras-ijacoco-time}{60.83}
\DefMacro{res-davidmoten-rxjava-extras-bjacoco-time}{102.53}
\DefMacro{res-davidmoten-rxjava-extras-ibjacoco-speedup-time}{1.69}
\DefMacro{res-davidmoten-rxjava-extras-max-ijacoco-time}{116.84}
\DefMacro{res-davidmoten-rxjava-extras-min-ijacoco-time}{6.42}
\DefMacro{res-davidmoten-rxjava-extras-max-bjacoco-time}{110.89}
\DefMacro{res-davidmoten-rxjava-extras-min-bjacoco-time}{97.38}
\DefMacro{res-davidmoten-rxjava-extras-total-ijacoco-time}{3,102.37}
\DefMacro{res-davidmoten-rxjava-extras-total-bjacoco-time}{5,228.95}
\DefMacro{res-davidmoten-rxjava-extras-ratio-time}{0.59}
\DefMacro{res-davidmoten-rxjava-extras-ijacoco-max-coverage}{69.3}
\DefMacro{res-davidmoten-rxjava-extras-bjacoco-max-coverage}{69.3}
\DefMacro{res-davidmoten-rxjava-extras-ijacoco-min-coverage}{67.9}
\DefMacro{res-davidmoten-rxjava-extras-bjacoco-min-coverage}{67.9}
\DefMacro{res-davidmoten-rxjava-extras-ijacoco-selected-rate}{47.06}
\DefMacro{res-finmath-finmath-lib-ijacoco-time}{487.13}
\DefMacro{res-finmath-finmath-lib-bjacoco-time}{3,992.06}
\DefMacro{res-finmath-finmath-lib-ibjacoco-speedup-time}{8.20}
\DefMacro{res-finmath-finmath-lib-max-ijacoco-time}{5,651.35}
\DefMacro{res-finmath-finmath-lib-min-ijacoco-time}{11.88}
\DefMacro{res-finmath-finmath-lib-max-bjacoco-time}{4,741.93}
\DefMacro{res-finmath-finmath-lib-min-bjacoco-time}{3,125.61}
\DefMacro{res-finmath-finmath-lib-total-ijacoco-time}{24,843.60}
\DefMacro{res-finmath-finmath-lib-total-bjacoco-time}{203,595.02}
\DefMacro{res-finmath-finmath-lib-ratio-time}{0.12}
\DefMacro{res-finmath-finmath-lib-ijacoco-max-coverage}{37.5}
\DefMacro{res-finmath-finmath-lib-bjacoco-max-coverage}{37.4}
\DefMacro{res-finmath-finmath-lib-ijacoco-min-coverage}{35.5}
\DefMacro{res-finmath-finmath-lib-bjacoco-min-coverage}{35.3}
\DefMacro{res-finmath-finmath-lib-ijacoco-selected-rate}{7.97}
\DefMacro{res-lmdbjava-lmdbjava-ijacoco-time}{59.16}
\DefMacro{res-lmdbjava-lmdbjava-bjacoco-time}{76.51}
\DefMacro{res-lmdbjava-lmdbjava-ibjacoco-speedup-time}{1.29}
\DefMacro{res-lmdbjava-lmdbjava-max-ijacoco-time}{88.43}
\DefMacro{res-lmdbjava-lmdbjava-min-ijacoco-time}{1.43}
\DefMacro{res-lmdbjava-lmdbjava-max-bjacoco-time}{85.38}
\DefMacro{res-lmdbjava-lmdbjava-min-bjacoco-time}{1.33}
\DefMacro{res-lmdbjava-lmdbjava-total-ijacoco-time}{3,017.30}
\DefMacro{res-lmdbjava-lmdbjava-total-bjacoco-time}{3,901.90}
\DefMacro{res-lmdbjava-lmdbjava-ratio-time}{0.77}
\DefMacro{res-lmdbjava-lmdbjava-ijacoco-max-coverage}{91.3}
\DefMacro{res-lmdbjava-lmdbjava-bjacoco-max-coverage}{91.3}
\DefMacro{res-lmdbjava-lmdbjava-ijacoco-min-coverage}{89.9}
\DefMacro{res-lmdbjava-lmdbjava-bjacoco-min-coverage}{89.9}
\DefMacro{res-lmdbjava-lmdbjava-ijacoco-selected-rate}{77.39}
\DefMacro{res-logic-ng-LogicNG-ijacoco-time}{797.87}
\DefMacro{res-logic-ng-LogicNG-bjacoco-time}{675.95}
\DefMacro{res-logic-ng-LogicNG-ibjacoco-speedup-time}{0.85}
\DefMacro{res-logic-ng-LogicNG-max-ijacoco-time}{954.33}
\DefMacro{res-logic-ng-LogicNG-min-ijacoco-time}{10.67}
\DefMacro{res-logic-ng-LogicNG-max-bjacoco-time}{750.20}
\DefMacro{res-logic-ng-LogicNG-min-bjacoco-time}{572.61}
\DefMacro{res-logic-ng-LogicNG-total-ijacoco-time}{40,691.25}
\DefMacro{res-logic-ng-LogicNG-total-bjacoco-time}{34,473.64}
\DefMacro{res-logic-ng-LogicNG-ratio-time}{1.18}
\DefMacro{res-logic-ng-LogicNG-ijacoco-max-coverage}{95.1}
\DefMacro{res-logic-ng-LogicNG-bjacoco-max-coverage}{95.1}
\DefMacro{res-logic-ng-LogicNG-ijacoco-min-coverage}{94.7}
\DefMacro{res-logic-ng-LogicNG-bjacoco-min-coverage}{94.7}
\DefMacro{res-logic-ng-LogicNG-ijacoco-selected-rate}{84.99}
\DefMacro{res-sonyxperiadev-gerrit-events-ijacoco-time}{31.71}
\DefMacro{res-sonyxperiadev-gerrit-events-bjacoco-time}{35.99}
\DefMacro{res-sonyxperiadev-gerrit-events-ibjacoco-speedup-time}{1.14}
\DefMacro{res-sonyxperiadev-gerrit-events-max-ijacoco-time}{52.70}
\DefMacro{res-sonyxperiadev-gerrit-events-min-ijacoco-time}{5.14}
\DefMacro{res-sonyxperiadev-gerrit-events-max-bjacoco-time}{45.77}
\DefMacro{res-sonyxperiadev-gerrit-events-min-bjacoco-time}{27.03}
\DefMacro{res-sonyxperiadev-gerrit-events-total-ijacoco-time}{1,617.05}
\DefMacro{res-sonyxperiadev-gerrit-events-total-bjacoco-time}{1,835.46}
\DefMacro{res-sonyxperiadev-gerrit-events-ratio-time}{0.88}
\DefMacro{res-sonyxperiadev-gerrit-events-ijacoco-max-coverage}{51.4}
\DefMacro{res-sonyxperiadev-gerrit-events-bjacoco-max-coverage}{51.4}
\DefMacro{res-sonyxperiadev-gerrit-events-ijacoco-min-coverage}{36.7}
\DefMacro{res-sonyxperiadev-gerrit-events-bjacoco-min-coverage}{36.7}
\DefMacro{res-sonyxperiadev-gerrit-events-ijacoco-selected-rate}{54.02}
\DefMacro{res-tabulapdf-tabula-java-ijacoco-time}{101.03}
\DefMacro{res-tabulapdf-tabula-java-bjacoco-time}{97.33}
\DefMacro{res-tabulapdf-tabula-java-ibjacoco-speedup-time}{0.96}
\DefMacro{res-tabulapdf-tabula-java-max-ijacoco-time}{105.79}
\DefMacro{res-tabulapdf-tabula-java-min-ijacoco-time}{98.89}
\DefMacro{res-tabulapdf-tabula-java-max-bjacoco-time}{101.05}
\DefMacro{res-tabulapdf-tabula-java-min-bjacoco-time}{95.32}
\DefMacro{res-tabulapdf-tabula-java-total-ijacoco-time}{5,152.66}
\DefMacro{res-tabulapdf-tabula-java-total-bjacoco-time}{4,963.84}
\DefMacro{res-tabulapdf-tabula-java-ratio-time}{1.04}
\DefMacro{res-tabulapdf-tabula-java-ijacoco-max-coverage}{83.7}
\DefMacro{res-tabulapdf-tabula-java-bjacoco-max-coverage}{83.7}
\DefMacro{res-tabulapdf-tabula-java-ijacoco-min-coverage}{82.2}
\DefMacro{res-tabulapdf-tabula-java-bjacoco-min-coverage}{82.2}
\DefMacro{res-tabulapdf-tabula-java-ijacoco-selected-rate}{95.38}
\DefMacro{res-apache-commons-lang-name}{commons-lang}
\DefMacro{res-apache-commons-lang-url}{https://github.com/apache/commons-lang.git}
\DefMacro{res-apache-commons-lang-head}{b41e9181}
\DefMacro{res-apache-commons-lang-num-ver}{51}
\DefMacro{res-apache-commons-lang-num-file}{333}
\DefMacro{res-apache-commons-lang-loc}{77,448}
\DefMacro{res-apache-commons-lang-num-class}{148}
\DefMacro{res-apache-commons-lang-num-method}{4,091}
\DefMacro{res-apache-commons-lang-time}{27.78}
\DefMacro{res-apache-commons-io-name}{commons-io}
\DefMacro{res-apache-commons-io-url}{https://github.com/apache/commons-io.git}
\DefMacro{res-apache-commons-io-head}{d3137782}
\DefMacro{res-apache-commons-io-num-ver}{51}
\DefMacro{res-apache-commons-io-num-file}{250}
\DefMacro{res-apache-commons-io-loc}{30,500}
\DefMacro{res-apache-commons-io-num-class}{106}
\DefMacro{res-apache-commons-io-num-method}{1,362}
\DefMacro{res-apache-commons-io-time}{46.51}
\DefMacro{res-apache-commons-dbcp-name}{commons-dbcp}
\DefMacro{res-apache-commons-dbcp-url}{https://github.com/apache/commons-dbcp.git}
\DefMacro{res-apache-commons-dbcp-head}{8e4a5652}
\DefMacro{res-apache-commons-dbcp-num-ver}{51}
\DefMacro{res-apache-commons-dbcp-num-file}{109}
\DefMacro{res-apache-commons-dbcp-loc}{23,111}
\DefMacro{res-apache-commons-dbcp-num-class}{32}
\DefMacro{res-apache-commons-dbcp-num-method}{580}
\DefMacro{res-apache-commons-dbcp-time}{102.36}
\DefMacro{res-apache-commons-net-name}{commons-net}
\DefMacro{res-apache-commons-net-url}{https://github.com/apache/commons-net.git}
\DefMacro{res-apache-commons-net-head}{f4bc1441}
\DefMacro{res-apache-commons-net-num-ver}{51}
\DefMacro{res-apache-commons-net-num-file}{273}
\DefMacro{res-apache-commons-net-loc}{28,246}
\DefMacro{res-apache-commons-net-num-class}{44}
\DefMacro{res-apache-commons-net-num-method}{301}
\DefMacro{res-apache-commons-net-time}{76.35}
\DefMacro{res-apache-commons-math-name}{commons-math}
\DefMacro{res-apache-commons-math-url}{https://github.com/apache/commons-math.git}
\DefMacro{res-apache-commons-math-head}{aeb21280}
\DefMacro{res-apache-commons-math-num-ver}{51}
\DefMacro{res-apache-commons-math-num-file}{1,488}
\DefMacro{res-apache-commons-math-loc}{191,967}
\DefMacro{res-apache-commons-math-num-class}{477}
\DefMacro{res-apache-commons-math-num-method}{6,557}
\DefMacro{res-apache-commons-math-time}{109.81}
\DefMacro{res-davidmoten-rxjava-extras-name}{rxjava-extras}
\DefMacro{res-davidmoten-rxjava-extras-url}{https://github.com/davidmoten/rxjava-extras.git}
\DefMacro{res-davidmoten-rxjava-extras-head}{11a083ed}
\DefMacro{res-davidmoten-rxjava-extras-num-ver}{51}
\DefMacro{res-davidmoten-rxjava-extras-num-file}{127}
\DefMacro{res-davidmoten-rxjava-extras-loc}{12,898}
\DefMacro{res-davidmoten-rxjava-extras-num-class}{48}
\DefMacro{res-davidmoten-rxjava-extras-num-method}{686}
\DefMacro{res-davidmoten-rxjava-extras-time}{94.13}
\DefMacro{res-sonyxperiadev-gerrit-events-name}{gerrit-events}
\DefMacro{res-sonyxperiadev-gerrit-events-url}{https://github.com/sonyxperiadev/gerrit-events.git}
\DefMacro{res-sonyxperiadev-gerrit-events-head}{5201a56f}
\DefMacro{res-sonyxperiadev-gerrit-events-num-ver}{51}
\DefMacro{res-sonyxperiadev-gerrit-events-num-file}{98}
\DefMacro{res-sonyxperiadev-gerrit-events-loc}{6,649}
\DefMacro{res-sonyxperiadev-gerrit-events-num-class}{21}
\DefMacro{res-sonyxperiadev-gerrit-events-num-method}{109}
\DefMacro{res-sonyxperiadev-gerrit-events-time}{26.44}
\DefMacro{res-apache-commons-collections-name}{commons-collections}
\DefMacro{res-apache-commons-collections-url}{https://github.com/apache/commons-collections.git}
\DefMacro{res-apache-commons-collections-head}{c46666c5}
\DefMacro{res-apache-commons-collections-num-ver}{51}
\DefMacro{res-apache-commons-collections-num-file}{536}
\DefMacro{res-apache-commons-collections-loc}{62,956}
\DefMacro{res-apache-commons-collections-num-class}{168}
\DefMacro{res-apache-commons-collections-num-method}{24,612}
\DefMacro{res-apache-commons-collections-time}{25.72}
\DefMacro{res-apache-commons-imaging-name}{commons-imaging}
\DefMacro{res-apache-commons-imaging-url}{https://github.com/apache/commons-imaging.git}
\DefMacro{res-apache-commons-imaging-head}{c5ca63fe}
\DefMacro{res-apache-commons-imaging-num-ver}{51}
\DefMacro{res-apache-commons-imaging-num-file}{490}
\DefMacro{res-apache-commons-imaging-loc}{38,664}
\DefMacro{res-apache-commons-imaging-num-class}{111}
\DefMacro{res-apache-commons-imaging-num-method}{573}
\DefMacro{res-apache-commons-imaging-time}{38.17}
\DefMacro{res-apache-commons-codec-name}{commons-codec}
\DefMacro{res-apache-commons-codec-url}{https://github.com/apache/commons-codec.git}
\DefMacro{res-apache-commons-codec-head}{3c212236}
\DefMacro{res-apache-commons-codec-num-ver}{51}
\DefMacro{res-apache-commons-codec-num-file}{133}
\DefMacro{res-apache-commons-codec-loc}{22,417}
\DefMacro{res-apache-commons-codec-num-class}{58}
\DefMacro{res-apache-commons-codec-num-method}{1,084}
\DefMacro{res-apache-commons-codec-time}{44.92}
\DefMacro{res-apache-commons-configuration-name}{commons-configuration}
\DefMacro{res-apache-commons-configuration-url}{https://github.com/apache/commons-configuration.git}
\DefMacro{res-apache-commons-configuration-head}{b84a0ef5}
\DefMacro{res-apache-commons-configuration-num-ver}{51}
\DefMacro{res-apache-commons-configuration-num-file}{467}
\DefMacro{res-apache-commons-configuration-loc}{69,625}
\DefMacro{res-apache-commons-configuration-num-class}{169}
\DefMacro{res-apache-commons-configuration-num-method}{2,831}
\DefMacro{res-apache-commons-configuration-time}{30.75}
\DefMacro{res-asterisk-java-asterisk-java-name}{asterisk-java}
\DefMacro{res-asterisk-java-asterisk-java-url}{https://github.com/asterisk-java/asterisk-java.git}
\DefMacro{res-asterisk-java-asterisk-java-head}{3576c01f}
\DefMacro{res-asterisk-java-asterisk-java-num-ver}{51}
\DefMacro{res-asterisk-java-asterisk-java-num-file}{814}
\DefMacro{res-asterisk-java-asterisk-java-loc}{57,208}
\DefMacro{res-asterisk-java-asterisk-java-num-class}{46}
\DefMacro{res-asterisk-java-asterisk-java-num-method}{252}
\DefMacro{res-asterisk-java-asterisk-java-time}{32.39}
\DefMacro{res-apache-commons-compress-name}{commons-compress}
\DefMacro{res-apache-commons-compress-url}{https://github.com/apache/commons-compress.git}
\DefMacro{res-apache-commons-compress-head}{2ed55677}
\DefMacro{res-apache-commons-compress-num-ver}{51}
\DefMacro{res-apache-commons-compress-num-file}{356}
\DefMacro{res-apache-commons-compress-loc}{44,911}
\DefMacro{res-apache-commons-compress-num-class}{137}
\DefMacro{res-apache-commons-compress-num-method}{1,065}
\DefMacro{res-apache-commons-compress-time}{24.19}
\DefMacro{res-alibaba-fastjson-name}{fastjson}
\DefMacro{res-alibaba-fastjson-url}{https://github.com/alibaba/fastjson.git}
\DefMacro{res-alibaba-fastjson-head}{30404ab3}
\DefMacro{res-alibaba-fastjson-num-ver}{51}
\DefMacro{res-alibaba-fastjson-num-file}{2,936}
\DefMacro{res-alibaba-fastjson-loc}{177,932}
\DefMacro{res-alibaba-fastjson-num-class}{2,278}
\DefMacro{res-alibaba-fastjson-num-method}{4,869}
\DefMacro{res-alibaba-fastjson-time}{83.82}
\DefMacro{res-lmdbjava-lmdbjava-name}{lmdbjava}
\DefMacro{res-lmdbjava-lmdbjava-url}{https://github.com/lmdbjava/lmdbjava.git}
\DefMacro{res-lmdbjava-lmdbjava-head}{7cf8f02f}
\DefMacro{res-lmdbjava-lmdbjava-num-ver}{51}
\DefMacro{res-lmdbjava-lmdbjava-num-file}{45}
\DefMacro{res-lmdbjava-lmdbjava-loc}{4,914}
\DefMacro{res-lmdbjava-lmdbjava-num-class}{8}
\DefMacro{res-lmdbjava-lmdbjava-num-method}{105}
\DefMacro{res-lmdbjava-lmdbjava-time}{32.60}
\DefMacro{res-brettwooldridge-HikariCP-name}{HikariCP}
\DefMacro{res-brettwooldridge-HikariCP-url}{https://github.com/brettwooldridge/HikariCP.git}
\DefMacro{res-brettwooldridge-HikariCP-head}{5d1ed1c6}
\DefMacro{res-brettwooldridge-HikariCP-num-ver}{51}
\DefMacro{res-brettwooldridge-HikariCP-num-file}{90}
\DefMacro{res-brettwooldridge-HikariCP-loc}{12,159}
\DefMacro{res-brettwooldridge-HikariCP-num-class}{37}
\DefMacro{res-brettwooldridge-HikariCP-num-method}{144}
\DefMacro{res-brettwooldridge-HikariCP-time}{112.40}
\DefMacro{res-bullhorn-sdk-rest-name}{sdk-rest}
\DefMacro{res-bullhorn-sdk-rest-url}{https://github.com/bullhorn/sdk-rest.git}
\DefMacro{res-bullhorn-sdk-rest-head}{a800949a}
\DefMacro{res-bullhorn-sdk-rest-num-ver}{51}
\DefMacro{res-bullhorn-sdk-rest-num-file}{626}
\DefMacro{res-bullhorn-sdk-rest-loc}{46,621}
\DefMacro{res-bullhorn-sdk-rest-num-class}{24}
\DefMacro{res-bullhorn-sdk-rest-num-method}{361}
\DefMacro{res-bullhorn-sdk-rest-time}{250.40}
\DefMacro{res-tabulapdf-tabula-java-name}{tabula-java}
\DefMacro{res-tabulapdf-tabula-java-url}{https://github.com/tabulapdf/tabula-java.git}
\DefMacro{res-tabulapdf-tabula-java-head}{8247954b}
\DefMacro{res-tabulapdf-tabula-java-num-ver}{51}
\DefMacro{res-tabulapdf-tabula-java-num-file}{52}
\DefMacro{res-tabulapdf-tabula-java-loc}{6,544}
\DefMacro{res-tabulapdf-tabula-java-num-class}{15}
\DefMacro{res-tabulapdf-tabula-java-num-method}{199}
\DefMacro{res-tabulapdf-tabula-java-time}{89.14}
\DefMacro{res-apache-commons-pool-name}{commons-pool}
\DefMacro{res-apache-commons-pool-url}{https://github.com/apache/commons-pool.git}
\DefMacro{res-apache-commons-pool-head}{1ff0aa0f}
\DefMacro{res-apache-commons-pool-num-ver}{51}
\DefMacro{res-apache-commons-pool-num-file}{92}
\DefMacro{res-apache-commons-pool-loc}{14,395}
\DefMacro{res-apache-commons-pool-num-class}{22}
\DefMacro{res-apache-commons-pool-num-method}{290}
\DefMacro{res-apache-commons-pool-time}{330.99}
\DefMacro{res-apache-commons-beanutils-name}{commons-beanutils}
\DefMacro{res-apache-commons-beanutils-url}{https://github.com/apache/commons-beanutils.git}
\DefMacro{res-apache-commons-beanutils-head}{55a786a3}
\DefMacro{res-apache-commons-beanutils-num-ver}{51}
\DefMacro{res-apache-commons-beanutils-num-file}{243}
\DefMacro{res-apache-commons-beanutils-loc}{31,831}
\DefMacro{res-apache-commons-beanutils-num-class}{94}
\DefMacro{res-apache-commons-beanutils-num-method}{1,273}
\DefMacro{res-apache-commons-beanutils-time}{9.84}
\DefMacro{res-finmath-finmath-lib-name}{finmath-lib}
\DefMacro{res-finmath-finmath-lib-url}{https://github.com/finmath/finmath-lib.git}
\DefMacro{res-finmath-finmath-lib-head}{fbcac3da}
\DefMacro{res-finmath-finmath-lib-num-ver}{51}
\DefMacro{res-finmath-finmath-lib-num-file}{1,166}
\DefMacro{res-finmath-finmath-lib-loc}{101,906}
\DefMacro{res-finmath-finmath-lib-num-class}{90}
\DefMacro{res-finmath-finmath-lib-num-method}{496}
\DefMacro{res-finmath-finmath-lib-time}{1,479.98}
\DefMacro{res-logic-ng-LogicNG-name}{LogicNG}
\DefMacro{res-logic-ng-LogicNG-url}{https://github.com/logic-ng/LogicNG.git}
\DefMacro{res-logic-ng-LogicNG-head}{60dcb918}
\DefMacro{res-logic-ng-LogicNG-num-ver}{51}
\DefMacro{res-logic-ng-LogicNG-num-file}{343}
\DefMacro{res-logic-ng-LogicNG-loc}{39,532}
\DefMacro{res-logic-ng-LogicNG-num-class}{116}
\DefMacro{res-logic-ng-LogicNG-num-method}{878}
\DefMacro{res-logic-ng-LogicNG-time}{452.85}
\DefMacro{apache-commons-lang-coverage-diff}{0.00}
\DefMacro{apache-commons-lang-coverage-bjacoco-std}{0.00}
\DefMacro{apache-commons-lang-coverage-exact-same}{\checkmark}
\DefMacro{apache-commons-lang-coverage-no-stat-sign-diff}{\checkmark}
\DefMacro{res-apache-commons-lang-ijacoco-coverage}{95.3}
\DefMacro{res-apache-commons-lang-bjacoco-coverage}{95.3}
\DefMacro{apache-commons-io-coverage-diff}{0.02}
\DefMacro{apache-commons-io-coverage-bjacoco-std}{0.02}
\DefMacro{apache-commons-io-coverage-exact-same}{\checkmark}
\DefMacro{apache-commons-io-coverage-no-stat-sign-diff}{\checkmark}
\DefMacro{res-apache-commons-io-ijacoco-coverage}{88.5}
\DefMacro{res-apache-commons-io-bjacoco-coverage}{88.5}
\DefMacro{apache-commons-dbcp-coverage-diff}{0.03}
\DefMacro{apache-commons-dbcp-coverage-bjacoco-std}{0.02}
\DefMacro{apache-commons-dbcp-coverage-exact-same}{\checkmark}
\DefMacro{apache-commons-dbcp-coverage-no-stat-sign-diff}{\checkmark}
\DefMacro{res-apache-commons-dbcp-ijacoco-coverage}{60.5}
\DefMacro{res-apache-commons-dbcp-bjacoco-coverage}{60.4}
\DefMacro{apache-commons-net-coverage-diff}{0.01}
\DefMacro{apache-commons-net-coverage-bjacoco-std}{0.02}
\DefMacro{apache-commons-net-coverage-exact-same}{\checkmark}
\DefMacro{apache-commons-net-coverage-no-stat-sign-diff}{\checkmark}
\DefMacro{res-apache-commons-net-ijacoco-coverage}{34.6}
\DefMacro{res-apache-commons-net-bjacoco-coverage}{34.6}
\DefMacro{apache-commons-math-coverage-diff}{0.00}
\DefMacro{apache-commons-math-coverage-bjacoco-std}{0.00}
\DefMacro{apache-commons-math-coverage-exact-same}{\checkmark}
\DefMacro{apache-commons-math-coverage-no-stat-sign-diff}{\checkmark}
\DefMacro{res-apache-commons-math-ijacoco-coverage}{90.5}
\DefMacro{res-apache-commons-math-bjacoco-coverage}{90.5}
\DefMacro{davidmoten-rxjava-extras-coverage-diff}{0.06}
\DefMacro{davidmoten-rxjava-extras-coverage-bjacoco-std}{0.03}
\DefMacro{davidmoten-rxjava-extras-coverage-exact-same}{\checkmark}
\DefMacro{davidmoten-rxjava-extras-coverage-no-stat-sign-diff}{\checkmark}
\DefMacro{res-davidmoten-rxjava-extras-ijacoco-coverage}{68.8}
\DefMacro{res-davidmoten-rxjava-extras-bjacoco-coverage}{68.8}
\DefMacro{sonyxperiadev-gerrit-events-coverage-diff}{0.03}
\DefMacro{sonyxperiadev-gerrit-events-coverage-bjacoco-std}{0.04}
\DefMacro{sonyxperiadev-gerrit-events-coverage-exact-same}{\checkmark}
\DefMacro{sonyxperiadev-gerrit-events-coverage-no-stat-sign-diff}{\checkmark}
\DefMacro{res-sonyxperiadev-gerrit-events-ijacoco-coverage}{41.7}
\DefMacro{res-sonyxperiadev-gerrit-events-bjacoco-coverage}{41.7}
\DefMacro{apache-commons-collections-coverage-diff}{0.01}
\DefMacro{apache-commons-collections-coverage-bjacoco-std}{0.01}
\DefMacro{apache-commons-collections-coverage-exact-same}{\checkmark}
\DefMacro{apache-commons-collections-coverage-no-stat-sign-diff}{\checkmark}
\DefMacro{res-apache-commons-collections-ijacoco-coverage}{88.1}
\DefMacro{res-apache-commons-collections-bjacoco-coverage}{88.1}
\DefMacro{apache-commons-imaging-coverage-diff}{0.00}
\DefMacro{apache-commons-imaging-coverage-bjacoco-std}{0.00}
\DefMacro{apache-commons-imaging-coverage-exact-same}{\checkmark}
\DefMacro{apache-commons-imaging-coverage-no-stat-sign-diff}{\checkmark}
\DefMacro{res-apache-commons-imaging-ijacoco-coverage}{73.9}
\DefMacro{res-apache-commons-imaging-bjacoco-coverage}{73.9}
\DefMacro{apache-commons-codec-coverage-diff}{0.00}
\DefMacro{apache-commons-codec-coverage-bjacoco-std}{0.00}
\DefMacro{apache-commons-codec-coverage-exact-same}{\checkmark}
\DefMacro{apache-commons-codec-coverage-no-stat-sign-diff}{\checkmark}
\DefMacro{res-apache-commons-codec-ijacoco-coverage}{94.2}
\DefMacro{res-apache-commons-codec-bjacoco-coverage}{94.2}
\DefMacro{apache-commons-configuration-coverage-diff}{0.00}
\DefMacro{apache-commons-configuration-coverage-bjacoco-std}{0.00}
\DefMacro{apache-commons-configuration-coverage-exact-same}{\checkmark}
\DefMacro{apache-commons-configuration-coverage-no-stat-sign-diff}{\checkmark}
\DefMacro{res-apache-commons-configuration-ijacoco-coverage}{89.3}
\DefMacro{res-apache-commons-configuration-bjacoco-coverage}{89.3}
\DefMacro{asterisk-java-asterisk-java-coverage-diff}{0.01}
\DefMacro{asterisk-java-asterisk-java-coverage-bjacoco-std}{0.00}
\DefMacro{asterisk-java-asterisk-java-coverage-exact-same}{\checkmark}
\DefMacro{asterisk-java-asterisk-java-coverage-no-stat-sign-diff}{\checkmark}
\DefMacro{res-asterisk-java-asterisk-java-ijacoco-coverage}{14.0}
\DefMacro{res-asterisk-java-asterisk-java-bjacoco-coverage}{14.0}
\DefMacro{apache-commons-compress-coverage-diff}{0.01}
\DefMacro{apache-commons-compress-coverage-bjacoco-std}{0.01}
\DefMacro{apache-commons-compress-coverage-exact-same}{\checkmark}
\DefMacro{apache-commons-compress-coverage-no-stat-sign-diff}{\checkmark}
\DefMacro{res-apache-commons-compress-ijacoco-coverage}{87.0}
\DefMacro{res-apache-commons-compress-bjacoco-coverage}{87.0}
\DefMacro{alibaba-fastjson-coverage-diff}{0.05}
\DefMacro{alibaba-fastjson-coverage-bjacoco-std}{0.24}
\DefMacro{alibaba-fastjson-coverage-exact-same}{\checkmark}
\DefMacro{alibaba-fastjson-coverage-no-stat-sign-diff}{\checkmark}
\DefMacro{res-alibaba-fastjson-ijacoco-coverage}{87.9}
\DefMacro{res-alibaba-fastjson-bjacoco-coverage}{87.9}
\DefMacro{lmdbjava-lmdbjava-coverage-diff}{0.00}
\DefMacro{lmdbjava-lmdbjava-coverage-bjacoco-std}{0.00}
\DefMacro{lmdbjava-lmdbjava-coverage-exact-same}{\checkmark}
\DefMacro{lmdbjava-lmdbjava-coverage-no-stat-sign-diff}{\checkmark}
\DefMacro{res-lmdbjava-lmdbjava-ijacoco-coverage}{90.6}
\DefMacro{res-lmdbjava-lmdbjava-bjacoco-coverage}{90.6}
\DefMacro{brettwooldridge-HikariCP-coverage-diff}{0.16}
\DefMacro{brettwooldridge-HikariCP-coverage-bjacoco-std}{0.12}
\DefMacro{brettwooldridge-HikariCP-coverage-exact-same}{}
\DefMacro{brettwooldridge-HikariCP-coverage-no-stat-sign-diff}{\checkmark}
\DefMacro{res-brettwooldridge-HikariCP-ijacoco-coverage}{80.5}
\DefMacro{res-brettwooldridge-HikariCP-bjacoco-coverage}{80.4}
\DefMacro{bullhorn-sdk-rest-coverage-diff}{0.01}
\DefMacro{bullhorn-sdk-rest-coverage-bjacoco-std}{0.00}
\DefMacro{bullhorn-sdk-rest-coverage-exact-same}{\checkmark}
\DefMacro{bullhorn-sdk-rest-coverage-no-stat-sign-diff}{\checkmark}
\DefMacro{res-bullhorn-sdk-rest-ijacoco-coverage}{65.2}
\DefMacro{res-bullhorn-sdk-rest-bjacoco-coverage}{65.2}
\DefMacro{tabulapdf-tabula-java-coverage-diff}{0.00}
\DefMacro{tabulapdf-tabula-java-coverage-bjacoco-std}{0.00}
\DefMacro{tabulapdf-tabula-java-coverage-exact-same}{\checkmark}
\DefMacro{tabulapdf-tabula-java-coverage-no-stat-sign-diff}{\checkmark}
\DefMacro{res-tabulapdf-tabula-java-ijacoco-coverage}{82.9}
\DefMacro{res-tabulapdf-tabula-java-bjacoco-coverage}{82.9}
\DefMacro{jenkinsci-email-ext-plugin-coverage-diff}{2.33}
\DefMacro{jenkinsci-email-ext-plugin-coverage-bjacoco-std}{0.01}
\DefMacro{jenkinsci-email-ext-plugin-coverage-exact-same}{}
\DefMacro{jenkinsci-email-ext-plugin-coverage-no-stat-sign-diff}{}
\DefMacro{res-jenkinsci-email-ext-plugin-ijacoco-coverage}{65.2}
\DefMacro{res-jenkinsci-email-ext-plugin-bjacoco-coverage}{67.5}
\DefMacro{apache-commons-pool-coverage-diff}{0.03}
\DefMacro{apache-commons-pool-coverage-bjacoco-std}{0.04}
\DefMacro{apache-commons-pool-coverage-exact-same}{\checkmark}
\DefMacro{apache-commons-pool-coverage-no-stat-sign-diff}{\checkmark}
\DefMacro{res-apache-commons-pool-ijacoco-coverage}{85.0}
\DefMacro{res-apache-commons-pool-bjacoco-coverage}{85.0}
\DefMacro{apache-commons-beanutils-coverage-diff}{0.01}
\DefMacro{apache-commons-beanutils-coverage-bjacoco-std}{0.01}
\DefMacro{apache-commons-beanutils-coverage-exact-same}{\checkmark}
\DefMacro{apache-commons-beanutils-coverage-no-stat-sign-diff}{\checkmark}
\DefMacro{res-apache-commons-beanutils-ijacoco-coverage}{71.6}
\DefMacro{res-apache-commons-beanutils-bjacoco-coverage}{71.6}
\DefMacro{finmath-finmath-lib-coverage-diff}{0.10}
\DefMacro{finmath-finmath-lib-coverage-bjacoco-std}{0.00}
\DefMacro{finmath-finmath-lib-coverage-exact-same}{\checkmark}
\DefMacro{finmath-finmath-lib-coverage-no-stat-sign-diff}{\checkmark}
\DefMacro{res-finmath-finmath-lib-ijacoco-coverage}{36.7}
\DefMacro{res-finmath-finmath-lib-bjacoco-coverage}{36.6}
\DefMacro{logic-ng-LogicNG-coverage-diff}{0.00}
\DefMacro{logic-ng-LogicNG-coverage-bjacoco-std}{0.00}
\DefMacro{logic-ng-LogicNG-coverage-exact-same}{\checkmark}
\DefMacro{logic-ng-LogicNG-coverage-no-stat-sign-diff}{\checkmark}
\DefMacro{res-logic-ng-LogicNG-ijacoco-coverage}{95.0}
\DefMacro{res-logic-ng-LogicNG-bjacoco-coverage}{95.0}
\DefMacro{res-apache-commons-lang-phase-compile}{10.47}
\DefMacro{res-apache-commons-lang-phase-analysis}{0.49}
\DefMacro{res-apache-commons-lang-phase-execution-collection}{4.65}
\DefMacro{res-apache-commons-lang-phase-report}{0.87}
\DefMacro{res-apache-commons-lang-phase-compile-percent}{63.5}
\DefMacro{res-apache-commons-lang-phase-analysis-percent}{3.0}
\DefMacro{res-apache-commons-lang-phase-execution-collection-percent}{28.2}
\DefMacro{res-apache-commons-lang-phase-report-percent}{5.3}
\DefMacro{res-apache-commons-io-phase-compile}{8.51}
\DefMacro{res-apache-commons-io-phase-analysis}{0.52}
\DefMacro{res-apache-commons-io-phase-execution-collection}{7.27}
\DefMacro{res-apache-commons-io-phase-report}{0.55}
\DefMacro{res-apache-commons-io-phase-compile-percent}{50.5}
\DefMacro{res-apache-commons-io-phase-analysis-percent}{3.1}
\DefMacro{res-apache-commons-io-phase-execution-collection-percent}{43.2}
\DefMacro{res-apache-commons-io-phase-report-percent}{3.3}
\DefMacro{res-apache-commons-dbcp-phase-compile}{8.21}
\DefMacro{res-apache-commons-dbcp-phase-analysis}{0.63}
\DefMacro{res-apache-commons-dbcp-phase-execution-collection}{51.48}
\DefMacro{res-apache-commons-dbcp-phase-report}{0.50}
\DefMacro{res-apache-commons-dbcp-phase-compile-percent}{13.5}
\DefMacro{res-apache-commons-dbcp-phase-analysis-percent}{1.0}
\DefMacro{res-apache-commons-dbcp-phase-execution-collection-percent}{84.6}
\DefMacro{res-apache-commons-dbcp-phase-report-percent}{0.8}
\DefMacro{res-apache-commons-net-phase-compile}{7.64}
\DefMacro{res-apache-commons-net-phase-analysis}{0.43}
\DefMacro{res-apache-commons-net-phase-execution-collection}{17.13}
\DefMacro{res-apache-commons-net-phase-report}{0.68}
\DefMacro{res-apache-commons-net-phase-compile-percent}{29.5}
\DefMacro{res-apache-commons-net-phase-analysis-percent}{1.7}
\DefMacro{res-apache-commons-net-phase-execution-collection-percent}{66.2}
\DefMacro{res-apache-commons-net-phase-report-percent}{2.6}
\DefMacro{res-apache-commons-math-phase-compile}{19.83}
\DefMacro{res-apache-commons-math-phase-analysis}{0.81}
\DefMacro{res-apache-commons-math-phase-execution-collection}{159.32}
\DefMacro{res-apache-commons-math-phase-report}{2.02}
\DefMacro{res-apache-commons-math-phase-compile-percent}{10.9}
\DefMacro{res-apache-commons-math-phase-analysis-percent}{0.4}
\DefMacro{res-apache-commons-math-phase-execution-collection-percent}{87.6}
\DefMacro{res-apache-commons-math-phase-report-percent}{1.1}
\DefMacro{res-davidmoten-rxjava-extras-phase-compile}{6.67}
\DefMacro{res-davidmoten-rxjava-extras-phase-analysis}{0.64}
\DefMacro{res-davidmoten-rxjava-extras-phase-execution-collection}{52.16}
\DefMacro{res-davidmoten-rxjava-extras-phase-report}{0.60}
\DefMacro{res-davidmoten-rxjava-extras-phase-compile-percent}{11.1}
\DefMacro{res-davidmoten-rxjava-extras-phase-analysis-percent}{1.1}
\DefMacro{res-davidmoten-rxjava-extras-phase-execution-collection-percent}{86.8}
\DefMacro{res-davidmoten-rxjava-extras-phase-report-percent}{1.0}
\DefMacro{res-sonyxperiadev-gerrit-events-phase-compile}{6.73}
\DefMacro{res-sonyxperiadev-gerrit-events-phase-analysis}{0.81}
\DefMacro{res-sonyxperiadev-gerrit-events-phase-execution-collection}{23.40}
\DefMacro{res-sonyxperiadev-gerrit-events-phase-report}{0.36}
\DefMacro{res-sonyxperiadev-gerrit-events-phase-compile-percent}{21.5}
\DefMacro{res-sonyxperiadev-gerrit-events-phase-analysis-percent}{2.6}
\DefMacro{res-sonyxperiadev-gerrit-events-phase-execution-collection-percent}{74.8}
\DefMacro{res-sonyxperiadev-gerrit-events-phase-report-percent}{1.2}
\DefMacro{res-apache-commons-collections-phase-compile}{11.09}
\DefMacro{res-apache-commons-collections-phase-analysis}{0.49}
\DefMacro{res-apache-commons-collections-phase-execution-collection}{7.21}
\DefMacro{res-apache-commons-collections-phase-report}{0.98}
\DefMacro{res-apache-commons-collections-phase-compile-percent}{56.1}
\DefMacro{res-apache-commons-collections-phase-analysis-percent}{2.5}
\DefMacro{res-apache-commons-collections-phase-execution-collection-percent}{36.5}
\DefMacro{res-apache-commons-collections-phase-report-percent}{5.0}
\DefMacro{res-apache-commons-imaging-phase-compile}{9.43}
\DefMacro{res-apache-commons-imaging-phase-analysis}{0.42}
\DefMacro{res-apache-commons-imaging-phase-execution-collection}{19.22}
\DefMacro{res-apache-commons-imaging-phase-report}{0.93}
\DefMacro{res-apache-commons-imaging-phase-compile-percent}{31.4}
\DefMacro{res-apache-commons-imaging-phase-analysis-percent}{1.4}
\DefMacro{res-apache-commons-imaging-phase-execution-collection-percent}{64.1}
\DefMacro{res-apache-commons-imaging-phase-report-percent}{3.1}
\DefMacro{res-apache-commons-codec-phase-compile}{7.70}
\DefMacro{res-apache-commons-codec-phase-analysis}{0.39}
\DefMacro{res-apache-commons-codec-phase-execution-collection}{10.00}
\DefMacro{res-apache-commons-codec-phase-report}{0.45}
\DefMacro{res-apache-commons-codec-phase-compile-percent}{41.5}
\DefMacro{res-apache-commons-codec-phase-analysis-percent}{2.1}
\DefMacro{res-apache-commons-codec-phase-execution-collection-percent}{53.9}
\DefMacro{res-apache-commons-codec-phase-report-percent}{2.4}
\DefMacro{res-apache-commons-configuration-phase-compile}{11.73}
\DefMacro{res-apache-commons-configuration-phase-analysis}{1.23}
\DefMacro{res-apache-commons-configuration-phase-execution-collection}{18.10}
\DefMacro{res-apache-commons-configuration-phase-report}{0.89}
\DefMacro{res-apache-commons-configuration-phase-compile-percent}{36.7}
\DefMacro{res-apache-commons-configuration-phase-analysis-percent}{3.9}
\DefMacro{res-apache-commons-configuration-phase-execution-collection-percent}{56.7}
\DefMacro{res-apache-commons-configuration-phase-report-percent}{2.8}
\DefMacro{res-asterisk-java-asterisk-java-phase-compile}{9.87}
\DefMacro{res-asterisk-java-asterisk-java-phase-analysis}{0.42}
\DefMacro{res-asterisk-java-asterisk-java-phase-execution-collection}{2.04}
\DefMacro{res-asterisk-java-asterisk-java-phase-report}{1.22}
\DefMacro{res-asterisk-java-asterisk-java-phase-compile-percent}{72.8}
\DefMacro{res-asterisk-java-asterisk-java-phase-analysis-percent}{3.1}
\DefMacro{res-asterisk-java-asterisk-java-phase-execution-collection-percent}{15.1}
\DefMacro{res-asterisk-java-asterisk-java-phase-report-percent}{9.0}
\DefMacro{res-apache-commons-compress-phase-compile}{10.29}
\DefMacro{res-apache-commons-compress-phase-analysis}{0.58}
\DefMacro{res-apache-commons-compress-phase-execution-collection}{34.32}
\DefMacro{res-apache-commons-compress-phase-report}{0.84}
\DefMacro{res-apache-commons-compress-phase-compile-percent}{22.4}
\DefMacro{res-apache-commons-compress-phase-analysis-percent}{1.3}
\DefMacro{res-apache-commons-compress-phase-execution-collection-percent}{74.6}
\DefMacro{res-apache-commons-compress-phase-report-percent}{1.8}
\DefMacro{res-alibaba-fastjson-phase-compile}{28.57}
\DefMacro{res-alibaba-fastjson-phase-analysis}{5.79}
\DefMacro{res-alibaba-fastjson-phase-execution-collection}{57.18}
\DefMacro{res-alibaba-fastjson-phase-report}{1.23}
\DefMacro{res-alibaba-fastjson-phase-compile-percent}{30.8}
\DefMacro{res-alibaba-fastjson-phase-analysis-percent}{6.2}
\DefMacro{res-alibaba-fastjson-phase-execution-collection-percent}{61.6}
\DefMacro{res-alibaba-fastjson-phase-report-percent}{1.3}
\DefMacro{res-lmdbjava-lmdbjava-phase-compile}{6.93}
\DefMacro{res-lmdbjava-lmdbjava-phase-analysis}{0.38}
\DefMacro{res-lmdbjava-lmdbjava-phase-execution-collection}{51.56}
\DefMacro{res-lmdbjava-lmdbjava-phase-report}{0.30}
\DefMacro{res-lmdbjava-lmdbjava-phase-compile-percent}{11.7}
\DefMacro{res-lmdbjava-lmdbjava-phase-analysis-percent}{0.6}
\DefMacro{res-lmdbjava-lmdbjava-phase-execution-collection-percent}{87.1}
\DefMacro{res-lmdbjava-lmdbjava-phase-report-percent}{0.5}
\DefMacro{res-brettwooldridge-HikariCP-phase-compile}{8.46}
\DefMacro{res-brettwooldridge-HikariCP-phase-analysis}{0.94}
\DefMacro{res-brettwooldridge-HikariCP-phase-execution-collection}{90.22}
\DefMacro{res-brettwooldridge-HikariCP-phase-report}{0.44}
\DefMacro{res-brettwooldridge-HikariCP-phase-compile-percent}{8.5}
\DefMacro{res-brettwooldridge-HikariCP-phase-analysis-percent}{0.9}
\DefMacro{res-brettwooldridge-HikariCP-phase-execution-collection-percent}{90.2}
\DefMacro{res-brettwooldridge-HikariCP-phase-report-percent}{0.4}
\DefMacro{res-bullhorn-sdk-rest-phase-compile}{10.79}
\DefMacro{res-bullhorn-sdk-rest-phase-analysis}{3.92}
\DefMacro{res-bullhorn-sdk-rest-phase-execution-collection}{371.25}
\DefMacro{res-bullhorn-sdk-rest-phase-report}{1.62}
\DefMacro{res-bullhorn-sdk-rest-phase-compile-percent}{2.8}
\DefMacro{res-bullhorn-sdk-rest-phase-analysis-percent}{1.0}
\DefMacro{res-bullhorn-sdk-rest-phase-execution-collection-percent}{95.8}
\DefMacro{res-bullhorn-sdk-rest-phase-report-percent}{0.4}
\DefMacro{res-tabulapdf-tabula-java-phase-compile}{6.63}
\DefMacro{res-tabulapdf-tabula-java-phase-analysis}{0.49}
\DefMacro{res-tabulapdf-tabula-java-phase-execution-collection}{87.59}
\DefMacro{res-tabulapdf-tabula-java-phase-report}{0.34}
\DefMacro{res-tabulapdf-tabula-java-phase-compile-percent}{7.0}
\DefMacro{res-tabulapdf-tabula-java-phase-analysis-percent}{0.5}
\DefMacro{res-tabulapdf-tabula-java-phase-execution-collection-percent}{92.2}
\DefMacro{res-tabulapdf-tabula-java-phase-report-percent}{0.4}
\DefMacro{res-jenkinsci-email-ext-plugin-phase-compile}{649.51}
\DefMacro{res-jenkinsci-email-ext-plugin-phase-analysis}{13.85}
\DefMacro{res-jenkinsci-email-ext-plugin-phase-execution-collection}{0.00}
\DefMacro{res-jenkinsci-email-ext-plugin-phase-report}{1.09}
\DefMacro{res-jenkinsci-email-ext-plugin-phase-compile-percent}{97.8}
\DefMacro{res-jenkinsci-email-ext-plugin-phase-analysis-percent}{2.1}
\DefMacro{res-jenkinsci-email-ext-plugin-phase-execution-collection-percent}{0.0}
\DefMacro{res-jenkinsci-email-ext-plugin-phase-report-percent}{0.2}
\DefMacro{res-apache-commons-pool-phase-compile}{7.71}
\DefMacro{res-apache-commons-pool-phase-analysis}{0.39}
\DefMacro{res-apache-commons-pool-phase-execution-collection}{141.78}
\DefMacro{res-apache-commons-pool-phase-report}{0.36}
\DefMacro{res-apache-commons-pool-phase-compile-percent}{5.1}
\DefMacro{res-apache-commons-pool-phase-analysis-percent}{0.3}
\DefMacro{res-apache-commons-pool-phase-execution-collection-percent}{94.4}
\DefMacro{res-apache-commons-pool-phase-report-percent}{0.2}
\DefMacro{res-apache-commons-beanutils-phase-compile}{9.49}
\DefMacro{res-apache-commons-beanutils-phase-analysis}{0.40}
\DefMacro{res-apache-commons-beanutils-phase-execution-collection}{2.22}
\DefMacro{res-apache-commons-beanutils-phase-report}{0.48}
\DefMacro{res-apache-commons-beanutils-phase-compile-percent}{75.4}
\DefMacro{res-apache-commons-beanutils-phase-analysis-percent}{3.2}
\DefMacro{res-apache-commons-beanutils-phase-execution-collection-percent}{17.6}
\DefMacro{res-apache-commons-beanutils-phase-report-percent}{3.8}
\DefMacro{res-finmath-finmath-lib-phase-compile}{9.91}
\DefMacro{res-finmath-finmath-lib-phase-analysis}{0.48}
\DefMacro{res-finmath-finmath-lib-phase-execution-collection}{424.00}
\DefMacro{res-finmath-finmath-lib-phase-report}{1.59}
\DefMacro{res-finmath-finmath-lib-phase-compile-percent}{2.3}
\DefMacro{res-finmath-finmath-lib-phase-analysis-percent}{0.1}
\DefMacro{res-finmath-finmath-lib-phase-execution-collection-percent}{97.3}
\DefMacro{res-finmath-finmath-lib-phase-report-percent}{0.4}
\DefMacro{res-logic-ng-LogicNG-phase-compile}{9.03}
\DefMacro{res-logic-ng-LogicNG-phase-analysis}{0.64}
\DefMacro{res-logic-ng-LogicNG-phase-execution-collection}{813.74}
\DefMacro{res-logic-ng-LogicNG-phase-report}{0.92}
\DefMacro{res-logic-ng-LogicNG-phase-compile-percent}{1.1}
\DefMacro{res-logic-ng-LogicNG-phase-analysis-percent}{0.1}
\DefMacro{res-logic-ng-LogicNG-phase-execution-collection-percent}{98.7}
\DefMacro{res-logic-ng-LogicNG-phase-report-percent}{0.1}
\DefMacro{res-phase-avg-compile}{38.05}
\DefMacro{res-phase-avg-analysis}{1.53}
\DefMacro{res-phase-avg-execution-collection}{106.34}
\DefMacro{res-phase-avg-report}{0.84}
\DefMacro{res-phase-avg-compile-percent}{25.9}
\DefMacro{res-phase-avg-analysis-percent}{1.0}
\DefMacro{res-phase-avg-execution-collection-percent}{72.5}
\DefMacro{res-phase-avg-report-percent}{0.6}

\DefMacro{retestall-sum-time}{3,794.82}
\DefMacro{retestall-avg-time}{172.49}
\DefMacro{retestall-avg-total-time}{8,797.07}
\DefMacro{retestall-sum-total-time}{193,535.57}
\DefMacro{ekstazi-sum-time}{1,709.48}
\DefMacro{ekstazi-avg-time}{77.70}
\DefMacro{ekstazi-avg-total-time}{3,962.89}
\DefMacro{ekstazi-sum-total-time}{87,183.68}
\DefMacro{res-ekstazi-selected-test-rate-mean}{20.63}
\DefMacro{ijacoco-avg-time}{126.47}
\DefMacro{ijacoco-sum-time}{2,782.44}
\DefMacro{bjacoco-avg-time}{300.57}
\DefMacro{bjacoco-sum-time}{6,612.62}
\DefMacro{ijacoco-avg-total-time}{6,450.21}
\DefMacro{ijacoco-sum-total-time}{141,904.69}
\DefMacro{bjacoco-avg-total-time}{15,329.25}
\DefMacro{bjacoco-sum-total-time}{337,243.55}
\DefMacro{ijacoco-avg-speedup}{1.86}
\DefMacro{ijacoco-sum-speedup}{40.92}
\DefMacro{ijacoco-max-speedup}{8.20}
\DefMacro{ijacoco-min-speedup}{0.75}
\DefMacro{ijacoco-cnt-positive-speedup}{17}
\DefMacro{ijacoco-cnt-negative-speedup}{5}
\DefMacro{res-ijacoco-selected-test-rate-mean}{47.34}
\DefMacro{num-revisions-per-repo}{51}
\DefMacro{num-repo}{22}
\DefMacro{res-projects-sum-num-ver}{1,122}
\DefMacro{res-projects-sum-num-file}{11,067}
\DefMacro{res-projects-sum-loc}{1,102,434}
\DefMacro{res-projects-sum-num-test-class}{4,249}
\DefMacro{res-projects-sum-num-test-method}{52,718}
\DefMacro{res-projects-sum-time}{3,521.54}

\DefMacro{num-repos-finerts}{23\xspace}
\DefMacro{sum-loc-approx}{1.1M\xspace}
\DefMacro{cov-diff-threshold}{0.10\%\xspace}

\begin{document}

\title{\Title}

\author{Jiale Amber Wang}
\authornote{Both authors contributed equally to this research.}
\affiliation{%
\institution{University of Waterloo}
\country{Canada}
}
\email{jiale.wang@uwaterloo.ca}
\author{Kaiyuan Wang}
\authornotemark[1] %
\affiliation{%
\institution{Google}
\country{USA}
}
\email{kaiyuanw@google.com}
\author{Pengyu Nie}
\affiliation{%
\institution{University of Waterloo}
\country{Canada}
}
\email{pynie@uwaterloo.ca}

\begin{abstract}

\Codecovanalysis has been widely adopted in the continuous integration
of open-source and industry software \repos to monitor the adequacy of
regression \testsuites.  However, computing \codecov can be costly, introducing
significant overhead during test execution.  Plus, re-collecting
\codecov for the entire test suite is usually unnecessary when only a
part of the \covdata is affected by code changes.  While regression
test selection (RTS) techniques exist to select a subset of \testcases
whose behaviors may be affected by code changes, they are not
compatible with \codecovanalysis techniques---that is, simply
executing RTS-selected \testcases leads to incorrect \codecov results.

In this paper, we present the first \icodecovanalysis technique, which
speeds up \codecovanalysis by executing a minimal
subset of \testcases to update the \covdata affected by code changes.
We implement our technique in a tool dubbed \ijacocoTool, which builds
on \ekstazi and \jacoco---the state-of-the-art RTS and
\codecovanalysis tools for \java.
We evaluate \ijacocoTool on \UseMacro{res-projects-sum-num-ver}
versions from \UseMacro{num-repo} open-source \repos and show that
\ijacocoTool can speed up \codecovanalysis time by an average of
\UseMacro{ijacoco-avg-speedup}$\times$ and up to
\UseMacro{ijacoco-max-speedup}$\times$ compared to \jacoco.

\end{abstract}

\begin{CCSXML}
<ccs2012>
<concept>
<concept_id>10011007.10011074.10011099.10011102.10011103</concept_id>
<concept_desc>Software and its engineering~Software testing and debugging</concept_desc>
<concept_significance>500</concept_significance>
</concept>
<concept>
<concept_id>10011007.10011074.10011111.10011113</concept_id>
<concept_desc>Software and its engineering~Software evolution</concept_desc>
<concept_significance>500</concept_significance>
</concept>
</ccs2012>
\end{CCSXML}

\ccsdesc[500]{Software and its engineering~Software testing and debugging}
\ccsdesc[500]{Software and its engineering~Software evolution}

\keywords{regression testing, code coverage analysis, regression test selection}

\maketitle

\section{Introduction}
\label{sec:intro}

Developers rely on high quality \tests to assess the correctness of software systems. By
setting up testing in continuous integration (CI), i.e., executing \testsuites on every
\revision, developers can quickly detect regressions~\cite{duvall2007continuous,
hilton2016usage}.  \emph{\Codecov}~\cite{frankl1993experimental,
PiwowarskiETAL93Coverage, zhu1997software, gligoric2015guidelines, kochhar2015code, hemmati2015effective, YangETAL06SurveyCoverage} is a
well-established measurement for the \testsuite adequacy. It is defined as the
portion of \codeelems (e.g., instructions, lines, branches) that are transitively executed during the execution of a
\testsuite~\cite{miller1963systematic}.

\Codecovanalysis is widely adopted in real-world software development process.  For
example, \jacoco~\cite{JacocoGithubRepo}, one of the most popular \codecovanalysis tools
for \java, is used by more than 395K open-source \repos on
\github~\cite{JacocoGithubDependents}.  Large industry companies such as Google and IBM
also see the importance of setting up \codecovanalysis as a part of their internal CI
pipelines~\cite{Kim03EfficientUseOfCodeCoverage, GoogleTestingBlogCoverage,
ivankovic2019code, adler2011code}.

However, \codecovanalysis can be time-consuming.  A typical \codecovanalysis technique
needs to instrument the \codebase to insert \probes into \codeelems, and then
execute the \testsuite to collect \emph{\covdata}---recording which \probes (and the
\codeelems between them) are covered.
Obviously, executing the \testsuites on each \revision is already
costly and time-intensive~\cite{micco2017state,
williams2009effectiveness}. Performing \codecovanalysis on top of it
adds more overhead~\cite{Kim03EfficientUseOfCodeCoverage,
ivankovic2019code, adler2011code}. This additional overhead adds up
over time as the \codebase evolves and scales.

To speed up regression testing during software evolution, researchers have proposed
\emph{regression test selection} (RTS)~\cite{engstrom2010systematic,
engstrom2008empirical, graves2001empirical, yoo2012regression, gligoric2015practical,
legunsen2016extensive}, which only executes selected \tests whose behaviors may be
affected by code changes.  Executing the other \tests that depend solely on unchanged
code is unnecessary, because their behaviors should not change.
Similarly, during software evolution, performing \codecov for the entire \codebase by
executing all \tests on each \revision is not only time-consuming but also unnecessary.
The \codecov for the part of the \codebase that solely depend on unchanged code should
not change either.

In this paper, we propose an efficient \emph{\icodecovanalysis} technique to collect
\codecov for regression \testsuites.  Given two software \revisions, the \covdata
collected on the old \revision, and the code changes, \icodecovanalysis selects a
minimal set of \tests that need to be executed to update the \covdata for the new
\revision.  Surprisingly, naively running RTS and
then running \codecovanalysis with RTS-selected \tests may not be efficient or correct.  The
reason is that collecting \covdata at the per-\test granularity has
performance and safety issues.  Thus, we propose to expand the set of RTS-selected \tests
in a way to ensure that the two \revisions' \covdata can be correctly merged (a detailed
explanation with example is given in \S\ref{sec:technique:analysis}).

To demonstrate the effectiveness of \icodecovanalysis, we implement \ijacocoTool, a tool
that \textbf{i}ncrementally collects \textbf{Ja}va source \textbf{co}de
\textbf{co}verage.  \ijacocoTool it built on top of \jacoco~\cite{JacocoGithubRepo}, a
widely-used \codecovanalysis tool for \java, and \ekstazi~\cite{gligoric2015ekstazi,
gligoric2015practical}, a state-of-the-art file-level dynamic RTS tool for \java.
\ijacocoTool is designed to have the same interface as \jacoco (i.e., can be used as a
command line tool and a \maven plugin), such that existing users of \jacoco can
seamlessly switch to \ijacocoTool.
We envision two usage scenarios of \ijacocoTool: (1)~speeding up \codecovanalysis for
each code \revision on CI; (2)~instantly providing code coverage feedback when
developers edit code on local machines.

We evaluated \ijacocoTool on a dataset of \UseMacro{res-projects-sum-num-ver} versions
from \UseMacro{num-repo} open-source \repos, with \UseMacro{sum-loc-approx} lines of
code in total.  Compared with \jacoco, \ijacocoTool can achieve an end-to-end time speedup of
\UseMacro{ijacoco-avg-speedup}$\times$ on average and up to \UseMacro{ijacoco-max-speedup}$\times$.
While \ijacocoTool incurs overhead on the first \revision of each \repo to build the
dependency graph, in most subsequent \revisions, \ijacocoTool only needs to run a small
subset of \tests to update \covdata.  Thus, \icodecovanalysis allows for lower CI cost
and fast feedback during software development.  We also compared the test selection rate
of \ijacocoTool with \ekstazi, and found that although \ijacocoTool needs to select
about twice as many \tests as \ekstazi to correctly compute \covdata,
the speedup of \ijacocoTool is still significant.

The \emph{correctness} of \icodecovanalysis is crucial, i.e., the \covdata collected
incrementally should match the \covdata collected from running all tests with the existing \codecovanalysis
technique.  We prove that our \icodecovanalysis technique is correct as long as the
underlying RTS technique is \emph{safe}. An RTS technique is safe if it does not miss any \test whose behavior
may be affected by a code change.  \ijacocoTool is built on top of \ekstazi, which is a safe RTS
tool for \java \repos~\cite{ZhuETAL19RTSCheck}.  Moreover, we confirmed that the
\codecov measured by \ijacocoTool and \jacoco are consistent across all \revisions in
our experiment.

\vspace{3pt}
\noindent
The main contributions of this work include:

\begin{itemize}[topsep=3pt,itemsep=1ex,partopsep=0ex,parsep=0ex,leftmargin=*]

\item \MyPara{Idea} We demonstrate that the idea of incremental
computation can speed up \codecovanalysis for regression
\testsuites.

\item \MyPara{Technique} We design the first \icodecovanalysis
technique that integrates \codecovanalysis and RTS techniques; note
that a non-trivial integration is needed to correctly update the
\covdata when executing a subset of \tests.

\item \MyPara{Implementation} We implement our technique as
\ijacocoTool, an industrial-level \icodecovanalysis tool for \java.

\item \MyPara{Evaluation} Our evaluation found that \ijacocoTool can
speed up \codecovanalysis end-to-end time by
\UseMacro{ijacoco-avg-speedup}$\times$ on average and up to
\UseMacro{ijacoco-max-speedup}$\times$ compared to \jacoco.

\end{itemize}

\noindent
The replication package of \ijacocoTool, including the tool, our experiment scripts, and
results, is open-sourced at:\\
\url{\ijacocoURL}

\section{Motivating Example}
\label{sec:example}

\DefMacro{example1-version-new}{\Pversion{9fb4f47f}}
\DefMacro{example1-version-old}{\Pversion{b1deb442}}
\DefMacro{example1-bjacoco-time}{29.75s\xspace}
\DefMacro{example1-ijacoco-time}{18.65s\xspace}
\DefMacro{example1-bjacoco-num-class}{149\xspace}
\DefMacro{example1-ijacoco-num-class}{4\xspace}
\DefMacro{example1-speedup}{1.60\xspace}

Figure~\ref{fig:example1} shows a real-world example from the
\Prepo{\UseMacro{res-apache-commons-lang-name}}
\repo\footnote{\url{\UseMacro{res-apache-commons-lang-url}}} where \icodecovanalysis is
helpful.  This \repo is configured to run \jacoco to collect \codecov on every new
\revision on \github's CI.  On \revision \UseMacro{example1-version-new}, \jacoco takes
\UseMacro{example1-bjacoco-time} to execute all \UseMacro{example1-bjacoco-num-class}
\tests and collect \codecov.
However, the code change between \revision \UseMacro{example1-version-new} and the
previous \revision \UseMacro{example1-version-old} is rather small: as shown in
Figure~\ref{fig:example1:diff}, the only change is merging two \CodeIn{catch} blocks
into one.  Re-computing \codecov for the entire \codebase is unnecessary because only a
small part of the \codebase is affected by the change.

Our proposed \icodecovanalysis technique first analyzes the code change, finds the part of the \codebase
whose \codecov may be affected by the change, and computes the \tests need to be executed
to collect the \covdata for the affected part.  When applying \ijacocoTool on this example, it
finds out that only \UseMacro{example1-ijacoco-num-class} \tests need to be executed, as
shown in Figure~\ref{fig:example1:tests-ijacoco}.  The end-to-end time for \ijacocoTool
to analyze code changes, execute the selected \tests, and collect \codecov is
\UseMacro{example1-ijacoco-time}, which is \UseMacro{example1-speedup} times faster than
\jacoco.

\begin{figure}[t]
\begin{center}
\begin{small}

\newsavebox\boxExampleI
\begin{lrbox}{\boxExampleI}
\begin{lstlisting}[language=java-pretty-no-number]
// org/apache/commons/lang3/reflect/FieldUtils.java
public static void removeFinalModifier(final Field field,
final boolean forceAccess) {
Validate.isTrue(field != null, "The field must not be null");
try { ...
} catch (final NoSuchFieldException ignored) {
} catch (final NoSuchFieldException | IllegalAccessException ignored) {
// The field class contains always a modifiers field
} catch (final IllegalAccessException ignored) {
// The modifiers field is made accessible
}
}
\end{lstlisting}
\end{lrbox}

\DefMacro*{h-line}{2.16ex}
\DefMacro*{w-line}{27em}

\begin{minipage}{.9\columnwidth}
\begin{tikzpicture}[
git-add/.style={minimum height=\UseMacro{h-line}, minimum width=\UseMacro{w-line}, text width=\UseMacro{w-line}, text height=0.3 * \UseMacro{h-line}, font=\scriptsize\ttfamily, fill=green!20},
git-del/.style={minimum height=\UseMacro{h-line}, minimum width=\UseMacro{w-line}, text width=\UseMacro{w-line}, text height=0.3 * \UseMacro{h-line}, font=\scriptsize\ttfamily, fill=red!20},
]{
\coordinate (c-nw) at (0, 0) {};
\node (b-del-6) [below right = 6 * \UseMacro{h-line} and 0 of c-nw] [git-del] {-};
\node (b-add-7) [below right = 7 * \UseMacro{h-line} and 0 of c-nw] [git-add] {+};
\node (b-del-9) [below right = 9.2 * \UseMacro{h-line} and 0 of c-nw] [git-del] {-};
\node (b-del-10) [below right = 10.2 * \UseMacro{h-line} and 0 of c-nw] [git-del] {-};
\node (b-code) [below right = 0 and 0 of c-nw] {\usebox{\boxExampleI}};
}
\end{tikzpicture}
\vspace{-15pt}
\subcaption{Code change between \revision \UseMacro{example1-version-old} and \UseMacro{example1-version-new}. \label{fig:example1:diff}}
\end{minipage}
\\
\begin{minipage}{.9\columnwidth}
\vspace{5pt}
\begin{lstlisting}[language=java-pretty-no-number, frame=single, xleftmargin=2pt, xrightmargin=4em]
org.apache.commons.lang3.builder.ReflectionDiffBuilderTest
org.apache.commons.lang3.reflect.FieldUtilsTest
org.apache.commons.lang3.reflect.TypeUtilsTest
org.apache.commons.lang3.time.StopWatchTest
\end{lstlisting}
\vspace{-3pt}
\subcaption{\Tests selected by \ijacocoTool. \label{fig:example1:tests-ijacoco}}
\end{minipage}

\caption{Example of using \ijacocoTool on \Prepo{\UseMacro{res-apache-commons-lang-name}}: when
computing the \codecov on \revision \UseMacro{example1-version-new}, \jacoco executes
all \UseMacro{example1-bjacoco-num-class} \tests and takes
\UseMacro{example1-bjacoco-time}, and \ijacocoTool only executes
\UseMacro{example1-ijacoco-num-class} \tests and takes \UseMacro{example1-ijacoco-time}
(speedup: \UseMacro{example1-speedup}{}$\times$).  \label{fig:example1}}
\end{small}
\end{center}
\end{figure}

\section{Background}
\label{sec:background}

In this section, we briefly introduce the
\codecovanalysis~(\S\ref{sec:background:coverage}) and RTS~(\S\ref{sec:background:rts})
techniques, which are the basis of our work.

\subsection{Code Coverage Analysis}
\label{sec:background:coverage}

\Codecov~\cite{Elmendorf69ControllingFunctionalTesting,ChilenskiAndMiller94ModifiedConditionDicisionCoverage,PiwowarskiETAL93Coverage,YangETAL06SurveyCoverage,ivankovic2019code,GoogleTestingBlogCoverage,frankl1993experimental,zhu1997software, gligoric2015guidelines, kochhar2015code, hemmati2015effective}
measures the adequacy of a \testsuite by quantifying the portion of \codeelems (e.g.,
lines) that have been covered (i.e., transitively executed) by the \tests.
The inputs to \codecovanalysis include a \codebase with a \testsuite, and the output is
\covdata, denoted as \aCovData, which records whether each \codeelem is covered.
\Codecovanalysis usually generates a human-readable report based on \aCovData, including
\codecov percentage at various granularities (e.g., line coverage, branch coverage), and
highlights the uncovered \codeelems.

A typical \codecovanalysis tool, such as \jacoco~\cite{JacocoGithubRepo}, maintains
\covdata in the format of a mapping from source code class (or file in non-object
oriented programming languages) to a set of \emph{probes} (denoted as $\aProbeSet =
\{\aProbe\}$) that are inserted into the class and then executed: $\aCovData = \{ \aClass \mapsto
\aProbeSet \}$.  A \probe \aProbe is an additional instruction instrumented by the tool
at the beginning of each basic block, which upon execution, adds itself to
$\aCovData[\aProbe.\text{class}]$.  Note that it is not necessary to place a \probe
before every \codeelem; instead, multiple \codeelems in the same basic block (i.e., on
the same execution path) can share one \probe.  If a \probe is executed during tests,
all the \codeelems in that basic block are considered as covered. A common
\codecovanalysis technique first instruments the \codebase to insert \probes, then
executes the \tests to collect the \covdata \aCovData, and finally generates a report
based on \aCovData.

\subsection{Regression Test Selection (RTS)}
\label{sec:background:rts}

Regression test
selection~\cite{gligoric2015ekstazi,gligoric2015practical,legunsen2017starts,legunsen2016extensive,liu2023more,zhang2018hybrid}
speeds up regression testing by only executing tests that are affected by code changes.
The inputs to RTS include two \revisions of a \codebase, the \testsuite on the new
\revision, and dependency graph from the old \revision.  The outputs are a subset of the
\testsuite whose behavior may change due to the code changes, and an updated dependency
graph to be used in the next \revision.
The workflow of an RTS tool has three phases: the \analysisphase selects \tests based on
code changes and the last \revision's dependency graph; the \executionphase executes the
selected \tests; and the \collectionphase collects the updated dependency graph.
RTS techniques vary by the granularity of the dependency graph and whether it is
collected statically or dynamically.  Prior work finds that using class-level (or
file-level) dependency graph is a ``sweet spot'' with low analysis overhead and decent
test selection
capability~\cite{gligoric2015ekstazi,gligoric2015practical,legunsen2017starts,legunsen2016extensive};
using more fine-grained dependency graph (e.g., method-level) helps RTS to be more
precise (i.e., avoid selecting \tests not affected by code changes) but usually at the
cost of higher analysis overhead and more engineering
effort~\cite{liu2023more,zhang2018hybrid}.
Dynamic RTS (i.e., collecting dependency graph via dynamic analysis) is usually more
precise and safer; static RTS (i.e., collecting dependency graph via static analysis)
can be faster and easier to perform offline (isolated from the
\executionphase)~\cite{legunsen2016extensive}.

In this work, we adopt the class-level dynamic RTS technique because its
good overall precision and safety.  We describe its three phases in more details.
Given two software \revisions, let \aClassSetOld be the classes and \aTestSetOld be the
\tests\footnote{We use ``\tests'' to refer to test classes in this paper.} on the old
\revision; let \aClassSetNew and \aTestSetNew be the classes and \tests on the new
\revision.
RTS requires the dependency graph collected from the old \revision $\aDepGraphOld =
\{\aTest \mapsto \aClass\}$, where: $\aTest \in \aTestSetOld$ is a \test, \aClass is a
(test or non-test) class that \aTest \emph{transitively} depends on; by definition,
$(\aTest \mapsto \aTest) \in \aDepGraphOld$.

The goal of the \analysisphase is to select a subset of \tests to be executed:
$\aTestSetNewSelectedRTS \subseteq \aTestSetNew$.  In this phase, RTS first figures out the
set of classes that have changed $\aClassSetChanged \subseteq \aClassSetOld$,
and then selects the \tests that (1)~depend on a changed class; or (2)~are added in the
new \revision:
$
\aTestSetNewSelectedRTS = \{\aTest \in \aTestSetNew \mid (\aTest
\mapsto \aClass) \in \aDepGraphOld \land \aClass \in \aClassSetChanged \} \sunion \aTestSetNew\setminus\aTestSetOld
$.
Specially, on the first \revision of using RTS, all test classes are selected
($\aTestSetNewSelectedRTS = \aTestSetNew$).

The \executionphase, which executes the selected test classes, and the \collectionphase,
which updates the dependency graph, are usually tightly integrated.  Specifically, RTS
performs instrumentation before executing tests to insert instructions that record the
dependency between (selected) tests and classes: $\aDepGraphDiff = \{\aTest \mapsto
\aClass\}$ for $\aTest \in \aTestSetNewSelectedRTS$.  Then, the updated dependency graph
can be represented as $\aDepGraphNew = \{(\aTest \mapsto \aClass) \in \aDepGraphOld \mid
\aTest \notin \aTestSetNewSelectedRTS\} \sunion \aDepGraphDiff$.  The updated dependency
graph will be used by RTS in the next \revision.

\section{Incremental Code Coverage Analysis}
\label{sec:technique}

\begin{figure}
\begin{center}
\scalebox{0.9}{

\DefMacro*{x-phase}{8em}
\DefMacro*{xsep-phase}{2em}
\DefMacro*{ysep-phase}{2ex}

\begin{tikzpicture}[
font=\small,
box/.style={rectangle, scale=1, inner sep=1pt, rounded corners=5pt, thick},
label/.style={scale=0.9, font={\bfseries\sc}},
anno/.style={font={\footnotesize}},
]

\coordinate (c-0) at (0,0) {};

\node (b-analysis) at (c-0)
[box, minimum width=14em, minimum height=30ex]
[draw=blue!50!black, fill=blue!10]
{};
\node (l-analysis) [below left = 0 and 0 of b-analysis.north east] [label] {\Analysis};

\node (b-execution) [below right = 2ex and \UseMacro{xsep-phase} of b-analysis.north east]
[box, minimum width=\UseMacro{x-phase}, minimum height=6ex]
[draw=red!50!black, fill=red!10]
{};
\node (l-execution) [below left = 0 and 0 of b-execution.north east] [label] {\Execution};

\node (b-collection) [below = \UseMacro{ysep-phase} of b-execution]
[box, minimum width=\UseMacro{x-phase}, minimum height=16ex]
[draw=yellow!50!black, fill=yellow!10]
{};
\node (l-collection) [below left = 0 and 0 of b-collection.north east] [label] {\Collection};

\coordinate (cy-tests) at (b-execution);
\coordinate (cy-dependency) at ($(b-collection.north) + (0, -4ex)$);
\coordinate (cy-covdata) at ($(b-collection.south) + (0, 4ex)$);

\coordinate (cx-input) at ($(b-analysis.west) + (2em, 0)$);

\coordinate (c-ClassSetChanged) at (cx-input |- cy-tests);
\node (l-ClassSetChanged) at (c-ClassSetChanged)
[anno]
{\aClassSetChanged};
\node (anno-ClassSetChanged) [below left = -0.5ex and 1em of l-ClassSetChanged.south west, anchor=north west]
[anno]
{change set};

\coordinate (c-DepGraphOld) at (cx-input |- cy-dependency);
\node (l-DepGraphOld) at (c-DepGraphOld)
[anno]
{\aDepGraphOld};
\node (anno-DepGraphOld) [below left = -0.5ex and 1em of l-DepGraphOld.south west, anchor=north west]
[anno]
{dependency graph};

\coordinate (c-CovDataOld) at (cx-input |- cy-covdata);
\node (l-CovDataOld) at (c-CovDataOld)
[anno]
{\aCovDataOld};
\node (anno-CovDataOld) [below left = -0.5ex and 1em of l-CovDataOld.south west, anchor=north west]
[anno]
{\covdata};

\coordinate (cx-analysis1) at ($(b-analysis.west)!0.35!(b-analysis.east)$);
\coordinate (cx-analysis2) at ($(b-analysis.west)!0.6!(b-analysis.east)$);
\coordinate (cx-analysis3) at ($(cx-analysis2) + (1.65em, 0)$);
\coordinate (cx-analysis4) at ($(b-analysis.west)!0.875!(b-analysis.east)$);

\coordinate (c-TestSetNewSelectedRTS) at (cx-analysis1 |- cy-tests);
\node (l-TestSetNewSelectedRTS) at (c-TestSetNewSelectedRTS)
[anno]
{\aTestSetNewSelectedRTS};

\coordinate (cy-tests-collection) at ($(b-execution)!0.5!(b-collection)$);
\coordinate (c-ClassSetCovInvalid) at (cx-analysis2 |- cy-tests-collection);
\node (l-ClassSetCovInvalid) at (c-ClassSetCovInvalid)
[anno]
{\aClassSetCovInvalid};

\node (l-RemoveCovData) at (cx-analysis3 |- cy-covdata)
[anno, scale=1.5, inner sep=0]
{$\ominus$};
\node (anno-RemoveCovData) [below = -0.5ex of l-RemoveCovData.south]
[anno]
{remove outdated};

\coordinate (c-TestSetNewSelectedI) at (cx-analysis4 |- cy-tests);
\node (l-TestSetNewSelectedI) at (c-TestSetNewSelectedI)
[anno]
{\aTestSetNewSelectedI};

\coordinate (cx-covdata-joint) at ($(b-collection.west)!0.15!(b-collection.east)$);
\coordinate (cx-collection-joint) at ($(cx-covdata-joint) + (1.5em, 0)$);
\coordinate (cx-collection-output) at ($(b-collection.west)!0.55!(b-collection.east)$);
\coordinate (cx-covdata-output1) at ($(b-collection.west)!0.55!(b-collection.east)$);
\coordinate (cx-covdata-output2) at ($(b-collection.west)!0.85!(b-collection.east)$);
\coordinate (cy-covdata-output2) at ($(cy-covdata)!0.6!(b-collection.south)$);

\node (l-CovDataDiff) at (cx-covdata-joint |- cy-tests)
[anno]
{\aCovDataDiff};

\node (l-DepGraphDiff) at (cx-collection-joint |- cy-tests)
[anno]
{\aDepGraphDiff};

\node (l-DepGraphNew) at (cx-collection-output |- cy-dependency)
[anno]
{\aDepGraphNew};

\node (l-CovDataNew) at (cx-covdata-output1 |- cy-covdata)
[anno]
{\aCovDataNew};

\node (l-ReportGen) at (cx-covdata-output1 |- b-analysis.south)
[anno, yshift=1ex]
{report generation};

\node (l-MergeDepGraph) at (cx-collection-joint |- cy-dependency)
[anno, scale=1.5, inner sep=0]
{$\oplus$};
\node (anno-MergeDepGraph) [below = -0.5ex of l-MergeDepGraph.south]
[anno]
{merge};

\node (l-MergeCovData) at (cx-covdata-joint |- cy-covdata)
[anno, scale=1.5, inner sep=0]
{$\oplus$};
\node (anno-MergeCovData) [below = -0.5ex of l-MergeCovData.south]
[anno]
{merge};

\coordinate (cx-lastv) at ($(b-analysis.west) + (-0.5em, 0)$);
\coordinate (cx-nextv) at ($(b-collection.east) + (0.5em, 0)$);
\coordinate (cy-dependency-covdata) at ($(cy-dependency)!0.5!(cy-covdata)$);

\node (l-lastv) at (cx-lastv |- cy-dependency-covdata)
[anno, rotate=270, yshift=-1em]
{from old \revision};
\node (l-nextv) at (cx-nextv |- cy-dependency-covdata)
[anno, rotate=270, yshift=1em]
{to next \revision};

\draw[->] (l-ClassSetChanged) -- (l-TestSetNewSelectedRTS);
\draw[->] (l-TestSetNewSelectedRTS) -- (l-TestSetNewSelectedI);
\draw[->] (l-TestSetNewSelectedI) -- (b-execution.west);
\draw[->] (l-DepGraphOld) -- (l-MergeDepGraph);
\draw[->] (l-MergeDepGraph) -- (l-DepGraphNew);
\draw[->] (l-CovDataOld) -- (l-RemoveCovData);
\draw[->] (l-RemoveCovData) -- (l-MergeCovData);
\draw[->] (l-MergeCovData) -- (l-CovDataNew);
\draw[->] (cx-lastv |- l-DepGraphOld) -- (l-DepGraphOld);
\draw[->] (cx-lastv |- l-CovDataOld) -- (l-CovDataOld);
\draw[->] (l-DepGraphNew) -- (cx-nextv |- l-DepGraphNew);
\draw[->] (l-CovDataNew) -- (cx-nextv |- l-CovDataNew);
\draw[->] (l-CovDataDiff) -- (l-MergeCovData);
\draw[->] (l-DepGraphDiff) -- (l-MergeDepGraph);

\draw[->, rounded corners=3pt] (l-DepGraphOld.east) -- ($(l-DepGraphOld.east) + (1em, 0)$) -- ($(l-ClassSetCovInvalid.west) + (-0.5em, 0)$) -- (l-ClassSetCovInvalid.west);
\draw[rounded corners=3pt] (l-TestSetNewSelectedRTS.east) -- ($(l-TestSetNewSelectedRTS.east) + (0.5em, 0)$) -- ($(l-ClassSetCovInvalid.west) + (-0.5em, 0)$) -- (l-ClassSetCovInvalid.west);

\draw[rounded corners=3pt] (l-ClassSetCovInvalid.east) -- ($(l-ClassSetCovInvalid.east) + (0.5em, 0)$) -- ($(l-TestSetNewSelectedI.west) + (-1em, 0)$) -- (l-TestSetNewSelectedI.west);

\draw[->] (l-CovDataNew.south) -- (l-ReportGen.north);

\draw[->, rounded corners=3pt] (l-ClassSetCovInvalid.east) -| (l-RemoveCovData.north);

\end{tikzpicture}

}
\end{center}
\caption{Workflow of \icodecovanalysis.} \label{fig:workflow}
\end{figure}

Figure~\ref{fig:workflow} shows the workflow of \icodecovanalysis, consisting of three
phases: \analysis, \execution, and \collection.  We first define the inputs and outputs
of the workflow (\S\ref{sec:technique:inout}), then describe the three phases
(\S\ref{sec:technique:analysis}--\S\ref{sec:technique:collection}), and finally prove
the correctness of our technique (\S\ref{sec:technique:correctness}).  We use symbols
without prime (e.g., \aDepGraphOld) to denote the data on the old \revision, and symbols
with prime (e.g., \aDepGraphNew) to denote the data on the new \revision.

\subsection{Inputs and Outputs}
\label{sec:technique:inout}

\Icodecovanalysis requires four inputs: the old and new \revisions of the \codebase,
the dependency graph collected from the old \revision \aDepGraphOld (similar to RTS),
and the \covdata collected from the old \revision \aCovDataOld. In this work, we focus
on integrating class-level dynamic RTS and \codecovanalysis which groups \covdata at
class-level.  Based on the findings in related work on
RTS~\cite{gligoric2015ekstazi,gligoric2015practical,legunsen2017starts,legunsen2016extensive}
and the fact that existing \codecovanalysis tools group \covdata at class-level (such as
\jacoco~\cite{JacocoGithubRepo}), we believe that integrating the two techniques at
class-level would lead to the best performance; we leave the exploration of other levels
of integration to future work.  Specifically, the \changeset $\aClassSetChanged = \{
\aClass \}$ is the set of classes that changed; the dependency graph $\aDepGraphOld =
\{\aTest \mapsto \aClass\}$ maps each \test to the (test or non-test) classes it
\emph{transitively} depends on; and the coverage data $\aCovDataOld = \{ \aClass \mapsto
\aProbeSet \}$ maps each class to the \probes in that class that are executed during
testing.

The outputs of \icodecovanalysis include: (1)~the updated dependency graph
\aDepGraphNew, which will be used by the \analysisphase in the \revision; (2)~the
\covdata \aCovDataNew, which is used to generate coverage reports and will also be used
by the next \revision.

\subsection{\Analysis Phase}
\label{sec:technique:analysis}

The goal of the \analysis phase is to select a subset of \tests that should be executed.
We start from the set of \tests selected by RTS, denoted by \aTestSetNewSelectedRTS:
\begin{flalign*}
\aTestSetNewSelectedRTS &= \aRTS(\aDepGraphOld, \aClassSetChanged) \\
&= \{\aTest \in \aTestSetNew \mid (\aTest
\mapsto \aClass) \in \aDepGraphOld \land \aClass \in \aClassSetChanged \} \sunion \aTestSetNew\setminus\aTestSetOld
\end{flalign*}
These \tests must be executed because their behaviors may change due to the \changeset.

However, for the purpose of updating \codecov, executing only RTS-selected \tests may be
insufficient.  Let's first consider the part of \covdata that should be updated after
the \changeset.  Since the \tests in \aTestSetNewSelectedRTS must be executed, the
\covdata for all the classes that they transitively depend on may change; we denote this
set of classes as \aClassSetCovInvalid which can be computed by looking up
\aDepGraphOld:
$$
\aClassSetCovInvalid = \bigcup_{\aTest \in \aTestSetNewSelectedRTS}{\aDepGraphOld[\aTest]}
$$

The old \covdata for the classes in \aClassSetCovInvalid should be discarded, because
the \tests in \aTestSetNewSelectedRTS may execute in different paths in the new
\revision and cover different \codeelems than before.  To collect the new \covdata for
the classes in \aClassSetCovInvalid, we need to execute all \tests that depend on them,
denoted as \aTestSetNewSelectedI (which is also the final set of selected \tests):
$$
\aTestSetNewSelectedI = \{ \aTest \mid (\aTest \mapsto \aClass) \in \aDepGraphOld \land \aClass \in \aClassSetCovInvalid \}
$$
Note that $\aTestSetNewSelectedRTS \subseteq \aTestSetNewSelectedI$, because for any
\test in \aTestSetNewSelectedRTS, there must be some classes in \aClassSetCovInvalid
that it depends on.

\begin{figure}[t]
\begin{center}
\begin{small}

\newsavebox{\lstExampleSi}
\begin{lrbox}{\lstExampleSi}
\begin{lstlisting}[language=java-pretty, numbersep=2pt]
class c1 {
c1() {}
void m() {return;}
}
\end{lstlisting}
\end{lrbox}

\newsavebox{\lstExampleSii}
\begin{lrbox}{\lstExampleSii}
\begin{lstlisting}[language=java-pretty, numbersep=2pt]
class c2 {
c2() {}
void m() {new c1().m();}
}
\end{lstlisting}
\end{lrbox}

\newsavebox{\lstExampleSiii}
\begin{lrbox}{\lstExampleSiii}
\begin{lstlisting}[language=java-pretty, numbersep=2pt]
class c3 {
c3() {}
void m() {return;}
}
\end{lstlisting}
\end{lrbox}

\newsavebox{\lstExampleTi}
\begin{lrbox}{\lstExampleTi}
\begin{lstlisting}[language=java-pretty, numbersep=2pt]
class t1 {
@Test void t() {
new c1();
}}
\end{lstlisting}
\end{lrbox}

\newsavebox{\lstExampleTii}
\begin{lrbox}{\lstExampleTii}
\begin{lstlisting}[language=java-pretty, numbersep=2pt]
class t2 {
@Test void t() {
new c2().m();
}}
\end{lstlisting}
\end{lrbox}

\newsavebox{\lstExampleTiii}
\begin{lrbox}{\lstExampleTiii}
\begin{lstlisting}[language=java-pretty, numbersep=2pt]
class t3 {
@Test void t() {
new c3();
}}
\end{lstlisting}
\end{lrbox}

\DefMacro*{xsep}{3em}
\DefMacro*{ysep}{2ex}

\begin{tikzpicture}[
font=\footnotesize,
codeblock/.style={rectangle, rounded corners, draw=black, minimum height=4ex, inner xsep=8pt},
test codeblock/.style={codeblock, text width=6em},
src codeblock/.style={codeblock, text width=9em},
anno/.style={font={\footnotesize}},
]

\coordinate (c-0) at (0, 0);

\coordinate (c-test) at (c-0);
\node (b-t1) [below right = 0 and 0 of c-test, anchor=north west]
[test codeblock]
{\usebox{\lstExampleTi}};

\node (b-t2) [below right = \UseMacro{ysep} and 0 of b-t1.south west, anchor=north west]
[test codeblock]
{\usebox{\lstExampleTii}};

\node (b-t3) [below right = \UseMacro{ysep} and 0 of b-t2.south west, anchor=north west]
[test codeblock]
{\usebox{\lstExampleTiii}};

\coordinate (c-src) at ($(c-test -| b-t1.east) + (\UseMacro{xsep}, 0)$);
\node (b-c1) [below right = 0 and 0 of c-src, anchor=north west]
[src codeblock]
{\usebox{\lstExampleSi}};

\node (b-c2) [below right = \UseMacro{ysep} and 0 of b-c1.south west, anchor=north west]
[src codeblock]
{\usebox{\lstExampleSii}};

\coordinate (c-diffs) at ($(b-c2.east) + (0, -2ex)$);
\node (b-diff1) [above right = 0.5ex and 1em of c-diffs]
[anno]
{\circled{a} \CodeIn{return;}};
\node (b-diff2) [below right = 0.5ex and 1em of c-diffs]
[anno]
{\circled{b} \CodeIn{new\ c3().m();}};
\node (anno-diffs) [above = 0.5ex of b-diff1.north]
[anno]
{\emph{possible \changeset}};

\node (b-c3) [below right = \UseMacro{ysep} and 0 of b-c2.south west, anchor=north west]
[src codeblock]
{\usebox{\lstExampleSiii}};

\draw[->] (b-t1) -- (b-c1);
\draw[->] (b-t2) -- (b-c2);
\draw[->] (b-t3) -- (b-c3);
\draw[->] (b-c2) -- (b-c1);

\draw[dotted, thick] (c-diffs) -- (b-diff1.west);
\draw[dotted, thick] (c-diffs) -- (b-diff2.west);

\end{tikzpicture}

\caption{Example showing the necessity for \icodecovanalysis selecting more test classes than RTS; directed edges represent ``depends on'' relationship.}\label{fig:selection-counter-example}
\end{small}
\end{center}
\end{figure}
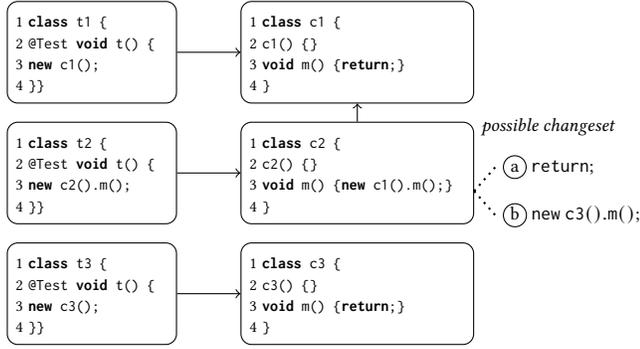

\MyPara{Necessity for selecting more \tests than RTS} One may wonder why we need to
select additional \tests in \aTestSetNewSelectedI compared to \aTestSetNewSelectedRTS.
Figure~\ref{fig:selection-counter-example} provides a counter-example where selecting
only \aTestSetNewSelectedRTS is not sufficient.  In this example, $\aTest_1, \aTest_2$
are \tests, and $\aClass_1, \aClass_2, \aClass_3$ are \nontest classes. The dependency
graph on the old \revision is:
$$
\aDepGraphOld = \biggl\{\makecell[c]{\aTest_1 \mapsto \aClass_1\\ \aTest_1 \mapsto \aTest_1}\quad
\makecell[c]{\aTest_2 \mapsto \aClass_1\\ \aTest_2 \mapsto \aClass_2\\ \aTest_2 \mapsto \aTest_2}\quad
\makecell[c]{\aTest_3 \mapsto \aClass_3\\ \aTest_3 \mapsto \aTest_3}\biggr\}
$$
On the shown (old) \revision, the covered lines include $\aClass_1$'s lines~2--3,
$\aClass_2$'s lines 2--3, and $\aClass_3$'s line~2.
Consider the \changeset \circled{a} which modifies $\aClass_2$'s line~3 to
``\CodeIn{return;}'', i.e., removing its dependency to $\aClass_1$. RTS will only select
$\aTestSetNewSelectedRTS = \{\aTest_2\}$. Executing $\aTest_2$ will still cover
$\aClass_2$'s lines, but no longer covers $\aClass_1$'s lines.  In fact, no \test would
cover $\aClass_1$'s line~3 after the change.  Since the change may results in a decrease
in $\aClass_1$'s coverage, without executing $\aTest_1$ or knowing which lines of
$\aClass_1$ are covered by $\aTest_1$ and $\aTest_2$, we cannot update the \covdata for
$\aClass_1$ correctly.  Thus, the final selected \tests should be $\aTestSetNewSelectedI
= \{\aTest_1, \aTest_2\}$.

An alternative approach, which has been explored in a prior work on using code coverage
information to assist regression testing
selection~\cite{ChittimalliAndHarrold09CoverageRegressionTesting}, is to collect
\covdata of each \test separately.
However, this approach is inefficient because it incurs overhead of
(1)~storing multiple copies of \covdata; and (2)~merging the \covdata from all \tests to
compute the union of the sets of \probes covered, increasing the time complexity from
$O(|\aClassSet|)$ to $O(|\aTestSet| |\aClassSet|)$; similar observations have been
reported in prior work~\cite{nie2020debugging}.
When making this design decision, we performed a preliminary study on our dataset and
found that performing \codecovanalysis for each \test separately (i.e., storing one
\covdata per \test but without merging them yet) causes an average of 76\% overhead when
compared to performing \codecovanalysis once for the entire \testsuite.

Moreover, collecting \covdata per \test may also result in missing \covdata of some
\tests, because certain \codeelems (such as static initializers) are only executed once
across the entire \testsuite, and thus will only be recorded by the first \test that
executes them.  Carefully handling such cases will require engineering effort and may
incur more overhead. As a consequence, we follow the common design decision of
\codecovanalysis tools to collect one copy of \covdata for the entire \testsuite.

\subsection{\Execution Phase}
\label{sec:technique:execution}

The \execution phase executes the selected \tests \aTestSetNewSelectedI, with
instrumentation required by both \codecovanalysis and RTS.  For \codecovanalysis, we
insert \probes to record which \codeelems are executed during execution; at the end of
the execution, we should get the incremental \covdata \aCovDataDiff for the selected \tests.
For RTS, we insert instructions to record all the classes that each \test transitively
depends on during execution; at the end, we should get the incremental dependency graph
\aDepGraphDiff for the selected \tests.
The two sets of instrumentation do not interfere with each other because RTS's
instrumentation never change the program execution path (e.g., it never adds, removes,
or updates branches).

\subsection{\Collection Phase}
\label{sec:technique:collection}

The \collection phase updates the dependency graph and merges the \covdata.
Namely, to be able to perform RTS in the next \revision, an updated dependency graph is
computed based on the information collected from the \execution phase:
$$
\aDepGraphNew = \{ (\aTest \mapsto \aClass) \in \aDepGraphOld \mid \aTest\notin \aTestSetNewSelectedI \} \sunion \aDepGraphDiff
$$
Although comparing to RTS (\S\ref{sec:background:rts}), we update the dependency graph
of more \tests (recall that $\aTestSetNewSelectedRTS \subseteq \aTestSetNewSelectedI$),
it will not affect the correctness of the updated dependency graph.

Then, we update the \covdata by integrating the incremental \covdata \aCovDataDiff
collected during the \execution phase into the old \covdata \aCovDataOld:
$$
\aCovDataNew = \{ (\aClass \mapsto \aElem) \in \aCovDataOld \mid \aClass\notin \aClassSetCovInvalid \} \sunion \aCovDataDiff
$$
Note that the \covdata for the classes in \aClassSetCovInvalid are completely
overwritten, and the \covdata for the other classes are \emph{merged}.  To illustrate
the merge process, consider the \changeset \circled{b} in
Figure~\ref{fig:selection-counter-example} which modifies $\aClass_1$'s line~2 to
``\CodeIn{new\ c3().m();}''.  Recall that the covered lines for $\aClass_3$ in the old
\revision is line~2, and the selected \tests $\aTestSetNewSelectedI = \{\aTest_1,
\aTest_2\}$. When executing $\aTest_2$, line~3 of $\aClass_3$ is also covered.  Since
there is no \test that depends on $\aClass_3$ whose execution paths may change due to the
\changeset, there will not be any reduction in $\aClass_3$'s coverage.  Therefore, the
old \covdata for $\aClass_3$ should be kept and merged with the incremental \covdata
collected during the \execution phase. In this case, $\aCovDataOld = \{ \aClass_3
\mapsto \{2\}, ...\}$, $\aCovDataDiff = \{ \aClass_3 \mapsto \{3\}, ... \}$, and thus
$\aCovDataNew = \{ \aClass_3 \mapsto \{2, 3\}, ...\}$.  This is consistent with the
result of re-executing all \tests in the new \revision.

\begin{table*}[t]
\begin{small}
\begin{center}
\caption{\Repos used in our evaluation. \label{tab:projects}}

\begin{tabular}{l|l|c|r|r|r|r|r}
\toprule
\multicolumn{1}{c|}{\multirow{2}{*}{\textbf{\UseMacro{TH-project}}}} & \multicolumn{1}{c|}{\multirow{2}{*}{\textbf{\UseMacro{TH-head-revision}}}} & \multicolumn{1}{c|}{\multirow{2}{*}{\textbf{\UseMacro{TH-num-revision}}}} & \multicolumn{2}{c|}{\textbf{\UseMacro{TH-project-size}}} & \multicolumn{3}{c}{\textbf{\UseMacro{TH-project-test}}} \\
& & & \multicolumn{1}{c}{\UseMacro{TH-num-file}} & \multicolumn{1}{c|}{\UseMacro{TH-loc}} & \multicolumn{1}{c}{\UseMacro{TH-num-test-class}} & \multicolumn{1}{c}{\UseMacro{TH-num-test-method}} & \multicolumn{1}{c}{\UseMacro{TH-test-time}} \\
\midrule
\href{\UseMacro{res-asterisk-java-asterisk-java-url}}{\UseMacro{res-asterisk-java-asterisk-java-name}} & \texttt{\UseMacro{res-asterisk-java-asterisk-java-head}} & \UseMacro{res-asterisk-java-asterisk-java-num-ver} & \UseMacro{res-asterisk-java-asterisk-java-num-file} & \UseMacro{res-asterisk-java-asterisk-java-loc} & \UseMacro{res-asterisk-java-asterisk-java-num-class} & \UseMacro{res-asterisk-java-asterisk-java-num-method} & \UseMacro{res-asterisk-java-asterisk-java-time}    \\
\href{\UseMacro{res-apache-commons-beanutils-url}}{\UseMacro{res-apache-commons-beanutils-name}} & \texttt{\UseMacro{res-apache-commons-beanutils-head}} & \UseMacro{res-apache-commons-beanutils-num-ver} & \UseMacro{res-apache-commons-beanutils-num-file} & \UseMacro{res-apache-commons-beanutils-loc} & \UseMacro{res-apache-commons-beanutils-num-class} & \UseMacro{res-apache-commons-beanutils-num-method} & \UseMacro{res-apache-commons-beanutils-time}    \\
\href{\UseMacro{res-apache-commons-codec-url}}{\UseMacro{res-apache-commons-codec-name}} & \texttt{\UseMacro{res-apache-commons-codec-head}} & \UseMacro{res-apache-commons-codec-num-ver} & \UseMacro{res-apache-commons-codec-num-file} & \UseMacro{res-apache-commons-codec-loc} & \UseMacro{res-apache-commons-codec-num-class} & \UseMacro{res-apache-commons-codec-num-method} & \UseMacro{res-apache-commons-codec-time}    \\
\href{\UseMacro{res-apache-commons-collections-url}}{\UseMacro{res-apache-commons-collections-name}} & \texttt{\UseMacro{res-apache-commons-collections-head}} & \UseMacro{res-apache-commons-collections-num-ver} & \UseMacro{res-apache-commons-collections-num-file} & \UseMacro{res-apache-commons-collections-loc} & \UseMacro{res-apache-commons-collections-num-class} & \UseMacro{res-apache-commons-collections-num-method} & \UseMacro{res-apache-commons-collections-time}    \\
\href{\UseMacro{res-apache-commons-compress-url}}{\UseMacro{res-apache-commons-compress-name}} & \texttt{\UseMacro{res-apache-commons-compress-head}} & \UseMacro{res-apache-commons-compress-num-ver} & \UseMacro{res-apache-commons-compress-num-file} & \UseMacro{res-apache-commons-compress-loc} & \UseMacro{res-apache-commons-compress-num-class} & \UseMacro{res-apache-commons-compress-num-method} & \UseMacro{res-apache-commons-compress-time}    \\
\href{\UseMacro{res-apache-commons-configuration-url}}{\UseMacro{res-apache-commons-configuration-name}} & \texttt{\UseMacro{res-apache-commons-configuration-head}} & \UseMacro{res-apache-commons-configuration-num-ver} & \UseMacro{res-apache-commons-configuration-num-file} & \UseMacro{res-apache-commons-configuration-loc} & \UseMacro{res-apache-commons-configuration-num-class} & \UseMacro{res-apache-commons-configuration-num-method} & \UseMacro{res-apache-commons-configuration-time}    \\
\href{\UseMacro{res-apache-commons-dbcp-url}}{\UseMacro{res-apache-commons-dbcp-name}} & \texttt{\UseMacro{res-apache-commons-dbcp-head}} & \UseMacro{res-apache-commons-dbcp-num-ver} & \UseMacro{res-apache-commons-dbcp-num-file} & \UseMacro{res-apache-commons-dbcp-loc} & \UseMacro{res-apache-commons-dbcp-num-class} & \UseMacro{res-apache-commons-dbcp-num-method} & \UseMacro{res-apache-commons-dbcp-time}    \\
\href{\UseMacro{res-apache-commons-imaging-url}}{\UseMacro{res-apache-commons-imaging-name}} & \texttt{\UseMacro{res-apache-commons-imaging-head}} & \UseMacro{res-apache-commons-imaging-num-ver} & \UseMacro{res-apache-commons-imaging-num-file} & \UseMacro{res-apache-commons-imaging-loc} & \UseMacro{res-apache-commons-imaging-num-class} & \UseMacro{res-apache-commons-imaging-num-method} & \UseMacro{res-apache-commons-imaging-time}    \\
\href{\UseMacro{res-apache-commons-io-url}}{\UseMacro{res-apache-commons-io-name}} & \texttt{\UseMacro{res-apache-commons-io-head}} & \UseMacro{res-apache-commons-io-num-ver} & \UseMacro{res-apache-commons-io-num-file} & \UseMacro{res-apache-commons-io-loc} & \UseMacro{res-apache-commons-io-num-class} & \UseMacro{res-apache-commons-io-num-method} & \UseMacro{res-apache-commons-io-time}    \\
\href{\UseMacro{res-apache-commons-lang-url}}{\UseMacro{res-apache-commons-lang-name}} & \texttt{\UseMacro{res-apache-commons-lang-head}} & \UseMacro{res-apache-commons-lang-num-ver} & \UseMacro{res-apache-commons-lang-num-file} & \UseMacro{res-apache-commons-lang-loc} & \UseMacro{res-apache-commons-lang-num-class} & \UseMacro{res-apache-commons-lang-num-method} & \UseMacro{res-apache-commons-lang-time}    \\
\href{\UseMacro{res-apache-commons-math-url}}{\UseMacro{res-apache-commons-math-name}} & \texttt{\UseMacro{res-apache-commons-math-head}} & \UseMacro{res-apache-commons-math-num-ver} & \UseMacro{res-apache-commons-math-num-file} & \UseMacro{res-apache-commons-math-loc} & \UseMacro{res-apache-commons-math-num-class} & \UseMacro{res-apache-commons-math-num-method} & \UseMacro{res-apache-commons-math-time}    \\
\href{\UseMacro{res-apache-commons-net-url}}{\UseMacro{res-apache-commons-net-name}} & \texttt{\UseMacro{res-apache-commons-net-head}} & \UseMacro{res-apache-commons-net-num-ver} & \UseMacro{res-apache-commons-net-num-file} & \UseMacro{res-apache-commons-net-loc} & \UseMacro{res-apache-commons-net-num-class} & \UseMacro{res-apache-commons-net-num-method} & \UseMacro{res-apache-commons-net-time}    \\
\href{\UseMacro{res-apache-commons-pool-url}}{\UseMacro{res-apache-commons-pool-name}} & \texttt{\UseMacro{res-apache-commons-pool-head}} & \UseMacro{res-apache-commons-pool-num-ver} & \UseMacro{res-apache-commons-pool-num-file} & \UseMacro{res-apache-commons-pool-loc} & \UseMacro{res-apache-commons-pool-num-class} & \UseMacro{res-apache-commons-pool-num-method} & \UseMacro{res-apache-commons-pool-time}    \\
\href{\UseMacro{res-alibaba-fastjson-url}}{\UseMacro{res-alibaba-fastjson-name}} & \texttt{\UseMacro{res-alibaba-fastjson-head}} & \UseMacro{res-alibaba-fastjson-num-ver} & \UseMacro{res-alibaba-fastjson-num-file} & \UseMacro{res-alibaba-fastjson-loc} & \UseMacro{res-alibaba-fastjson-num-class} & \UseMacro{res-alibaba-fastjson-num-method} & \UseMacro{res-alibaba-fastjson-time}    \\
\href{\UseMacro{res-finmath-finmath-lib-url}}{\UseMacro{res-finmath-finmath-lib-name}} & \texttt{\UseMacro{res-finmath-finmath-lib-head}} & \UseMacro{res-finmath-finmath-lib-num-ver} & \UseMacro{res-finmath-finmath-lib-num-file} & \UseMacro{res-finmath-finmath-lib-loc} & \UseMacro{res-finmath-finmath-lib-num-class} & \UseMacro{res-finmath-finmath-lib-num-method} & \UseMacro{res-finmath-finmath-lib-time}    \\
\href{\UseMacro{res-sonyxperiadev-gerrit-events-url}}{\UseMacro{res-sonyxperiadev-gerrit-events-name}} & \texttt{\UseMacro{res-sonyxperiadev-gerrit-events-head}} & \UseMacro{res-sonyxperiadev-gerrit-events-num-ver} & \UseMacro{res-sonyxperiadev-gerrit-events-num-file} & \UseMacro{res-sonyxperiadev-gerrit-events-loc} & \UseMacro{res-sonyxperiadev-gerrit-events-num-class} & \UseMacro{res-sonyxperiadev-gerrit-events-num-method} & \UseMacro{res-sonyxperiadev-gerrit-events-time}    \\
\href{\UseMacro{res-brettwooldridge-HikariCP-url}}{\UseMacro{res-brettwooldridge-HikariCP-name}} & \texttt{\UseMacro{res-brettwooldridge-HikariCP-head}} & \UseMacro{res-brettwooldridge-HikariCP-num-ver} & \UseMacro{res-brettwooldridge-HikariCP-num-file} & \UseMacro{res-brettwooldridge-HikariCP-loc} & \UseMacro{res-brettwooldridge-HikariCP-num-class} & \UseMacro{res-brettwooldridge-HikariCP-num-method} & \UseMacro{res-brettwooldridge-HikariCP-time}    \\
\href{\UseMacro{res-lmdbjava-lmdbjava-url}}{\UseMacro{res-lmdbjava-lmdbjava-name}} & \texttt{\UseMacro{res-lmdbjava-lmdbjava-head}} & \UseMacro{res-lmdbjava-lmdbjava-num-ver} & \UseMacro{res-lmdbjava-lmdbjava-num-file} & \UseMacro{res-lmdbjava-lmdbjava-loc} & \UseMacro{res-lmdbjava-lmdbjava-num-class} & \UseMacro{res-lmdbjava-lmdbjava-num-method} & \UseMacro{res-lmdbjava-lmdbjava-time}    \\
\href{\UseMacro{res-logic-ng-LogicNG-url}}{\UseMacro{res-logic-ng-LogicNG-name}} & \texttt{\UseMacro{res-logic-ng-LogicNG-head}} & \UseMacro{res-logic-ng-LogicNG-num-ver} & \UseMacro{res-logic-ng-LogicNG-num-file} & \UseMacro{res-logic-ng-LogicNG-loc} & \UseMacro{res-logic-ng-LogicNG-num-class} & \UseMacro{res-logic-ng-LogicNG-num-method} & \UseMacro{res-logic-ng-LogicNG-time}    \\
\href{\UseMacro{res-davidmoten-rxjava-extras-url}}{\UseMacro{res-davidmoten-rxjava-extras-name}} & \texttt{\UseMacro{res-davidmoten-rxjava-extras-head}} & \UseMacro{res-davidmoten-rxjava-extras-num-ver} & \UseMacro{res-davidmoten-rxjava-extras-num-file} & \UseMacro{res-davidmoten-rxjava-extras-loc} & \UseMacro{res-davidmoten-rxjava-extras-num-class} & \UseMacro{res-davidmoten-rxjava-extras-num-method} & \UseMacro{res-davidmoten-rxjava-extras-time}    \\
\href{\UseMacro{res-bullhorn-sdk-rest-url}}{\UseMacro{res-bullhorn-sdk-rest-name}} & \texttt{\UseMacro{res-bullhorn-sdk-rest-head}} & \UseMacro{res-bullhorn-sdk-rest-num-ver} & \UseMacro{res-bullhorn-sdk-rest-num-file} & \UseMacro{res-bullhorn-sdk-rest-loc} & \UseMacro{res-bullhorn-sdk-rest-num-class} & \UseMacro{res-bullhorn-sdk-rest-num-method} & \UseMacro{res-bullhorn-sdk-rest-time}    \\
\href{\UseMacro{res-tabulapdf-tabula-java-url}}{\UseMacro{res-tabulapdf-tabula-java-name}} & \texttt{\UseMacro{res-tabulapdf-tabula-java-head}} & \UseMacro{res-tabulapdf-tabula-java-num-ver} & \UseMacro{res-tabulapdf-tabula-java-num-file} & \UseMacro{res-tabulapdf-tabula-java-loc} & \UseMacro{res-tabulapdf-tabula-java-num-class} & \UseMacro{res-tabulapdf-tabula-java-num-method} & \UseMacro{res-tabulapdf-tabula-java-time}    \\
\midrule
\multicolumn{1}{c|}{\UseMacro{TH-sum}} & \UseMacro{TH-na} & \UseMacro{TH-na} & \UseMacro{res-projects-sum-num-file} & \UseMacro{res-projects-sum-loc} & \UseMacro{res-projects-sum-num-test-class} & \UseMacro{res-projects-sum-num-test-method} & \UseMacro{res-projects-sum-time} \\
\bottomrule \end{tabular}

\end{center}
\end{small}
\end{table*}

\subsection{Correctness}
\label{sec:technique:correctness}

In this subsection, we prove the correctness of \icodecovanalysis.  Our baseline is traditional
\codecovanalysis (\S\ref{sec:background:coverage}), which collects \codecov by executing
all \tests.  Correctness in our case means that the \covdata collected by
\icodecovanalysis is the same as the one collected by the baseline.  The correctness of
our technique depends on the safety of the RTS technique.  A safe RTS selects all
\tests whose behaviors might be affected by the \changeset.

\MyPara{Theorem} \Icodecovanalysis is correct if the underlying RTS technique is safe.

\MyPara{Proof}
Given the \changeset \aClassSetChanged, a safe RTS technique should select all \tests
\aTestSetNewSelectedRTS whose behaviors may change.  This means that in the extreme
case, all classes in \aClassSetCovInvalid (recall this is the set of classes any \test
in \aTestSetNewSelectedRTS transitively depends on) may have their coverage increased or
decreased.  That is why we need to execute all \tests in \aTestSetNewSelectedI to
completely overwrite the \covdata for classes in \aClassSetCovInvalid.

Now assume there is a class $\aClass \notin \aClassSetCovInvalid$ that is a transitive
dependent of some \tests in \aTestSetNewSelectedI.  We assert that \covdata for \aClass
can only increase due to \aClassSetChanged (and thus is safe to be merged with
\aCovDataOld from the old \revision).  This is because if the \changeset result in any
decrease in \aClass's coverage, then \changeset depends on \aClass, which indicates that
$\aClass \in \aClassSetCovInvalid$.  This contradicts with our assumption. \hfill $\square$

\section{\ijacocoTool Implementation}
\label{sec:impl}

\ijacocoTool is built on top of \jacoco and \ekstazi as the underlying \codecovanalysis and RTS
tools, respectively.  In this section, we first describe how each phase of \ijacocoTool
is implemented in \S\ref{sec:impl:analysis}--\S\ref{sec:impl:collection}, but focus on
integration details rather than repeating the techniques already described in
\S\ref{sec:technique}.  Then, \S\ref{sec:impl:maven} describes the usage of \ijacocoTool
as a plugin to the \maven build system.

\subsection{Analysis Phase}
\label{sec:impl:analysis}

The \analysisphase starts with performing RTS.  \ijacocoTool makes no change to
\ekstazi's dependency graph format, which is a list of checksums of all dependent
classes for each \test.  When computing the checksum of a class, \ekstazi would remove
all debugging information from the class file such that code changes that do not affect
test execution are ignored (e.g., renaming variables or updating comments).  However,
since \jacoco identifies \probes by line numbers, it is important to include the line
number table into the checksum computation.  Once \ekstazi has selected the \tests
\aTestSetNewSelectedRTS, \ijacocoTool follows the steps in
\S\ref{sec:technique:analysis} to compute the set of \tests (\aTestSetNewSelectedI) that need to be executed.

\subsection{Execution Phase}
\label{sec:impl:execution}

Both \ekstazi and \jacoco needs to instrument the \codebase at the time of test
execution using the \CodeIn{javaagent} mechanism~\cite{JavaLangInstrument}.  As
described in \S\ref{sec:technique:execution}, the two sets of instrumentation do not
interfere with each other and thus their ordering does not matter.

\subsection{Collection Phase}
\label{sec:impl:collection}

The \collectionphase is where the dependency graph is updated and \covdata is collected.
\ijacocoTool also makes no change to \jacoco's \covdata format (i.e., a hash map stored
as the \CodeIn{jacoco.exec} file).  To correctly compute the \covdata, \ijacocoTool
first loads the old \revision's \covdata from the file system (if available), removes
the entries that belongs to the classes (\aClassSetCovInvalid) whose coverage may be affected,
and then unions it with the collected \covdata diff (\aCovDataDiff) during
execution.

\MyPara{Coverage report generation}
Since the format of \covdata is unchanged, \ijacocoTool reuses \jacoco's report
generation functionalities.  Specifically, \jacoco's report generation takes as inputs
the \covdata \aCovDataNew and the \codebase.  The \codebase is required to recover the
mapping between \probes and \codeelems, such that the set of executed \probes can be
translated into the set of \codeelems covered, denoted as \aElemSetCoveredNew.  \Codecov
metrics can be computed as the percentage of covered \codeelems out of all \codeelems:
$
\aCoverageNew = \frac{|\aElemSetCoveredNew|}{|\aElemSet|}
$.
\jacoco computes \codecov at line-level, instruction-level, branch-level, and
method-level.  The report also includes the \codecov metrics over the entire \repo, or
within a given package or class.
Moreover, for each source code file, the report also includes a visualization (in HTML
format) to annotate which lines and branches are covered.

\subsection{Maven Plugin}
\label{sec:impl:maven}

\ijacocoTool is shipped with a \maven plugin, just like \ekstazi and \jacoco, so that it can
be simply integrated into \repos using the \maven build system.  The \ijacocoTool \maven
plugin analyzes the tests to select \aTestSetNewSelectedI at the
\CodeIn{process\text{-}test\text{-}classes}
phase\footnote{https://maven.apache.org/guides/introduction/introduction-to-the-lifecycle.html},
and sets the \CodeIn{excludes} property of the \surefire
plugin\footnote{https://maven.apache.org/surefire/maven-surefire-plugin/examples/inclusion-exclusion.html}
to skip the unselected tests.
\ijacocoTool's execution phase happens at the \CodeIn{test} phase in the \maven
lifecycle, where instrumentation is performed by adding the \CodeIn{javaagent} argument
to the \surefire plugin's configuration.
\ijacocoTool's collection phase happens as a shutdown hook at the end of test execution.
When report generation is needed, \ijacocoTool invokes the corresponding \jacoco
functionalities at the \CodeIn{verify} phase in the \maven lifecycle.

\section{Evaluation}
\label{sec:eval}

In this section, we study the following three research questions to assess
\ijacocoTool's performance and correctness:

\DefMacro{rq-vsbjacoco}{RQ1}
\noindent\MyPara{\UseMacro{rq-vsbjacoco}} How much \codecovanalysis time speedup can we get from \ijacocoTool compared with \jacoco?

\DefMacro{rq-vsekstazi}{RQ2}
\noindent\MyPara{\UseMacro{rq-vsekstazi}} How does \ijacocoTool's test selection rate compare to \ekstazi?

\DefMacro{rq-phase}{RQ3}
\noindent\MyPara{\UseMacro{rq-phase}} How much time does each phase of \ijacocoTool take?

\DefMacro{rq-correctness}{RQ4}
\noindent\MyPara{\UseMacro{rq-correctness}} Is \ijacocoTool correct, i.e., producing the
same \codecov results as \jacoco?

We first describe the subject \repos used in our evaluation (\S\ref{sec:eval:subjects}),
then the experimental setup (\S\ref{sec:eval:setup}), and finally present the results
and answer the research questions (\S\ref{sec:eval:results}).

\subsection{Subjects}
\label{sec:eval:subjects}

We reused the list of \repos and \revisions in the evaluation of a recent related work
on RTS~\cite{liu2023more}, which includes \UseMacro{num-repos-finerts} open-source Java
\repos and \UseMacro{num-revisions-per-repo} \revisions per \repo.  However, we found
that \Prepo{email-ext-plugin} uses a mocking library that was incompatible with \ekstazi
(and thus \ijacocoTool which is built on top of it) in half of its \revisions, and thus we
excluded this \repo.  Table~\ref{tab:projects} lists the remaining \UseMacro{num-repo}
subject \repos used in our evaluation, as well as their first (oldest) \revisions, and
metrics of their \codebase and \testsuites.  Note that the
\UseMacro{num-revisions-per-repo} \revisions for each \repo were selected such that
there exists at least a code change at bytecode level (e.g., excluding simple comment
changes) between two \revisions~\cite{liu2023more}.
In total, our evaluation subject set involves \UseMacro{res-projects-sum-num-ver}
\revisions, \UseMacro{sum-loc-approx} lines of code, and
\UseMacro{res-projects-sum-num-test-class} test classes.  All of the \repos are using
the \maven build system.

\subsection{Experimental Setup}
\label{sec:eval:setup}

For each \repo, we clone it from \github, and then for each \revision in the selected
\UseMacro{num-revisions-per-repo} \revisions, we checkout to that \revision, enable one
of \{\ijacocoTool, \jacoco, \ekstazi{}\} or none of them (which we call \retestall), and
then execute tests using the \maven command \CodeIn{mvn\ clean\ test}.
We measure the \emph{end-to-end time} of the \maven command, which includes all three
phases of \ijacocoTool and \ekstazi and all steps of \jacoco.  For \ijacocoTool and
\jacoco, we store the \covdata at each \revision and report \emph{line-level \codecov}
(line coverage) metric in this paper as a representation of the \codecov results.  For
\ijacocoTool and \ekstazi, we record the number of \tests they selected to compute their
\emph{test selection rate} as the number of selected \tests divided by the total number
of \tests.

We enabled \ijacocoTool, \jacoco, and \ekstazi by using their \maven plugins.
Specifically, we insert a \maven profile into the \repo's build configuration file
(\CodeIn{pom.xml}) that adds the corresponding plugin to the \maven build lifecycle.  To
fairly compare the performance of \ijacocoTool and \jacoco, we forked the \jacoco tool
at the same \revision as the one \ijacocoTool was built on, and renamed the tool name to
\CodeIn{bjacoco} to avoid interference with existing \jacoco configurations; then, we
disable any existing \jacoco plugin if the \repo has it set up in the build
configuration file.   For all three tools and \retestall, we disabled \maven plugins
that are not relevant for our experiments, e.g., \CodeIn{checkstyle} and
\CodeIn{javadoc}.

\begin{table*}[t]
\begin{small}
\begin{center}
\caption{End-to-end time of \retestall, \ekstazi, \jacoco, and \ijacocoTool, which are averaged and summed over \UseMacro{num-revisions-per-repo} \revisions; and speedup of \ijacocoTool compared to \jacoco. All the time differences between \jacoco and \ijacocoTool are statistically significant with 95\% confidence level.} \label{tab:results-performance}

\begin{tabular}{l| r|r| r|r| r|r| r|r| r}
\toprule
\multicolumn{1}{c|}{\multirow{3}{*}{\textbf{\UseMacro{TH-project}}}} & \multicolumn{8}{c|}{\textbf{\UseMacro{TH-test-time}}} & \multicolumn{1}{c}{\textbf{\UseMacro{TH-speedup}}} \\
& \multicolumn{2}{c|}{\retestall} & \multicolumn{2}{c|}{\ekstazi} & \multicolumn{2}{c|}{\bjacoco} & \multicolumn{2}{c|}{\ijacocoTool} & \ijacocoTool \\
& \multicolumn{1}{c|}{\UseMacro{TH-avg}} & \multicolumn{1}{c|}{\UseMacro{TH-sum}} & \multicolumn{1}{c|}{\UseMacro{TH-avg}} & \multicolumn{1}{c|}{\UseMacro{TH-sum}} & \multicolumn{1}{c|}{\UseMacro{TH-avg}} & \multicolumn{1}{c|}{\UseMacro{TH-sum}} & \multicolumn{1}{c|}{\UseMacro{TH-avg}} & \multicolumn{1}{c|}{\UseMacro{TH-sum}} & \multicolumn{1}{c}{\UseMacro{TH-avg}}\\
\midrule
\UseMacro{res-asterisk-java-asterisk-java-name} & \UseMacro{res-asterisk-java-asterisk-java-retestall-time} & \UseMacro{res-asterisk-java-asterisk-java-retestall-total-time} & \UseMacro{res-asterisk-java-asterisk-java-ekstazi-time} & \UseMacro{res-asterisk-java-asterisk-java-ekstazi-total-time} & \UseMacro{res-asterisk-java-asterisk-java-bjacoco-time} & \UseMacro{res-asterisk-java-asterisk-java-total-bjacoco-time}  & \UseMacro{res-asterisk-java-asterisk-java-ijacoco-time} & \UseMacro{res-asterisk-java-asterisk-java-total-ijacoco-time} & \UseMacro{res-asterisk-java-asterisk-java-ibjacoco-speedup-time} \\
\UseMacro{res-apache-commons-beanutils-name} & \UseMacro{res-apache-commons-beanutils-retestall-time} & \UseMacro{res-apache-commons-beanutils-retestall-total-time} & \UseMacro{res-apache-commons-beanutils-ekstazi-time} & \UseMacro{res-apache-commons-beanutils-ekstazi-total-time} & \UseMacro{res-apache-commons-beanutils-bjacoco-time} & \UseMacro{res-apache-commons-beanutils-total-bjacoco-time}  & \UseMacro{res-apache-commons-beanutils-ijacoco-time} & \UseMacro{res-apache-commons-beanutils-total-ijacoco-time} & \UseMacro{res-apache-commons-beanutils-ibjacoco-speedup-time} \\
\UseMacro{res-apache-commons-codec-name} & \UseMacro{res-apache-commons-codec-retestall-time} & \UseMacro{res-apache-commons-codec-retestall-total-time} & \UseMacro{res-apache-commons-codec-ekstazi-time} & \UseMacro{res-apache-commons-codec-ekstazi-total-time} & \UseMacro{res-apache-commons-codec-bjacoco-time} & \UseMacro{res-apache-commons-codec-total-bjacoco-time}  & \UseMacro{res-apache-commons-codec-ijacoco-time} & \UseMacro{res-apache-commons-codec-total-ijacoco-time} & \UseMacro{res-apache-commons-codec-ibjacoco-speedup-time} \\
\UseMacro{res-apache-commons-collections-name} & \UseMacro{res-apache-commons-collections-retestall-time} & \UseMacro{res-apache-commons-collections-retestall-total-time} & \UseMacro{res-apache-commons-collections-ekstazi-time} & \UseMacro{res-apache-commons-collections-ekstazi-total-time} & \UseMacro{res-apache-commons-collections-bjacoco-time} & \UseMacro{res-apache-commons-collections-total-bjacoco-time}  & \UseMacro{res-apache-commons-collections-ijacoco-time} & \UseMacro{res-apache-commons-collections-total-ijacoco-time} & \UseMacro{res-apache-commons-collections-ibjacoco-speedup-time} \\
\UseMacro{res-apache-commons-compress-name} & \UseMacro{res-apache-commons-compress-retestall-time} & \UseMacro{res-apache-commons-compress-retestall-total-time} & \UseMacro{res-apache-commons-compress-ekstazi-time} & \UseMacro{res-apache-commons-compress-ekstazi-total-time} & \UseMacro{res-apache-commons-compress-bjacoco-time} & \UseMacro{res-apache-commons-compress-total-bjacoco-time}  & \UseMacro{res-apache-commons-compress-ijacoco-time} & \UseMacro{res-apache-commons-compress-total-ijacoco-time} & \UseMacro{res-apache-commons-compress-ibjacoco-speedup-time} \\
\UseMacro{res-apache-commons-configuration-name} & \UseMacro{res-apache-commons-configuration-retestall-time} & \UseMacro{res-apache-commons-configuration-retestall-total-time} & \UseMacro{res-apache-commons-configuration-ekstazi-time} & \UseMacro{res-apache-commons-configuration-ekstazi-total-time} & \UseMacro{res-apache-commons-configuration-bjacoco-time} & \UseMacro{res-apache-commons-configuration-total-bjacoco-time}  & \UseMacro{res-apache-commons-configuration-ijacoco-time} & \UseMacro{res-apache-commons-configuration-total-ijacoco-time} & \UseMacro{res-apache-commons-configuration-ibjacoco-speedup-time} \\
\UseMacro{res-apache-commons-dbcp-name} & \UseMacro{res-apache-commons-dbcp-retestall-time} & \UseMacro{res-apache-commons-dbcp-retestall-total-time} & \UseMacro{res-apache-commons-dbcp-ekstazi-time} & \UseMacro{res-apache-commons-dbcp-ekstazi-total-time} & \UseMacro{res-apache-commons-dbcp-bjacoco-time} & \UseMacro{res-apache-commons-dbcp-total-bjacoco-time}  & \UseMacro{res-apache-commons-dbcp-ijacoco-time} & \UseMacro{res-apache-commons-dbcp-total-ijacoco-time} & \UseMacro{res-apache-commons-dbcp-ibjacoco-speedup-time} \\
\UseMacro{res-apache-commons-imaging-name} & \UseMacro{res-apache-commons-imaging-retestall-time} & \UseMacro{res-apache-commons-imaging-retestall-total-time} & \UseMacro{res-apache-commons-imaging-ekstazi-time} & \UseMacro{res-apache-commons-imaging-ekstazi-total-time} & \UseMacro{res-apache-commons-imaging-bjacoco-time} & \UseMacro{res-apache-commons-imaging-total-bjacoco-time}  & \UseMacro{res-apache-commons-imaging-ijacoco-time} & \UseMacro{res-apache-commons-imaging-total-ijacoco-time} & \UseMacro{res-apache-commons-imaging-ibjacoco-speedup-time} \\
\UseMacro{res-apache-commons-io-name} & \UseMacro{res-apache-commons-io-retestall-time} & \UseMacro{res-apache-commons-io-retestall-total-time} & \UseMacro{res-apache-commons-io-ekstazi-time} & \UseMacro{res-apache-commons-io-ekstazi-total-time} & \UseMacro{res-apache-commons-io-bjacoco-time} & \UseMacro{res-apache-commons-io-total-bjacoco-time}  & \UseMacro{res-apache-commons-io-ijacoco-time} & \UseMacro{res-apache-commons-io-total-ijacoco-time} & \UseMacro{res-apache-commons-io-ibjacoco-speedup-time} \\
\UseMacro{res-apache-commons-lang-name} & \UseMacro{res-apache-commons-lang-retestall-time} & \UseMacro{res-apache-commons-lang-retestall-total-time} & \UseMacro{res-apache-commons-lang-ekstazi-time} & \UseMacro{res-apache-commons-lang-ekstazi-total-time} & \UseMacro{res-apache-commons-lang-bjacoco-time} & \UseMacro{res-apache-commons-lang-total-bjacoco-time}  & \UseMacro{res-apache-commons-lang-ijacoco-time} & \UseMacro{res-apache-commons-lang-total-ijacoco-time} & \UseMacro{res-apache-commons-lang-ibjacoco-speedup-time} \\
\UseMacro{res-apache-commons-math-name} & \UseMacro{res-apache-commons-math-retestall-time} & \UseMacro{res-apache-commons-math-retestall-total-time} & \UseMacro{res-apache-commons-math-ekstazi-time} & \UseMacro{res-apache-commons-math-ekstazi-total-time} & \UseMacro{res-apache-commons-math-bjacoco-time} & \UseMacro{res-apache-commons-math-total-bjacoco-time}  & \UseMacro{res-apache-commons-math-ijacoco-time} & \UseMacro{res-apache-commons-math-total-ijacoco-time} & \UseMacro{res-apache-commons-math-ibjacoco-speedup-time} \\
\UseMacro{res-apache-commons-net-name} & \UseMacro{res-apache-commons-net-retestall-time} & \UseMacro{res-apache-commons-net-retestall-total-time} & \UseMacro{res-apache-commons-net-ekstazi-time} & \UseMacro{res-apache-commons-net-ekstazi-total-time} & \UseMacro{res-apache-commons-net-bjacoco-time} & \UseMacro{res-apache-commons-net-total-bjacoco-time}  & \UseMacro{res-apache-commons-net-ijacoco-time} & \UseMacro{res-apache-commons-net-total-ijacoco-time} & \UseMacro{res-apache-commons-net-ibjacoco-speedup-time} \\
\UseMacro{res-apache-commons-pool-name} & \UseMacro{res-apache-commons-pool-retestall-time} & \UseMacro{res-apache-commons-pool-retestall-total-time} & \UseMacro{res-apache-commons-pool-ekstazi-time} & \UseMacro{res-apache-commons-pool-ekstazi-total-time} & \UseMacro{res-apache-commons-pool-bjacoco-time} & \UseMacro{res-apache-commons-pool-total-bjacoco-time}  & \UseMacro{res-apache-commons-pool-ijacoco-time} & \UseMacro{res-apache-commons-pool-total-ijacoco-time} & \UseMacro{res-apache-commons-pool-ibjacoco-speedup-time} \\
\UseMacro{res-alibaba-fastjson-name} & \UseMacro{res-alibaba-fastjson-retestall-time} & \UseMacro{res-alibaba-fastjson-retestall-total-time} & \UseMacro{res-alibaba-fastjson-ekstazi-time} & \UseMacro{res-alibaba-fastjson-ekstazi-total-time} & \UseMacro{res-alibaba-fastjson-bjacoco-time} & \UseMacro{res-alibaba-fastjson-total-bjacoco-time}  & \UseMacro{res-alibaba-fastjson-ijacoco-time} & \UseMacro{res-alibaba-fastjson-total-ijacoco-time} & \UseMacro{res-alibaba-fastjson-ibjacoco-speedup-time} \\
\UseMacro{res-finmath-finmath-lib-name} & \UseMacro{res-finmath-finmath-lib-retestall-time} & \UseMacro{res-finmath-finmath-lib-retestall-total-time} & \UseMacro{res-finmath-finmath-lib-ekstazi-time} & \UseMacro{res-finmath-finmath-lib-ekstazi-total-time} & \UseMacro{res-finmath-finmath-lib-bjacoco-time} & \UseMacro{res-finmath-finmath-lib-total-bjacoco-time}  & \UseMacro{res-finmath-finmath-lib-ijacoco-time} & \UseMacro{res-finmath-finmath-lib-total-ijacoco-time} & \UseMacro{res-finmath-finmath-lib-ibjacoco-speedup-time} \\
\UseMacro{res-sonyxperiadev-gerrit-events-name} & \UseMacro{res-sonyxperiadev-gerrit-events-retestall-time} & \UseMacro{res-sonyxperiadev-gerrit-events-retestall-total-time} & \UseMacro{res-sonyxperiadev-gerrit-events-ekstazi-time} & \UseMacro{res-sonyxperiadev-gerrit-events-ekstazi-total-time} & \UseMacro{res-sonyxperiadev-gerrit-events-bjacoco-time} & \UseMacro{res-sonyxperiadev-gerrit-events-total-bjacoco-time}  & \UseMacro{res-sonyxperiadev-gerrit-events-ijacoco-time} & \UseMacro{res-sonyxperiadev-gerrit-events-total-ijacoco-time} & \UseMacro{res-sonyxperiadev-gerrit-events-ibjacoco-speedup-time} \\
\UseMacro{res-brettwooldridge-HikariCP-name} & \UseMacro{res-brettwooldridge-HikariCP-retestall-time} & \UseMacro{res-brettwooldridge-HikariCP-retestall-total-time} & \UseMacro{res-brettwooldridge-HikariCP-ekstazi-time} & \UseMacro{res-brettwooldridge-HikariCP-ekstazi-total-time} & \UseMacro{res-brettwooldridge-HikariCP-bjacoco-time} & \UseMacro{res-brettwooldridge-HikariCP-total-bjacoco-time}  & \UseMacro{res-brettwooldridge-HikariCP-ijacoco-time} & \UseMacro{res-brettwooldridge-HikariCP-total-ijacoco-time} & \UseMacro{res-brettwooldridge-HikariCP-ibjacoco-speedup-time} \\
\UseMacro{res-lmdbjava-lmdbjava-name} & \UseMacro{res-lmdbjava-lmdbjava-retestall-time} & \UseMacro{res-lmdbjava-lmdbjava-retestall-total-time} & \UseMacro{res-lmdbjava-lmdbjava-ekstazi-time} & \UseMacro{res-lmdbjava-lmdbjava-ekstazi-total-time} & \UseMacro{res-lmdbjava-lmdbjava-bjacoco-time} & \UseMacro{res-lmdbjava-lmdbjava-total-bjacoco-time}  & \UseMacro{res-lmdbjava-lmdbjava-ijacoco-time} & \UseMacro{res-lmdbjava-lmdbjava-total-ijacoco-time} & \UseMacro{res-lmdbjava-lmdbjava-ibjacoco-speedup-time} \\
\UseMacro{res-logic-ng-LogicNG-name} & \UseMacro{res-logic-ng-LogicNG-retestall-time} & \UseMacro{res-logic-ng-LogicNG-retestall-total-time} & \UseMacro{res-logic-ng-LogicNG-ekstazi-time} & \UseMacro{res-logic-ng-LogicNG-ekstazi-total-time} & \UseMacro{res-logic-ng-LogicNG-bjacoco-time} & \UseMacro{res-logic-ng-LogicNG-total-bjacoco-time}  & \UseMacro{res-logic-ng-LogicNG-ijacoco-time} & \UseMacro{res-logic-ng-LogicNG-total-ijacoco-time} & \UseMacro{res-logic-ng-LogicNG-ibjacoco-speedup-time} \\
\UseMacro{res-davidmoten-rxjava-extras-name} & \UseMacro{res-davidmoten-rxjava-extras-retestall-time} & \UseMacro{res-davidmoten-rxjava-extras-retestall-total-time} & \UseMacro{res-davidmoten-rxjava-extras-ekstazi-time} & \UseMacro{res-davidmoten-rxjava-extras-ekstazi-total-time} & \UseMacro{res-davidmoten-rxjava-extras-bjacoco-time} & \UseMacro{res-davidmoten-rxjava-extras-total-bjacoco-time}  & \UseMacro{res-davidmoten-rxjava-extras-ijacoco-time} & \UseMacro{res-davidmoten-rxjava-extras-total-ijacoco-time} & \UseMacro{res-davidmoten-rxjava-extras-ibjacoco-speedup-time} \\
\UseMacro{res-bullhorn-sdk-rest-name} & \UseMacro{res-bullhorn-sdk-rest-retestall-time} & \UseMacro{res-bullhorn-sdk-rest-retestall-total-time} & \UseMacro{res-bullhorn-sdk-rest-ekstazi-time} & \UseMacro{res-bullhorn-sdk-rest-ekstazi-total-time} & \UseMacro{res-bullhorn-sdk-rest-bjacoco-time} & \UseMacro{res-bullhorn-sdk-rest-total-bjacoco-time}  & \UseMacro{res-bullhorn-sdk-rest-ijacoco-time} & \UseMacro{res-bullhorn-sdk-rest-total-ijacoco-time} & \UseMacro{res-bullhorn-sdk-rest-ibjacoco-speedup-time} \\
\UseMacro{res-tabulapdf-tabula-java-name} & \UseMacro{res-tabulapdf-tabula-java-retestall-time} & \UseMacro{res-tabulapdf-tabula-java-retestall-total-time} & \UseMacro{res-tabulapdf-tabula-java-ekstazi-time} & \UseMacro{res-tabulapdf-tabula-java-ekstazi-total-time} & \UseMacro{res-tabulapdf-tabula-java-bjacoco-time} & \UseMacro{res-tabulapdf-tabula-java-total-bjacoco-time}  & \UseMacro{res-tabulapdf-tabula-java-ijacoco-time} & \UseMacro{res-tabulapdf-tabula-java-total-ijacoco-time} & \UseMacro{res-tabulapdf-tabula-java-ibjacoco-speedup-time} \\
\midrule
\multicolumn{1}{c|}{\textbf{\UseMacro{TH-avg}}} &  \UseMacro{retestall-avg-time} & \UseMacro{retestall-avg-total-time} & \UseMacro{ekstazi-avg-time} & \UseMacro{ekstazi-avg-total-time} & \UseMacro{bjacoco-avg-time} & \UseMacro{bjacoco-avg-total-time}  & \UseMacro{ijacoco-avg-time} & \UseMacro{ijacoco-avg-total-time} & \UseMacro{ijacoco-avg-speedup} \\
\bottomrule \end{tabular}

\end{center}
\end{small}
\end{table*}

Due to inevitable test flakiness~\cite{luo2014empirical, parry2021survey,
eck2019understanding, lam2020study, lam2019idflakies} in complicated \codebase, the run time
and \codecov metrics may vary across runs (i.e., the execution of \tests
may take different execution path and result in slightly different \codecov).  To
mitigate this, we run each experiment 5 times and report the average run time.
Moreover, we excluded the flaky tests whose outcomes change or \codecov fluctuate
dramatically (e.g., due to their pre-conditions being undeterministically met or not)
across the runs.
When comparing the \codecov and run time of \jacoco and \ijacocoTool, we conducted
statistical significance tests using bootstrap tests~\cite{Berg-Kirkpatrick12StatSign}
with a 95\% confidence level.

\MyPara{Environment}  We run all experiments on servers with 16 Intel Xeon vCPU cores
@2.5GHz, 60GB memory, and running Ubuntu 22.04.  We use \java version 8.0.392 and \maven
version 3.9.6.

\subsection{Results}
\label{sec:eval:results}

Table~\ref{tab:results-performance} shows the end-to-end time of \retestall and the
three tools used in our evaluation, \ekstazi, \jacoco, and \ijacocoTool.  There are two
columns for \retestall or each tool: the \UseMacro{TH-avg} column shows the average time
across the \UseMacro{num-revisions-per-repo} \revisions of each \repo, and the
\UseMacro{TH-sum} column shows the total time of all \revisions.  The last column of the
table computes the speedup of \ijacocoTool compared to \jacoco, which is \jacoco's
end-to-end time divided by \ijacocoTool's end-to-end time, averaged across
\UseMacro{num-revisions-per-repo} \revisions.  The last row computes the average across
all \repos.

We can observe that \ijacocoTool achieves an average speedup of
\UseMacro{ijacoco-avg-speedup}$\times$.  The highest speedup of
\UseMacro{ijacoco-max-speedup}$\times$ was achieved on the
\Prepo{\UseMacro{res-finmath-finmath-lib-name}} \repo.  This demonstrates the
effectiveness of \icodecovanalysis.  Notably, we can see that performing
\codecovanalysis incurs significant overhead, bringing the average end-to-end time from
\retestall's \UseMacro{retestall-avg-time}s to \jacoco's \UseMacro{bjacoco-avg-time}s;
but \ijacocoTool is able to reduce this time to \UseMacro{ijacoco-avg-time}s, which is
even shorter than \retestall.  This indicates that adopting \icodecovanalysis can
greatly reduce the CI cost for nowadays software \repos, where computing \codecov on
each \revision is a common practice.

\ijacocoTool is overall faster than \jacoco on \UseMacro{ijacoco-cnt-positive-speedup}
out of \UseMacro{num-repo} \repos.  We inspected the remaining
\UseMacro{ijacoco-cnt-negative-speedup} \repos and found that \ijacocoTool did not
achieve good performance for them due to either (1)~RTS selecting a large number of
\tests (\Prepo{\UseMacro{res-apache-commons-beanutils-name}} and
\Prepo{\UseMacro{res-bullhorn-sdk-rest-name}}),
(2)~many \tests depend on the same non-test class, so that changing any \test requires
executing other \tests depending on that class, resulting in much more \tests being
selected by \ijacocoTool than \ekstazi (\Prepo{\UseMacro{res-apache-commons-math-name}},
\Prepo{\UseMacro{res-logic-ng-LogicNG-name}}, and
\Prepo{\UseMacro{res-tabulapdf-tabula-java-name}}).
Nevertheless, the overhead of applying \ijacocoTool for these \repos (and \revisions) is
not large.  Future work can investigate how to reduce the overhead, potentially by
turning off incremental analysis when certain patterns of code changes are detected.

\begin{figure*}
\begin{center}
\begin{small}
\begin{minipage}[b]{0.23\textwidth}
\includegraphics[width=\textwidth]{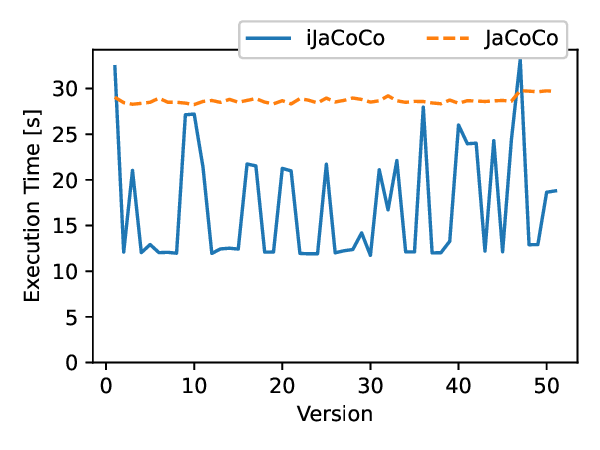}
\subcaption{end-to-end time}\label{fig:results-apache-commons-lang:time}
\end{minipage}
\begin{minipage}[b]{0.23\textwidth}
\includegraphics[width=\textwidth]{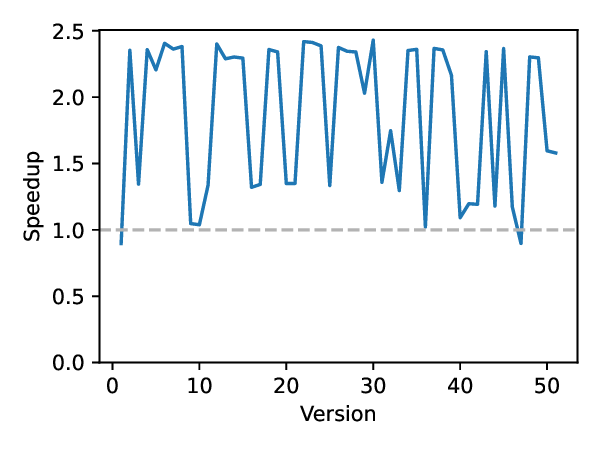}
\subcaption{\ijacocoTool speedup}\label{fig:results-apache-commons-lang:speedup}
\end{minipage}
\begin{minipage}[b]{0.23\textwidth}
\includegraphics[width=\textwidth]{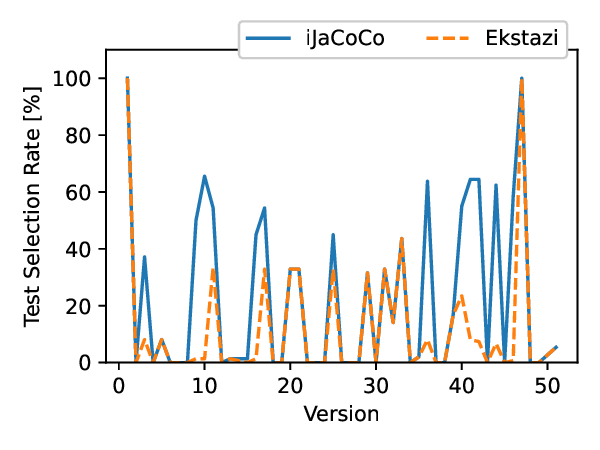}
\subcaption{test selection rate}\label{fig:results-apache-commons-lang:test-selection}
\end{minipage}
\begin{minipage}[b]{0.23\textwidth}
\includegraphics[width=\textwidth]{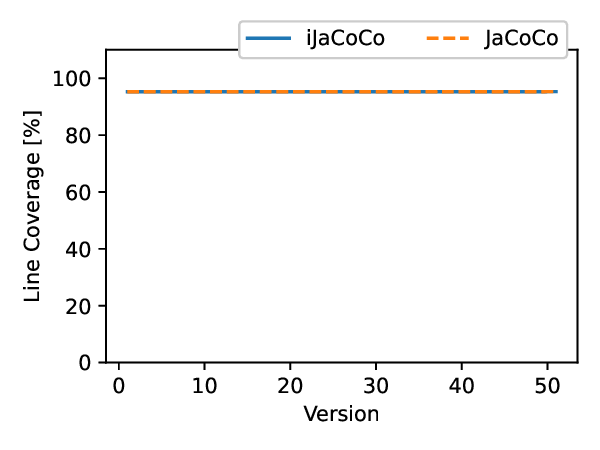}
\subcaption{\codecov}\label{fig:results-apache-commons-lang:coverage}
\end{minipage}
\caption{Experiment results for \UseMacro{res-apache-commons-lang-name}. \label{fig:results-apache-commons-lang}}
\end{small}
\end{center}
\end{figure*}

\begin{figure*}
\begin{center}
\begin{small}
\begin{minipage}[b]{0.23\textwidth}
\includegraphics[width=\textwidth]{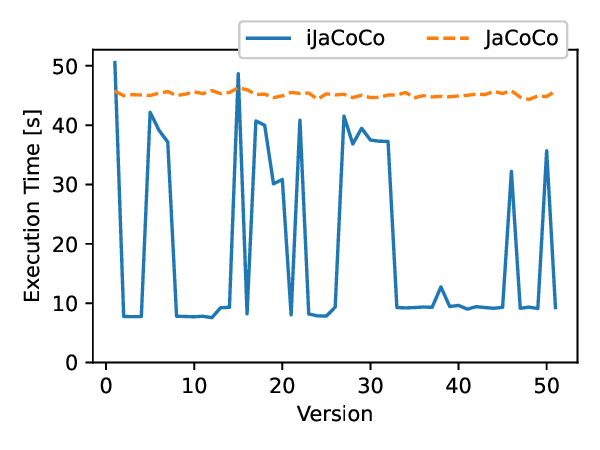}
\subcaption{end-to-end time}\label{fig:results-apache-commons-codec:time}
\end{minipage}
\begin{minipage}[b]{0.23\textwidth}
\includegraphics[width=\textwidth]{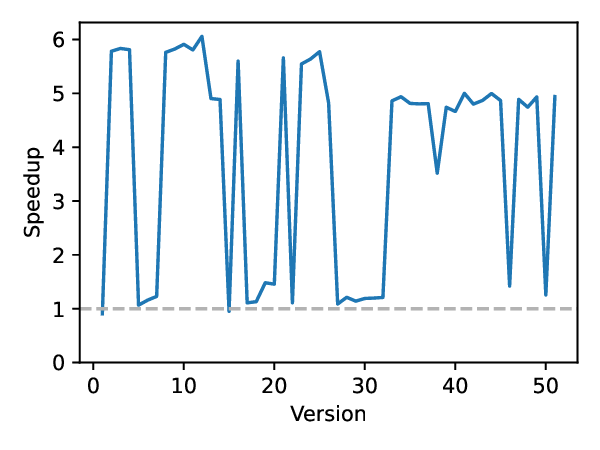}
\subcaption{\ijacocoTool speedup}\label{fig:results-apache-commons-codec:speedup}
\end{minipage}
\begin{minipage}[b]{0.23\textwidth}
\includegraphics[width=\textwidth]{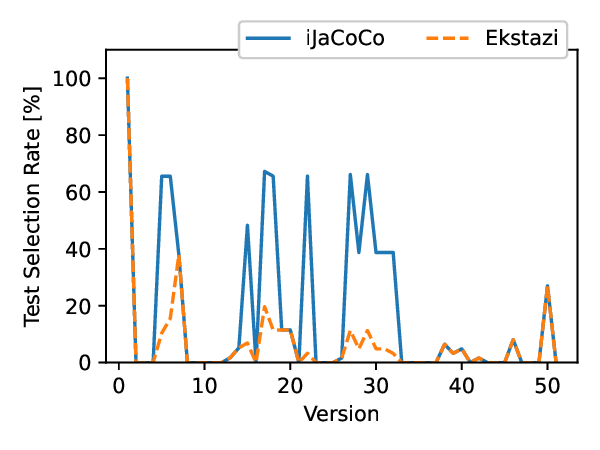}
\subcaption{test selection rate}\label{fig:results-apache-commons-codec:test-selection}
\end{minipage}
\begin{minipage}[b]{0.23\textwidth}
\includegraphics[width=\textwidth]{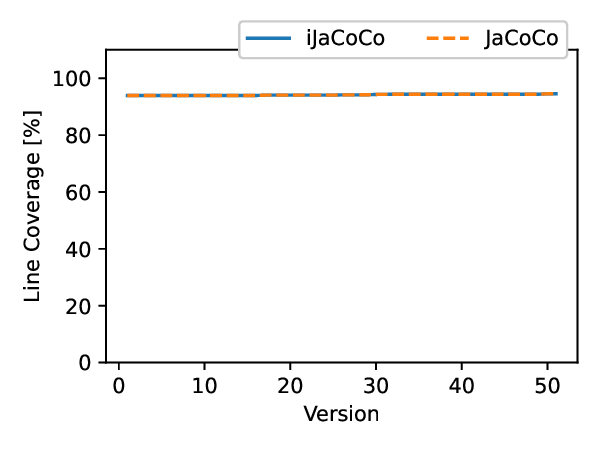}
\subcaption{\codecov}\label{fig:results-apache-commons-codec:coverage}
\end{minipage}
\caption{Experiment results for \UseMacro{res-apache-commons-codec-name}. \label{fig:results-apache-commons-codec}}
\end{small}
\end{center}
\end{figure*}

We also studied the evolution of \ijacocoTool's performance across \revisions.  Due to
the page limitations, we highlight the results on a few \repos; the full results can be
found in our replication package.  Figure~\ref{fig:results-apache-commons-lang:time}
and Figure~\ref{fig:results-apache-commons-lang:speedup} shows the evolution of
\ijacocoTool vs. \jacoco end-to-end time and \ijacocoTool's speedup across
\UseMacro{num-revisions-per-repo} \revisions of
\Prepo{\UseMacro{res-apache-commons-lang-name}}, respectively.  We can see that
\jacoco's time is quite stable across \revisions, while \ijacocoTool's time varies
(but mostly shorter than \jacoco).  On \revisions where the code change is small,
\ijacocoTool's time is also shorter; on the first \revision and one of the later
\revisions where the code change was large, \ijacocoTool needs to execute almost all
\tests and build dependency graphs, leading to a bit longer time than \jacoco.
Figure~\ref{fig:results-apache-commons-codec:time} and
Figure~\ref{fig:results-apache-commons-codec:speedup} show the same plots for
\Prepo{\UseMacro{res-apache-commons-codec-name}}, and we can see a very similar trend.

\begin{figure*}
\begin{center}
\begin{small}
\begin{minipage}[b]{0.23\textwidth}
\includegraphics[width=\textwidth]{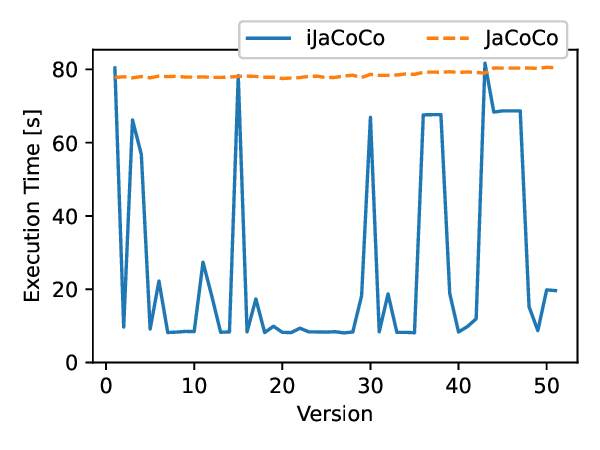}
\subcaption{end-to-end time}\label{fig:results-apache-commons-net:time}
\end{minipage}
\begin{minipage}[b]{0.23\textwidth}
\includegraphics[width=\textwidth]{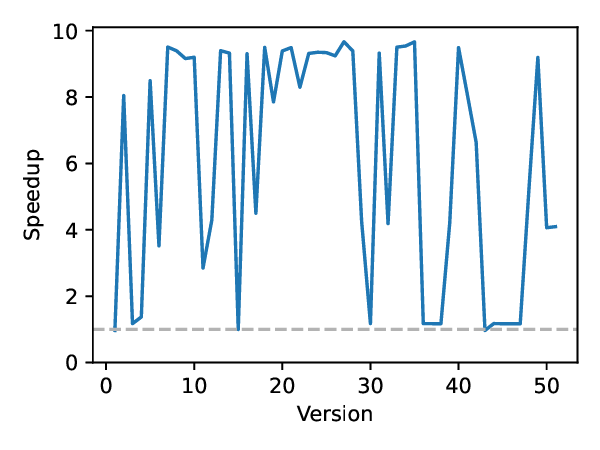}
\subcaption{\ijacocoTool speedup}\label{fig:results-apache-commons-net:speedup}
\end{minipage}
\begin{minipage}[b]{0.23\textwidth}
\includegraphics[width=\textwidth]{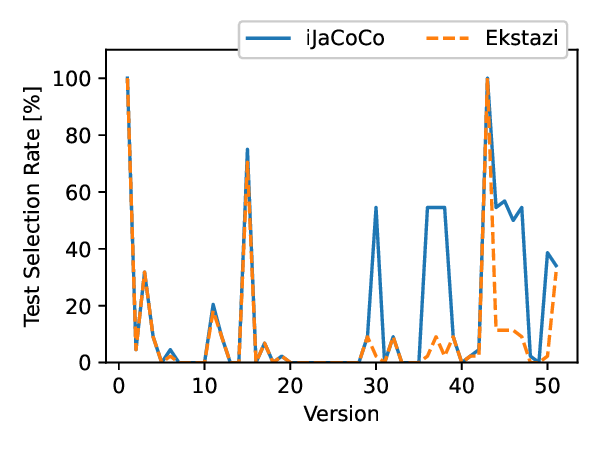}
\subcaption{test selection rate}\label{fig:results-apache-commons-net:test-selection}
\end{minipage}
\begin{minipage}[b]{0.23\textwidth}
\includegraphics[width=\textwidth]{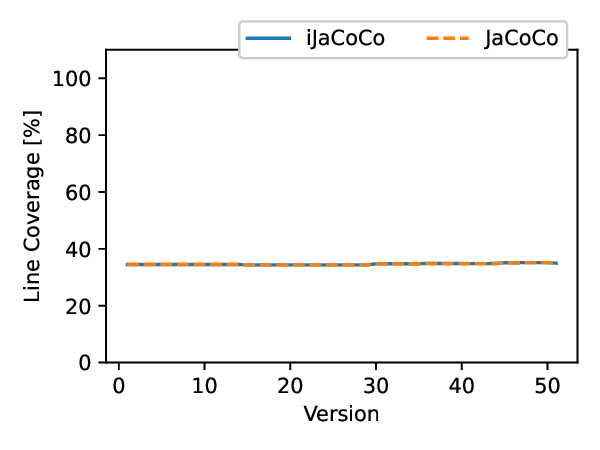}
\subcaption{\codecov}\label{fig:results-apache-commons-net:coverage}
\end{minipage}
\caption{Experiment results for \UseMacro{res-apache-commons-net-name}. \label{fig:results-apache-commons-net}}
\end{small}
\end{center}
\end{figure*}

\begin{figure*}
\begin{center}
\begin{small}
\begin{minipage}[b]{0.23\textwidth}
\includegraphics[width=\textwidth]{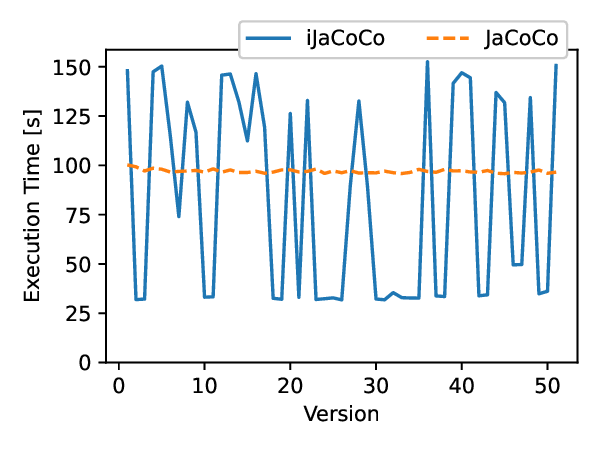}
\subcaption{end-to-end time}\label{fig:results-alibaba-fastjson:time}
\end{minipage}
\begin{minipage}[b]{0.23\textwidth}
\includegraphics[width=\textwidth]{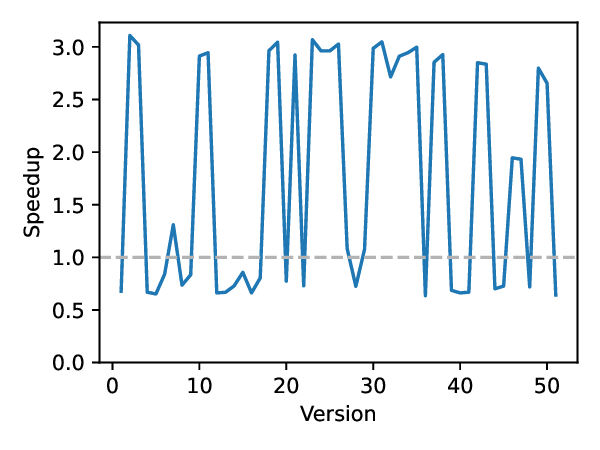}
\subcaption{\ijacocoTool speedup}\label{fig:results-alibaba-fastjson:speedup}
\end{minipage}
\begin{minipage}[b]{0.23\textwidth}
\includegraphics[width=\textwidth]{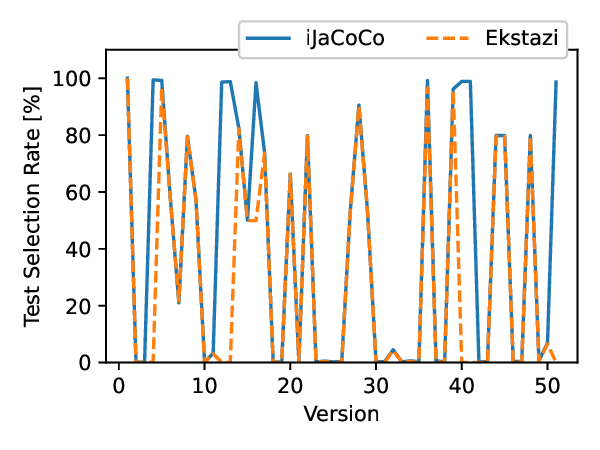}
\subcaption{test selection rate}\label{fig:results-alibaba-fastjson:test-selection}
\end{minipage}
\begin{minipage}[b]{0.23\textwidth}
\includegraphics[width=\textwidth]{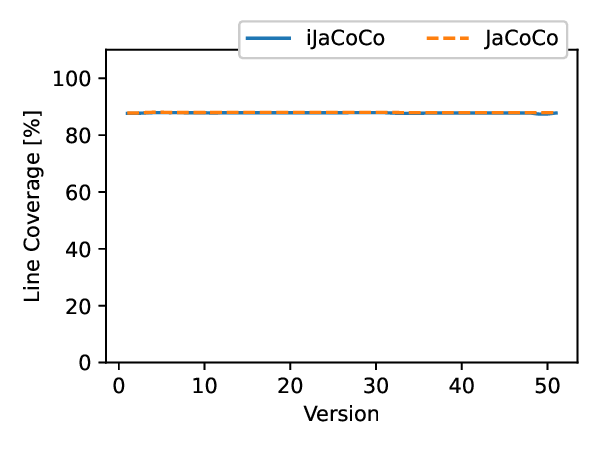}
\subcaption{\codecov}\label{fig:results-alibaba-fastjson:coverage}
\end{minipage}
\caption{Experiment results for \UseMacro{res-alibaba-fastjson-name}. \label{fig:results-alibaba-fastjson}}
\end{small}
\end{center}
\end{figure*}

In \Prepo{\UseMacro{res-apache-commons-net-name}} as shown in
Figure~\ref{fig:results-apache-commons-net:time} and
Figure~\ref{fig:results-apache-commons-net:speedup}, on some of the \revisions,
\ijacocoTool does not need to execute any \test at all (e.g., due to the class with code
change not being covered by any \tests), leading to a very high speedup of around
10$\times$.
Another interesting case we studied is \Prepo{\UseMacro{res-alibaba-fastjson-name}},
shown in Figure~\ref{fig:results-alibaba-fastjson:time} and
Figure~\ref{fig:results-alibaba-fastjson:speedup}, where \ijacocoTool has quite high
overhead on several \revisions (but still achieve an overall speedup of
\UseMacro{res-alibaba-fastjson-ibjacoco-speedup-time}$\times$).  A closer inspection of
those \revisions reveals that when the code change involves adding a \test, for example
at its 4th \revision \Pversion{36e03222}\footnote{Code diff at
\url{https://github.com/alibaba/fastjson/compare/12d92f61...36e03222}}, \ijacocoTool
needs to execute almost all the \tests as the \covdata of many classes are affected.

\rqanswer{\UseMacro{rq-vsbjacoco}}{\ijacocoTool can greatly speed up \codecovanalysis
when compared to \jacoco, achieving an average speedup of
\UseMacro{ijacoco-avg-speedup}$\times$ and up to \UseMacro{ijacoco-max-speedup}$\times$.

Our main observations include: (1)~the speedup of \ijacocoTool over \jacoco is
correlated with the speedup of \ekstazi over \retestall; (2)~\ijacocoTool incurs
overhead for collecting dependency graph on the first version (same as RTS) but is
usually faster than \jacoco on subsequent versions; and (3)~\ijacocoTool is more
effective when the code change is small.
}

\begin{table*}[t]
\begin{small}
\begin{center}
\caption{Columns 2--3: test selection rate of \ekstazi and \ijacocoTool. Columns 4--7: the time of different phases when using \ijacocoTool, and their percentage of \ijacocoTool's total end-to-end time.} \label{tab:results-selected-test-and-phase-time}

\begin{tabular}{l|r|r|r|r|r|r}
\toprule
\multicolumn{1}{c|}{\multirow{3}{*}{\textbf{\UseMacro{TH-project}}}} & \multicolumn{2}{c|}{\textbf{\UseMacro{TH-selected-test-ratio}}} & \multicolumn{4}{c}{\textbf{\UseMacro{TH-phase-time}}} \\
& \multicolumn{1}{c|}{\ekstazi} & \multicolumn{1}{c|}{\ijacocoTool} & \multicolumn{1}{c|}{\UseMacro{TH-compile-time}} & \multicolumn{1}{c|}{\UseMacro{TH-analysis-time}} & \multicolumn{1}{c|}{\UseMacro{TH-execution-collection-time}} & \multicolumn{1}{c}{\UseMacro{TH-report-time}} \\
\midrule
\UseMacro{res-asterisk-java-asterisk-java-name} & \UseMacro{res-asterisk-java-asterisk-java-ekstazi-selected-rate} & \UseMacro{res-asterisk-java-asterisk-java-ijacoco-selected-rate} & \UseMacro{res-asterisk-java-asterisk-java-phase-compile} (\UseMacro{res-asterisk-java-asterisk-java-phase-compile-percent}\%) & \UseMacro{res-asterisk-java-asterisk-java-phase-analysis} (\UseMacro{res-asterisk-java-asterisk-java-phase-analysis-percent}\%) & \UseMacro{res-asterisk-java-asterisk-java-phase-execution-collection} (\UseMacro{res-asterisk-java-asterisk-java-phase-execution-collection-percent}\%) & \UseMacro{res-asterisk-java-asterisk-java-phase-report} (\UseMacro{res-asterisk-java-asterisk-java-phase-report-percent}\%)\\
\UseMacro{res-apache-commons-beanutils-name} & \UseMacro{res-apache-commons-beanutils-ekstazi-selected-rate} & \UseMacro{res-apache-commons-beanutils-ijacoco-selected-rate} & \UseMacro{res-apache-commons-beanutils-phase-compile} (\UseMacro{res-apache-commons-beanutils-phase-compile-percent}\%) & \UseMacro{res-apache-commons-beanutils-phase-analysis} (\UseMacro{res-apache-commons-beanutils-phase-analysis-percent}\%) & \UseMacro{res-apache-commons-beanutils-phase-execution-collection} (\UseMacro{res-apache-commons-beanutils-phase-execution-collection-percent}\%) & \UseMacro{res-apache-commons-beanutils-phase-report} (\UseMacro{res-apache-commons-beanutils-phase-report-percent}\%)\\
\UseMacro{res-apache-commons-codec-name} & \UseMacro{res-apache-commons-codec-ekstazi-selected-rate} & \UseMacro{res-apache-commons-codec-ijacoco-selected-rate} & \UseMacro{res-apache-commons-codec-phase-compile} (\UseMacro{res-apache-commons-codec-phase-compile-percent}\%) & \UseMacro{res-apache-commons-codec-phase-analysis} (\UseMacro{res-apache-commons-codec-phase-analysis-percent}\%) & \UseMacro{res-apache-commons-codec-phase-execution-collection} (\UseMacro{res-apache-commons-codec-phase-execution-collection-percent}\%) & \UseMacro{res-apache-commons-codec-phase-report} (\UseMacro{res-apache-commons-codec-phase-report-percent}\%)\\
\UseMacro{res-apache-commons-collections-name} & \UseMacro{res-apache-commons-collections-ekstazi-selected-rate} & \UseMacro{res-apache-commons-collections-ijacoco-selected-rate} & \UseMacro{res-apache-commons-collections-phase-compile} (\UseMacro{res-apache-commons-collections-phase-compile-percent}\%) & \UseMacro{res-apache-commons-collections-phase-analysis} (\UseMacro{res-apache-commons-collections-phase-analysis-percent}\%) & \UseMacro{res-apache-commons-collections-phase-execution-collection} (\UseMacro{res-apache-commons-collections-phase-execution-collection-percent}\%) & \UseMacro{res-apache-commons-collections-phase-report} (\UseMacro{res-apache-commons-collections-phase-report-percent}\%)\\
\UseMacro{res-apache-commons-compress-name} & \UseMacro{res-apache-commons-compress-ekstazi-selected-rate} & \UseMacro{res-apache-commons-compress-ijacoco-selected-rate} & \UseMacro{res-apache-commons-compress-phase-compile} (\UseMacro{res-apache-commons-compress-phase-compile-percent}\%) & \UseMacro{res-apache-commons-compress-phase-analysis} (\UseMacro{res-apache-commons-compress-phase-analysis-percent}\%) & \UseMacro{res-apache-commons-compress-phase-execution-collection} (\UseMacro{res-apache-commons-compress-phase-execution-collection-percent}\%) & \UseMacro{res-apache-commons-compress-phase-report} (\UseMacro{res-apache-commons-compress-phase-report-percent}\%)\\
\UseMacro{res-apache-commons-configuration-name} & \UseMacro{res-apache-commons-configuration-ekstazi-selected-rate} & \UseMacro{res-apache-commons-configuration-ijacoco-selected-rate} & \UseMacro{res-apache-commons-configuration-phase-compile} (\UseMacro{res-apache-commons-configuration-phase-compile-percent}\%) & \UseMacro{res-apache-commons-configuration-phase-analysis} (\UseMacro{res-apache-commons-configuration-phase-analysis-percent}\%) & \UseMacro{res-apache-commons-configuration-phase-execution-collection} (\UseMacro{res-apache-commons-configuration-phase-execution-collection-percent}\%) & \UseMacro{res-apache-commons-configuration-phase-report} (\UseMacro{res-apache-commons-configuration-phase-report-percent}\%)\\
\UseMacro{res-apache-commons-dbcp-name} & \UseMacro{res-apache-commons-dbcp-ekstazi-selected-rate} & \UseMacro{res-apache-commons-dbcp-ijacoco-selected-rate} & \UseMacro{res-apache-commons-dbcp-phase-compile} (\UseMacro{res-apache-commons-dbcp-phase-compile-percent}\%) & \UseMacro{res-apache-commons-dbcp-phase-analysis} (\UseMacro{res-apache-commons-dbcp-phase-analysis-percent}\%) & \UseMacro{res-apache-commons-dbcp-phase-execution-collection} (\UseMacro{res-apache-commons-dbcp-phase-execution-collection-percent}\%) & \UseMacro{res-apache-commons-dbcp-phase-report} (\UseMacro{res-apache-commons-dbcp-phase-report-percent}\%)\\
\UseMacro{res-apache-commons-imaging-name} & \UseMacro{res-apache-commons-imaging-ekstazi-selected-rate} & \UseMacro{res-apache-commons-imaging-ijacoco-selected-rate} & \UseMacro{res-apache-commons-imaging-phase-compile} (\UseMacro{res-apache-commons-imaging-phase-compile-percent}\%) & \UseMacro{res-apache-commons-imaging-phase-analysis} (\UseMacro{res-apache-commons-imaging-phase-analysis-percent}\%) & \UseMacro{res-apache-commons-imaging-phase-execution-collection} (\UseMacro{res-apache-commons-imaging-phase-execution-collection-percent}\%) & \UseMacro{res-apache-commons-imaging-phase-report} (\UseMacro{res-apache-commons-imaging-phase-report-percent}\%)\\
\UseMacro{res-apache-commons-io-name} & \UseMacro{res-apache-commons-io-ekstazi-selected-rate} & \UseMacro{res-apache-commons-io-ijacoco-selected-rate} & \UseMacro{res-apache-commons-io-phase-compile} (\UseMacro{res-apache-commons-io-phase-compile-percent}\%) & \UseMacro{res-apache-commons-io-phase-analysis} (\UseMacro{res-apache-commons-io-phase-analysis-percent}\%) & \UseMacro{res-apache-commons-io-phase-execution-collection} (\UseMacro{res-apache-commons-io-phase-execution-collection-percent}\%) & \UseMacro{res-apache-commons-io-phase-report} (\UseMacro{res-apache-commons-io-phase-report-percent}\%)\\
\UseMacro{res-apache-commons-lang-name} & \UseMacro{res-apache-commons-lang-ekstazi-selected-rate} & \UseMacro{res-apache-commons-lang-ijacoco-selected-rate} & \UseMacro{res-apache-commons-lang-phase-compile} (\UseMacro{res-apache-commons-lang-phase-compile-percent}\%) & \UseMacro{res-apache-commons-lang-phase-analysis} (\UseMacro{res-apache-commons-lang-phase-analysis-percent}\%) & \UseMacro{res-apache-commons-lang-phase-execution-collection} (\UseMacro{res-apache-commons-lang-phase-execution-collection-percent}\%) & \UseMacro{res-apache-commons-lang-phase-report} (\UseMacro{res-apache-commons-lang-phase-report-percent}\%)\\
\UseMacro{res-apache-commons-math-name} & \UseMacro{res-apache-commons-math-ekstazi-selected-rate} & \UseMacro{res-apache-commons-math-ijacoco-selected-rate} & \UseMacro{res-apache-commons-math-phase-compile} (\UseMacro{res-apache-commons-math-phase-compile-percent}\%) & \UseMacro{res-apache-commons-math-phase-analysis} (\UseMacro{res-apache-commons-math-phase-analysis-percent}\%) & \UseMacro{res-apache-commons-math-phase-execution-collection} (\UseMacro{res-apache-commons-math-phase-execution-collection-percent}\%) & \UseMacro{res-apache-commons-math-phase-report} (\UseMacro{res-apache-commons-math-phase-report-percent}\%)\\
\UseMacro{res-apache-commons-net-name} & \UseMacro{res-apache-commons-net-ekstazi-selected-rate} & \UseMacro{res-apache-commons-net-ijacoco-selected-rate} & \UseMacro{res-apache-commons-net-phase-compile} (\UseMacro{res-apache-commons-net-phase-compile-percent}\%) & \UseMacro{res-apache-commons-net-phase-analysis} (\UseMacro{res-apache-commons-net-phase-analysis-percent}\%) & \UseMacro{res-apache-commons-net-phase-execution-collection} (\UseMacro{res-apache-commons-net-phase-execution-collection-percent}\%) & \UseMacro{res-apache-commons-net-phase-report} (\UseMacro{res-apache-commons-net-phase-report-percent}\%)\\
\UseMacro{res-apache-commons-pool-name} & \UseMacro{res-apache-commons-pool-ekstazi-selected-rate} & \UseMacro{res-apache-commons-pool-ijacoco-selected-rate} & \UseMacro{res-apache-commons-pool-phase-compile} (\UseMacro{res-apache-commons-pool-phase-compile-percent}\%) & \UseMacro{res-apache-commons-pool-phase-analysis} (\UseMacro{res-apache-commons-pool-phase-analysis-percent}\%) & \UseMacro{res-apache-commons-pool-phase-execution-collection} (\UseMacro{res-apache-commons-pool-phase-execution-collection-percent}\%) & \UseMacro{res-apache-commons-pool-phase-report} (\UseMacro{res-apache-commons-pool-phase-report-percent}\%)\\
\UseMacro{res-alibaba-fastjson-name} & \UseMacro{res-alibaba-fastjson-ekstazi-selected-rate} & \UseMacro{res-alibaba-fastjson-ijacoco-selected-rate} & \UseMacro{res-alibaba-fastjson-phase-compile} (\UseMacro{res-alibaba-fastjson-phase-compile-percent}\%) & \UseMacro{res-alibaba-fastjson-phase-analysis} (\UseMacro{res-alibaba-fastjson-phase-analysis-percent}\%) & \UseMacro{res-alibaba-fastjson-phase-execution-collection} (\UseMacro{res-alibaba-fastjson-phase-execution-collection-percent}\%) & \UseMacro{res-alibaba-fastjson-phase-report} (\UseMacro{res-alibaba-fastjson-phase-report-percent}\%)\\
\UseMacro{res-finmath-finmath-lib-name} & \UseMacro{res-finmath-finmath-lib-ekstazi-selected-rate} & \UseMacro{res-finmath-finmath-lib-ijacoco-selected-rate} & \UseMacro{res-finmath-finmath-lib-phase-compile} (\UseMacro{res-finmath-finmath-lib-phase-compile-percent}\%) & \UseMacro{res-finmath-finmath-lib-phase-analysis} (\UseMacro{res-finmath-finmath-lib-phase-analysis-percent}\%) & \UseMacro{res-finmath-finmath-lib-phase-execution-collection} (\UseMacro{res-finmath-finmath-lib-phase-execution-collection-percent}\%) & \UseMacro{res-finmath-finmath-lib-phase-report} (\UseMacro{res-finmath-finmath-lib-phase-report-percent}\%)\\
\UseMacro{res-sonyxperiadev-gerrit-events-name} & \UseMacro{res-sonyxperiadev-gerrit-events-ekstazi-selected-rate} & \UseMacro{res-sonyxperiadev-gerrit-events-ijacoco-selected-rate} & \UseMacro{res-sonyxperiadev-gerrit-events-phase-compile} (\UseMacro{res-sonyxperiadev-gerrit-events-phase-compile-percent}\%) & \UseMacro{res-sonyxperiadev-gerrit-events-phase-analysis} (\UseMacro{res-sonyxperiadev-gerrit-events-phase-analysis-percent}\%) & \UseMacro{res-sonyxperiadev-gerrit-events-phase-execution-collection} (\UseMacro{res-sonyxperiadev-gerrit-events-phase-execution-collection-percent}\%) & \UseMacro{res-sonyxperiadev-gerrit-events-phase-report} (\UseMacro{res-sonyxperiadev-gerrit-events-phase-report-percent}\%)\\
\UseMacro{res-brettwooldridge-HikariCP-name} & \UseMacro{res-brettwooldridge-HikariCP-ekstazi-selected-rate} & \UseMacro{res-brettwooldridge-HikariCP-ijacoco-selected-rate} & \UseMacro{res-brettwooldridge-HikariCP-phase-compile} (\UseMacro{res-brettwooldridge-HikariCP-phase-compile-percent}\%) & \UseMacro{res-brettwooldridge-HikariCP-phase-analysis} (\UseMacro{res-brettwooldridge-HikariCP-phase-analysis-percent}\%) & \UseMacro{res-brettwooldridge-HikariCP-phase-execution-collection} (\UseMacro{res-brettwooldridge-HikariCP-phase-execution-collection-percent}\%) & \UseMacro{res-brettwooldridge-HikariCP-phase-report} (\UseMacro{res-brettwooldridge-HikariCP-phase-report-percent}\%)\\
\UseMacro{res-lmdbjava-lmdbjava-name} & \UseMacro{res-lmdbjava-lmdbjava-ekstazi-selected-rate} & \UseMacro{res-lmdbjava-lmdbjava-ijacoco-selected-rate} & \UseMacro{res-lmdbjava-lmdbjava-phase-compile} (\UseMacro{res-lmdbjava-lmdbjava-phase-compile-percent}\%) & \UseMacro{res-lmdbjava-lmdbjava-phase-analysis} (\UseMacro{res-lmdbjava-lmdbjava-phase-analysis-percent}\%) & \UseMacro{res-lmdbjava-lmdbjava-phase-execution-collection} (\UseMacro{res-lmdbjava-lmdbjava-phase-execution-collection-percent}\%) & \UseMacro{res-lmdbjava-lmdbjava-phase-report} (\UseMacro{res-lmdbjava-lmdbjava-phase-report-percent}\%)\\
\UseMacro{res-logic-ng-LogicNG-name} & \UseMacro{res-logic-ng-LogicNG-ekstazi-selected-rate} & \UseMacro{res-logic-ng-LogicNG-ijacoco-selected-rate} & \UseMacro{res-logic-ng-LogicNG-phase-compile} (\UseMacro{res-logic-ng-LogicNG-phase-compile-percent}\%) & \UseMacro{res-logic-ng-LogicNG-phase-analysis} (\UseMacro{res-logic-ng-LogicNG-phase-analysis-percent}\%) & \UseMacro{res-logic-ng-LogicNG-phase-execution-collection} (\UseMacro{res-logic-ng-LogicNG-phase-execution-collection-percent}\%) & \UseMacro{res-logic-ng-LogicNG-phase-report} (\UseMacro{res-logic-ng-LogicNG-phase-report-percent}\%)\\
\UseMacro{res-davidmoten-rxjava-extras-name} & \UseMacro{res-davidmoten-rxjava-extras-ekstazi-selected-rate} & \UseMacro{res-davidmoten-rxjava-extras-ijacoco-selected-rate} & \UseMacro{res-davidmoten-rxjava-extras-phase-compile} (\UseMacro{res-davidmoten-rxjava-extras-phase-compile-percent}\%) & \UseMacro{res-davidmoten-rxjava-extras-phase-analysis} (\UseMacro{res-davidmoten-rxjava-extras-phase-analysis-percent}\%) & \UseMacro{res-davidmoten-rxjava-extras-phase-execution-collection} (\UseMacro{res-davidmoten-rxjava-extras-phase-execution-collection-percent}\%) & \UseMacro{res-davidmoten-rxjava-extras-phase-report} (\UseMacro{res-davidmoten-rxjava-extras-phase-report-percent}\%)\\
\UseMacro{res-bullhorn-sdk-rest-name} & \UseMacro{res-bullhorn-sdk-rest-ekstazi-selected-rate} & \UseMacro{res-bullhorn-sdk-rest-ijacoco-selected-rate} & \UseMacro{res-bullhorn-sdk-rest-phase-compile} (\UseMacro{res-bullhorn-sdk-rest-phase-compile-percent}\%) & \UseMacro{res-bullhorn-sdk-rest-phase-analysis} (\UseMacro{res-bullhorn-sdk-rest-phase-analysis-percent}\%) & \UseMacro{res-bullhorn-sdk-rest-phase-execution-collection} (\UseMacro{res-bullhorn-sdk-rest-phase-execution-collection-percent}\%) & \UseMacro{res-bullhorn-sdk-rest-phase-report} (\UseMacro{res-bullhorn-sdk-rest-phase-report-percent}\%)\\
\UseMacro{res-tabulapdf-tabula-java-name} & \UseMacro{res-tabulapdf-tabula-java-ekstazi-selected-rate} & \UseMacro{res-tabulapdf-tabula-java-ijacoco-selected-rate} & \UseMacro{res-tabulapdf-tabula-java-phase-compile} (\UseMacro{res-tabulapdf-tabula-java-phase-compile-percent}\%) & \UseMacro{res-tabulapdf-tabula-java-phase-analysis} (\UseMacro{res-tabulapdf-tabula-java-phase-analysis-percent}\%) & \UseMacro{res-tabulapdf-tabula-java-phase-execution-collection} (\UseMacro{res-tabulapdf-tabula-java-phase-execution-collection-percent}\%) & \UseMacro{res-tabulapdf-tabula-java-phase-report} (\UseMacro{res-tabulapdf-tabula-java-phase-report-percent}\%)\\
\midrule
\multicolumn{1}{c|}{\textbf{\UseMacro{TH-avg}}} & \UseMacro{res-ekstazi-selected-test-rate-mean} & \UseMacro{res-ijacoco-selected-test-rate-mean} & \UseMacro{res-phase-avg-compile} (\UseMacro{res-phase-avg-compile-percent}\%) & \UseMacro{res-phase-avg-analysis} (\UseMacro{res-phase-avg-analysis-percent}\%) & \UseMacro{res-phase-avg-execution-collection} (\UseMacro{res-phase-avg-execution-collection-percent}\%) & \UseMacro{res-phase-avg-report} (\UseMacro{res-phase-avg-report-percent}\%) \\
\bottomrule\end{tabular}

\end{center}
\end{small}
\end{table*}

The columns 2--3 of Table~\ref{tab:results-selected-test-and-phase-time} shows the test selection rate of \ekstazi and
\ijacocoTool, averaged across all \revisions.  As explained in \S\ref{sec:technique},
\ijacocoTool needs to select more \tests than \ekstazi to support the correct collection
of \covdata.  On average, \ijacocoTool selects
\UseMacro{res-ijacoco-selected-test-rate-mean}\% of the \tests, and \ekstazi selects
\UseMacro{res-ekstazi-selected-test-rate-mean}\% of the \tests.
Figures~\ref{fig:results-apache-commons-lang:test-selection},
\ref{fig:results-apache-commons-codec:test-selection},
\ref{fig:results-apache-commons-net:test-selection}, and
\ref{fig:results-alibaba-fastjson:test-selection} show the evolution of test selection
rate of \ijacocoTool vs. \ekstazi on the four \repos we highlighted earlier.  We can see
that on some \revisions, \ijacocoTool selects the same number of \tests as \ekstazi,
while on others it needs to select more \tests.

\rqanswer{\UseMacro{rq-vsekstazi}}{\ijacocoTool selects
\UseMacro{res-ijacoco-selected-test-rate-mean}\% tests on average, while \ekstazi
selects \UseMacro{res-ekstazi-selected-test-rate-mean}\%; \ijacocoTool needs to select
approximately twice as many tests as \ekstazi to support the correct collection of
\covdata.}

The columns 4--7 of Table~\ref{tab:results-selected-test-and-phase-time} shows the time
taken by different phases when using \ijacocoTool.  Specifically, the end-to-end time
can be broken down into four phases: (1)~the compilation phase of the build system (for
compiling the source code and \tests), which takes \UseMacro{res-phase-avg-compile}s or
\UseMacro{res-phase-avg-compile-percent}\% on average; (2)~the analysis phase of
\ijacocoTool (\S\ref{sec:impl:analysis}), which takes \UseMacro{res-phase-avg-analysis}s
or \UseMacro{res-phase-avg-analysis-percent}\% on average; (3)~the execution and
collection phases of \ijacocoTool (\S\ref{sec:impl:execution} and
\S\ref{sec:impl:collection}; the two phases are interleaved and thus can only be
measured together), which takes \UseMacro{res-phase-avg-execution-collection}s or
\UseMacro{res-phase-avg-execution-collection-percent}\% on average; and (4)~the coverage
report generation phase (\S\ref{sec:impl:collection}), which takes
\UseMacro{res-phase-avg-report}s or \UseMacro{res-phase-avg-report-percent}\% on
average.

\rqanswer{\UseMacro{rq-phase}}{The majority of \ijacocoTool's end-to-end time is spent
on compilation (\UseMacro{res-phase-avg-compile-percent}\%) and the execution and
collection phases (\UseMacro{res-phase-avg-execution-collection-percent}\%). The
analysis phase (\UseMacro{res-phase-avg-analysis-percent}\%) and the report generation
(\UseMacro{res-phase-avg-report-percent}\%) introduce a small overhead.}

\begin{table*}[t]
\begin{small}
\begin{center}
\caption{The line coverage of \jacoco and \ijacocoTool and the correctness of
\ijacocoTool. Columns 2--7: the minimum, maximum, and average line coverage of \jacoco
and \ijacocoTool across \UseMacro{num-revisions-per-repo} \revisions;
\UseMacro{TH-diff}: the average difference between \jacoco's and \ijacocoTool's
coverage; \UseMacro{THx-exact-same}: whether \jacoco and \ijacocoTool is exactly the same across
all \revisions; \UseMacro{THx-no-stat-sign-diff}: whether the difference between \jacoco's and \ijacocoTool's
coverage is not statistically significant.}
\label{tab:results-coverage}

\begin{tabular}{l |r|r|r |r|r|r |r |r|r}
\toprule
\multicolumn{1}{c|}{\multirow{3}{*}{\textbf{\UseMacro{TH-project}}}} & \multicolumn{7}{c|}{\textbf{\UseMacro{TH-coverage}}} & \multicolumn{2}{c}{\textbf{\UseMacro{TH-correctness}}} \\
& \multicolumn{3}{c|}{\bjacoco} & \multicolumn{3}{c|}{\ijacocoTool} &  & \multirow{2}{*}{\UseMacro{TH-exact-same}} & \multirow{2}{*}{\UseMacro{TH-no-stat-sign-diff}}\\
& \UseMacro{TH-min} & \UseMacro{TH-max} & \UseMacro{TH-avg} & \UseMacro{TH-min} & \UseMacro{TH-max} & \UseMacro{TH-avg} & \multicolumn{1}{c|}{\UseMacro{TH-diff}} &  &\\
\midrule
\UseMacro{res-asterisk-java-asterisk-java-name} & \UseMacro{res-asterisk-java-asterisk-java-bjacoco-min-coverage} & \UseMacro{res-asterisk-java-asterisk-java-bjacoco-max-coverage} & \UseMacro{res-asterisk-java-asterisk-java-bjacoco-coverage} & \UseMacro{res-asterisk-java-asterisk-java-ijacoco-min-coverage} & \UseMacro{res-asterisk-java-asterisk-java-ijacoco-max-coverage} & \UseMacro{res-asterisk-java-asterisk-java-ijacoco-coverage} &  \UseMacro{asterisk-java-asterisk-java-coverage-diff} & \UseMacro{asterisk-java-asterisk-java-coverage-exact-same} & \UseMacro{asterisk-java-asterisk-java-coverage-no-stat-sign-diff} \\
\UseMacro{res-apache-commons-beanutils-name} & \UseMacro{res-apache-commons-beanutils-bjacoco-min-coverage} & \UseMacro{res-apache-commons-beanutils-bjacoco-max-coverage} & \UseMacro{res-apache-commons-beanutils-bjacoco-coverage} & \UseMacro{res-apache-commons-beanutils-ijacoco-min-coverage} & \UseMacro{res-apache-commons-beanutils-ijacoco-max-coverage} & \UseMacro{res-apache-commons-beanutils-ijacoco-coverage} &  \UseMacro{apache-commons-beanutils-coverage-diff} & \UseMacro{apache-commons-beanutils-coverage-exact-same} & \UseMacro{apache-commons-beanutils-coverage-no-stat-sign-diff} \\
\UseMacro{res-apache-commons-codec-name} & \UseMacro{res-apache-commons-codec-bjacoco-min-coverage} & \UseMacro{res-apache-commons-codec-bjacoco-max-coverage} & \UseMacro{res-apache-commons-codec-bjacoco-coverage} & \UseMacro{res-apache-commons-codec-ijacoco-min-coverage} & \UseMacro{res-apache-commons-codec-ijacoco-max-coverage} & \UseMacro{res-apache-commons-codec-ijacoco-coverage} &  \UseMacro{apache-commons-codec-coverage-diff} & \UseMacro{apache-commons-codec-coverage-exact-same} & \UseMacro{apache-commons-codec-coverage-no-stat-sign-diff} \\
\UseMacro{res-apache-commons-collections-name} & \UseMacro{res-apache-commons-collections-bjacoco-min-coverage} & \UseMacro{res-apache-commons-collections-bjacoco-max-coverage} & \UseMacro{res-apache-commons-collections-bjacoco-coverage} & \UseMacro{res-apache-commons-collections-ijacoco-min-coverage} & \UseMacro{res-apache-commons-collections-ijacoco-max-coverage} & \UseMacro{res-apache-commons-collections-ijacoco-coverage} &  \UseMacro{apache-commons-collections-coverage-diff} & \UseMacro{apache-commons-collections-coverage-exact-same} & \UseMacro{apache-commons-collections-coverage-no-stat-sign-diff} \\
\UseMacro{res-apache-commons-compress-name} & \UseMacro{res-apache-commons-compress-bjacoco-min-coverage} & \UseMacro{res-apache-commons-compress-bjacoco-max-coverage} & \UseMacro{res-apache-commons-compress-bjacoco-coverage} & \UseMacro{res-apache-commons-compress-ijacoco-min-coverage} & \UseMacro{res-apache-commons-compress-ijacoco-max-coverage} & \UseMacro{res-apache-commons-compress-ijacoco-coverage} &  \UseMacro{apache-commons-compress-coverage-diff} & \UseMacro{apache-commons-compress-coverage-exact-same} & \UseMacro{apache-commons-compress-coverage-no-stat-sign-diff} \\
\UseMacro{res-apache-commons-configuration-name} & \UseMacro{res-apache-commons-configuration-bjacoco-min-coverage} & \UseMacro{res-apache-commons-configuration-bjacoco-max-coverage} & \UseMacro{res-apache-commons-configuration-bjacoco-coverage} & \UseMacro{res-apache-commons-configuration-ijacoco-min-coverage} & \UseMacro{res-apache-commons-configuration-ijacoco-max-coverage} & \UseMacro{res-apache-commons-configuration-ijacoco-coverage} &  \UseMacro{apache-commons-configuration-coverage-diff} & \UseMacro{apache-commons-configuration-coverage-exact-same} & \UseMacro{apache-commons-configuration-coverage-no-stat-sign-diff} \\
\UseMacro{res-apache-commons-dbcp-name} & \UseMacro{res-apache-commons-dbcp-bjacoco-min-coverage} & \UseMacro{res-apache-commons-dbcp-bjacoco-max-coverage} & \UseMacro{res-apache-commons-dbcp-bjacoco-coverage} & \UseMacro{res-apache-commons-dbcp-ijacoco-min-coverage} & \UseMacro{res-apache-commons-dbcp-ijacoco-max-coverage} & \UseMacro{res-apache-commons-dbcp-ijacoco-coverage} &  \UseMacro{apache-commons-dbcp-coverage-diff} & \UseMacro{apache-commons-dbcp-coverage-exact-same} & \UseMacro{apache-commons-dbcp-coverage-no-stat-sign-diff} \\
\UseMacro{res-apache-commons-imaging-name} & \UseMacro{res-apache-commons-imaging-bjacoco-min-coverage} & \UseMacro{res-apache-commons-imaging-bjacoco-max-coverage} & \UseMacro{res-apache-commons-imaging-bjacoco-coverage} & \UseMacro{res-apache-commons-imaging-ijacoco-min-coverage} & \UseMacro{res-apache-commons-imaging-ijacoco-max-coverage} & \UseMacro{res-apache-commons-imaging-ijacoco-coverage} &  \UseMacro{apache-commons-imaging-coverage-diff} & \UseMacro{apache-commons-imaging-coverage-exact-same} & \UseMacro{apache-commons-imaging-coverage-no-stat-sign-diff} \\
\UseMacro{res-apache-commons-io-name} & \UseMacro{res-apache-commons-io-bjacoco-min-coverage} & \UseMacro{res-apache-commons-io-bjacoco-max-coverage} & \UseMacro{res-apache-commons-io-bjacoco-coverage} & \UseMacro{res-apache-commons-io-ijacoco-min-coverage} & \UseMacro{res-apache-commons-io-ijacoco-max-coverage} & \UseMacro{res-apache-commons-io-ijacoco-coverage} &  \UseMacro{apache-commons-io-coverage-diff} & \UseMacro{apache-commons-io-coverage-exact-same} & \UseMacro{apache-commons-io-coverage-no-stat-sign-diff} \\
\UseMacro{res-apache-commons-lang-name} & \UseMacro{res-apache-commons-lang-bjacoco-min-coverage} & \UseMacro{res-apache-commons-lang-bjacoco-max-coverage} & \UseMacro{res-apache-commons-lang-bjacoco-coverage} & \UseMacro{res-apache-commons-lang-ijacoco-min-coverage} & \UseMacro{res-apache-commons-lang-ijacoco-max-coverage} & \UseMacro{res-apache-commons-lang-ijacoco-coverage} &  \UseMacro{apache-commons-lang-coverage-diff} & \UseMacro{apache-commons-lang-coverage-exact-same} & \UseMacro{apache-commons-lang-coverage-no-stat-sign-diff} \\
\UseMacro{res-apache-commons-math-name} & \UseMacro{res-apache-commons-math-bjacoco-min-coverage} & \UseMacro{res-apache-commons-math-bjacoco-max-coverage} & \UseMacro{res-apache-commons-math-bjacoco-coverage} & \UseMacro{res-apache-commons-math-ijacoco-min-coverage} & \UseMacro{res-apache-commons-math-ijacoco-max-coverage} & \UseMacro{res-apache-commons-math-ijacoco-coverage} &  \UseMacro{apache-commons-math-coverage-diff} & \UseMacro{apache-commons-math-coverage-exact-same} & \UseMacro{apache-commons-math-coverage-no-stat-sign-diff} \\
\UseMacro{res-apache-commons-net-name} & \UseMacro{res-apache-commons-net-bjacoco-min-coverage} & \UseMacro{res-apache-commons-net-bjacoco-max-coverage} & \UseMacro{res-apache-commons-net-bjacoco-coverage} & \UseMacro{res-apache-commons-net-ijacoco-min-coverage} & \UseMacro{res-apache-commons-net-ijacoco-max-coverage} & \UseMacro{res-apache-commons-net-ijacoco-coverage} &  \UseMacro{apache-commons-net-coverage-diff} & \UseMacro{apache-commons-net-coverage-exact-same} & \UseMacro{apache-commons-net-coverage-no-stat-sign-diff} \\
\UseMacro{res-apache-commons-pool-name} & \UseMacro{res-apache-commons-pool-bjacoco-min-coverage} & \UseMacro{res-apache-commons-pool-bjacoco-max-coverage} & \UseMacro{res-apache-commons-pool-bjacoco-coverage} & \UseMacro{res-apache-commons-pool-ijacoco-min-coverage} & \UseMacro{res-apache-commons-pool-ijacoco-max-coverage} & \UseMacro{res-apache-commons-pool-ijacoco-coverage} &  \UseMacro{apache-commons-pool-coverage-diff} & \UseMacro{apache-commons-pool-coverage-exact-same} & \UseMacro{apache-commons-pool-coverage-no-stat-sign-diff} \\
\UseMacro{res-alibaba-fastjson-name} & \UseMacro{res-alibaba-fastjson-bjacoco-min-coverage} & \UseMacro{res-alibaba-fastjson-bjacoco-max-coverage} & \UseMacro{res-alibaba-fastjson-bjacoco-coverage} & \UseMacro{res-alibaba-fastjson-ijacoco-min-coverage} & \UseMacro{res-alibaba-fastjson-ijacoco-max-coverage} & \UseMacro{res-alibaba-fastjson-ijacoco-coverage} &  \UseMacro{alibaba-fastjson-coverage-diff} & \UseMacro{alibaba-fastjson-coverage-exact-same} & \UseMacro{alibaba-fastjson-coverage-no-stat-sign-diff} \\
\UseMacro{res-finmath-finmath-lib-name} & \UseMacro{res-finmath-finmath-lib-bjacoco-min-coverage} & \UseMacro{res-finmath-finmath-lib-bjacoco-max-coverage} & \UseMacro{res-finmath-finmath-lib-bjacoco-coverage} & \UseMacro{res-finmath-finmath-lib-ijacoco-min-coverage} & \UseMacro{res-finmath-finmath-lib-ijacoco-max-coverage} & \UseMacro{res-finmath-finmath-lib-ijacoco-coverage} &  \UseMacro{finmath-finmath-lib-coverage-diff} & \UseMacro{finmath-finmath-lib-coverage-exact-same} & \UseMacro{finmath-finmath-lib-coverage-no-stat-sign-diff} \\
\UseMacro{res-sonyxperiadev-gerrit-events-name} & \UseMacro{res-sonyxperiadev-gerrit-events-bjacoco-min-coverage} & \UseMacro{res-sonyxperiadev-gerrit-events-bjacoco-max-coverage} & \UseMacro{res-sonyxperiadev-gerrit-events-bjacoco-coverage} & \UseMacro{res-sonyxperiadev-gerrit-events-ijacoco-min-coverage} & \UseMacro{res-sonyxperiadev-gerrit-events-ijacoco-max-coverage} & \UseMacro{res-sonyxperiadev-gerrit-events-ijacoco-coverage} &  \UseMacro{sonyxperiadev-gerrit-events-coverage-diff} & \UseMacro{sonyxperiadev-gerrit-events-coverage-exact-same} & \UseMacro{sonyxperiadev-gerrit-events-coverage-no-stat-sign-diff} \\
\UseMacro{res-brettwooldridge-HikariCP-name} & \UseMacro{res-brettwooldridge-HikariCP-bjacoco-min-coverage} & \UseMacro{res-brettwooldridge-HikariCP-bjacoco-max-coverage} & \UseMacro{res-brettwooldridge-HikariCP-bjacoco-coverage} & \UseMacro{res-brettwooldridge-HikariCP-ijacoco-min-coverage} & \UseMacro{res-brettwooldridge-HikariCP-ijacoco-max-coverage} & \UseMacro{res-brettwooldridge-HikariCP-ijacoco-coverage} &  \UseMacro{brettwooldridge-HikariCP-coverage-diff} & \UseMacro{brettwooldridge-HikariCP-coverage-exact-same} & \UseMacro{brettwooldridge-HikariCP-coverage-no-stat-sign-diff} \\
\UseMacro{res-lmdbjava-lmdbjava-name} & \UseMacro{res-lmdbjava-lmdbjava-bjacoco-min-coverage} & \UseMacro{res-lmdbjava-lmdbjava-bjacoco-max-coverage} & \UseMacro{res-lmdbjava-lmdbjava-bjacoco-coverage} & \UseMacro{res-lmdbjava-lmdbjava-ijacoco-min-coverage} & \UseMacro{res-lmdbjava-lmdbjava-ijacoco-max-coverage} & \UseMacro{res-lmdbjava-lmdbjava-ijacoco-coverage} &  \UseMacro{lmdbjava-lmdbjava-coverage-diff} & \UseMacro{lmdbjava-lmdbjava-coverage-exact-same} & \UseMacro{lmdbjava-lmdbjava-coverage-no-stat-sign-diff} \\
\UseMacro{res-logic-ng-LogicNG-name} & \UseMacro{res-logic-ng-LogicNG-bjacoco-min-coverage} & \UseMacro{res-logic-ng-LogicNG-bjacoco-max-coverage} & \UseMacro{res-logic-ng-LogicNG-bjacoco-coverage} & \UseMacro{res-logic-ng-LogicNG-ijacoco-min-coverage} & \UseMacro{res-logic-ng-LogicNG-ijacoco-max-coverage} & \UseMacro{res-logic-ng-LogicNG-ijacoco-coverage} &  \UseMacro{logic-ng-LogicNG-coverage-diff} & \UseMacro{logic-ng-LogicNG-coverage-exact-same} & \UseMacro{logic-ng-LogicNG-coverage-no-stat-sign-diff} \\
\UseMacro{res-davidmoten-rxjava-extras-name} & \UseMacro{res-davidmoten-rxjava-extras-bjacoco-min-coverage} & \UseMacro{res-davidmoten-rxjava-extras-bjacoco-max-coverage} & \UseMacro{res-davidmoten-rxjava-extras-bjacoco-coverage} & \UseMacro{res-davidmoten-rxjava-extras-ijacoco-min-coverage} & \UseMacro{res-davidmoten-rxjava-extras-ijacoco-max-coverage} & \UseMacro{res-davidmoten-rxjava-extras-ijacoco-coverage} &  \UseMacro{davidmoten-rxjava-extras-coverage-diff} & \UseMacro{davidmoten-rxjava-extras-coverage-exact-same} & \UseMacro{davidmoten-rxjava-extras-coverage-no-stat-sign-diff} \\
\UseMacro{res-bullhorn-sdk-rest-name} & \UseMacro{res-bullhorn-sdk-rest-bjacoco-min-coverage} & \UseMacro{res-bullhorn-sdk-rest-bjacoco-max-coverage} & \UseMacro{res-bullhorn-sdk-rest-bjacoco-coverage} & \UseMacro{res-bullhorn-sdk-rest-ijacoco-min-coverage} & \UseMacro{res-bullhorn-sdk-rest-ijacoco-max-coverage} & \UseMacro{res-bullhorn-sdk-rest-ijacoco-coverage} &  \UseMacro{bullhorn-sdk-rest-coverage-diff} & \UseMacro{bullhorn-sdk-rest-coverage-exact-same} & \UseMacro{bullhorn-sdk-rest-coverage-no-stat-sign-diff} \\
\UseMacro{res-tabulapdf-tabula-java-name} & \UseMacro{res-tabulapdf-tabula-java-bjacoco-min-coverage} & \UseMacro{res-tabulapdf-tabula-java-bjacoco-max-coverage} & \UseMacro{res-tabulapdf-tabula-java-bjacoco-coverage} & \UseMacro{res-tabulapdf-tabula-java-ijacoco-min-coverage} & \UseMacro{res-tabulapdf-tabula-java-ijacoco-max-coverage} & \UseMacro{res-tabulapdf-tabula-java-ijacoco-coverage} &  \UseMacro{tabulapdf-tabula-java-coverage-diff} & \UseMacro{tabulapdf-tabula-java-coverage-exact-same} & \UseMacro{tabulapdf-tabula-java-coverage-no-stat-sign-diff} \\
\bottomrule \end{tabular}

\end{center}
\end{small}
\end{table*}

Table~\ref{tab:results-coverage} compares the line coverage of \jacoco and \ijacocoTool.
We report the minimum, maximum, and average line coverage of \jacoco and \ijacocoTool
across \UseMacro{num-revisions-per-repo} \revisions of each \repo, as they represent an aggregated
distribution of the coverage results.  As we can see, most of the numbers are the same
or very close.  The \UseMacro{TH-diff} column shows the difference between \jacoco's and
\ijacocoTool's average line coverage.  Most of the differences are below 0.10\%, and the
largest difference of 0.16\% happens on the
\Prepo{\UseMacro{res-brettwooldridge-HikariCP-name}} \repo.

These small differences do not necessarily indicate that \ijacocoTool is incorrect, as
\codecov may change if test execution took different paths due to test flakiness and
test order dependency.  For example, in \Prepo{\UseMacro{res-alibaba-fastjson-name}}
\revision \Pversion{7ffa2a01}, \ijacocoTool reports that \CodeIn{SerializeWriter.java}'s
lines 1977--1998 are covered when running the selected 483 \tests, while \jacoco reports
that the same lines of code are not covered when running all 2,278 \tests.  To determine
whether the coverage difference is due to the different set of \tests or bugs in
\ijacocoTool, we collected coverage again with \jacoco but only on the selected \tests,
and found that those lines of code change from ``not covered'' to ``covered''.  Thus,
the coverage difference is due to a test order dependency, where the selected \tests
would execute more code if the other \tests are not executed.  If developers write
high-quality \tests that do not have such test order dependency, \ijacocoTool will
report the exact same \codecov as \jacoco.

To determine \ijacocoTool's correctness, we considered two criteria:
(1)~\UseMacro{THx-exact-same}, \ijacocoTool is correct if for all \revisions, the
coverage difference between \jacoco and \ijacocoTool is smaller than a threshold of
\UseMacro{cov-diff-threshold}; (2)~\UseMacro{THx-no-stat-sign-diff}, \ijacocoTool is
correct if the difference between \jacoco's and \ijacocoTool's coverage is not
statistically significant.
We found that \ijacocoTool satisfies the first criterion for all but one \repo
(\Prepo{\UseMacro{res-brettwooldridge-HikariCP-name}}), and satisfies the second
criterion for all \repos.

In addition, we also generated line plots to illustrate the evolution of \jacoco's and
\ijacocoTool's coverage across \revisions, shown in
figures~\ref{fig:results-apache-commons-lang:coverage},
\ref{fig:results-apache-commons-codec:coverage},
\ref{fig:results-apache-commons-net:coverage}, and
\ref{fig:results-alibaba-fastjson:coverage} for four of the \repos, and included in the
replication package for all \repos.  We can see that the lines of \jacoco and
\ijacocoTool are overlapping, indicating that they produce almost the same \codecov
results across all \revisions.

\rqanswer{\UseMacro{rq-correctness}}{We performed rigorous examination and confirmed
that \ijacocoTool is correct, i.e., it produces the same \codecov results as \jacoco.}

\section{Threats to Validity}
\label{sec:threats}

\MyPara{External}
We have extensively evaluated \ijacocoTool on a dataset of
\UseMacro{res-projects-sum-num-ver} \revisions from \UseMacro{num-repo} open-source
\java \repos.  However, this dataset may not be representative of all \java \repos.  To
mitigate this threat, we used the same \repos and \revisions that was used in a prior
work on RTS~\cite{liu2023more}; many \repos used in our study have been used in other
prior work on RTS~\cite{gligoric2015ekstazi, gligoric2015practical, legunsen2017starts,
legunsen2016extensive}.

Flaky tests~\cite{luo2014empirical, parry2021survey, eck2019understanding, lam2020study,
lam2019idflakies}, i.e., the tests that may pass or fail without changing the \codebase,
and tests that may take different execution paths without changing the \codebase are
very common in practice.
Any \codecovanalysis tool, including \ijacocoTool and \jacoco, may produce different
\codecov results across different runs due to test flakiness.
To minimize the impact of flaky tests on our evaluation, we have repeated all our
experiments 5 times and reported the average results.  The standard deviation of line
coverage across the 5 runs is less than 1\% for most of the \repos in our evaluation.

Existing RTS techniques might be unsafe in certain scenarios~\cite{ZhuETAL19RTSCheck}.
Our proposed \icodecovanalysis technique is correct if the underlying RTS technique is
safe (\S\ref{sec:technique:correctness}).  When the underlying RTS is unsafe, our
technique may miss updating the \covdata for some affected classes, leading to incorrect
\codecov results.  To mitigate this threat, we implement \ijacocoTool with \ekstazi as
the underlying RTS tool, which has no known safety issues for software executed within
the \java virtual machine.

\MyPara{Internal} Our implementation of \ijacocoTool may contain bugs that could impact
our conclusions.  To mitigate this threat, we tested \ijacocoTool against \jacoco as the
baseline and confirmed that they produce the same \codecov results modulo the impact of
test flakiness.  We also performed many sanity checks and manual inspections on our code
and scripts.

\MyPara{Construct} An alternative way to realize \icodecovanalysis is to change the
\covdata to be collected per \test instead of for the entire \testsuite, such to avoid
selecting more \tests than RTS would select.  We have explained why this approach may
result in much higher analysis overhead in \S\ref{sec:technique:analysis}, and performed
preliminary experiments to confirm this overhead.

The sets of instrumentation performed by \codecovanalysis to insert \probes and by RTS
to track dependencies have similar functionalities and could be combined in theory.
That is to say, we could further optimize \ijacocoTool by only performing one set of
instrumentation to track both \codecov and dependencies.  We leave this as future work.

\section{Related Work}
\label{sec:related}

\MyPara{Change impact analysis} Change impact analysis
(CIA)~\cite{li2013survey, lehnert2011taxonomy, kretsou2021change,
alam2015impact} is a technique to identify the potential effects of a
change in software. CIA can be used for regression testing by
selecting tests that cover the changes~\cite{ren2004chianti,
orso2003leveraging, ren2007change, zhang2008celadon,
martin2007automated, pourgalehdari2008meta, wang2008regression}.
\ijacocoTool's can be seen as a type of CIA but focuses on identifying the affected
\codecov by a code change.

\MyPara{Regression test selection} Regression test selection
(RTS)~\cite{engstrom2010systematic, engstrom2008empirical, graves2001empirical,
yoo2012regression} is a technique that selects a subset of tests affected by changes in
software, thereby reducing the time and cost of regression testing. RTS's dependency
analysis can be performed at different granularities, such as coarser-grained, at the
target/module-level~\cite{elbaum2014techniques, esfahani2016cloudbuild, memon2017taming,
shi2017optimizing} or finer-grained, at the
class/method/statement-level~\cite{rothermel1994framework, rothermel1997safe,
harrold2001regression, zaber2021towards, gligoric2015practical}. Many analysis-based RTS
tools~\cite{zhang2018hybrid, zaber2021towards, liu2023more} are proposed for \java
projects, among them, \ekstazi~\cite{gligoric2015ekstazi, gligoric2015practical} and
STARTS~\cite{legunsen2017starts, legunsen2016extensive} are popular tools that track
class-level dependencies. \ekstazi tracks dependencies dynamically, whereas STARTS
tracks dependencies statically.
Recently, researchers have also proposed machine learning (ML)-based RTS
techniques~\cite{zhang2022comparing, elsner2021empirically, lundsten2019ealrts,
pan2022test, bertolino2020learning}, which uses data-driven models to predict which
tests to select instead of analyzing the dependencies.

\ijacocoTool is built on \ekstazi as its RTS component.  In theory, any RTS tool can be
used to support \icodecovanalysis, but only safe RTS tools can ensure the correctness of
the collected \codecov results.  Thus, ML-based RTS tools, which are by nature not safe,
are not suitable to be used in \icodecovanalysis.

\MyPara{Applications of RTS} Aside from speeding up regression testing, RTS tools can be
used in various contexts. DeFlaker~\cite{bell2018deflaker} leveraged RTS to identify
flaky tests by marking as flaky the tests that fail but are not affected by the changes.
\citet{li2022evolution} proposed IncIDFlakies, a technique that analyzes code changes to
detect order-dependent flaky tests from newly-introduced code with the help of RTS.
\citet{chen2018speeding} sped up mutation testing by only recollecting the mutation
testing results of the affected tests.  Genetic improvement~\cite{guizzo2021enhancing}
can also benefit from RTS by only running the affected tests to evaluate the generated
patches.  \citet{legunsen2019techniques} leveraged RTS-like technique to speed up
runtime verification in evolving software systems.  \citet{celik2017icoq} used the idea
of RTS to perform regression proof selection in verification projects written in Coq.
Our work is the first to apply RTS in the context of speeding up \codecovanalysis.

Chittimalli and Harrold~\cite{ChittimalliAndHarrold09CoverageRegressionTesting} explored
integrating RTS and code coverage analysis, but for a different purpose than our work:
their goal was to speedup RTS by only instrumenting the necessary probes to the code
elements covered by the tests.  Since reporting \codecov is not their focus, the
\covdata is not in the typical format that \codecovanalysis techniques use.
Specifically, they maintained the set of executed probes per test, which we discussed in
\S\ref{sec:technique:analysis} why this is not desirable due to the overhead of merging
\covdata.

\MyPara{Code coverage analysis} Code coverage measures the quality of
tests and shows the percentage of code executed by the
tests~\cite{hemmati2015effective, kochhar2015code}.  Code coverage
analysis tools~\cite{YangETAL06SurveyCoverage, shahid2011evaluation} capture the
coverage and generate reports by instrumenting the code and tracking
the execution. However, the overhead of instrumentation and analyzing
coverage data to generate reports can be high, especially for
large-scale code base~\cite{ivankovic2019code, adler2011code}.
We proposed to speed up \codecovanalysis by incrementally computing the \covdata only
for the part of \codebase affected by the changes, and implemented our idea as
\ijacocoTool based on a popular industry-level code coverage analysis tool for \java,
\jacoco~\cite{JacocoGithubRepo}.

\section{Conclusion}
\label{sec:conclusion}

We proposed the first \icodecovanalysis technique and its implementation, \ijacocoTool,
for collecting \codecov in the context of software evolution. The key difference between
\ijacocoTool and prior work is that it selects a minimal set of \tests, saving the
costly test execution time, and provides an accurate coverage report upon code changes.
We proved the correctness of \icodecovanalysis and demonstrated the correctness of
\ijacocoTool. Our evaluation showed that \ijacocoTool can significantly speed up
\codecovanalysis by \UseMacro{ijacoco-avg-speedup}$\times$ compared to the industrial
standard code coverage tool \jacoco.

\section*{Acknowledgments}
We thank Milos Gligoric, Yu Liu, and the anonymous reviewers for their comments and feedback.
This work is enabled in part by support provided by Compute Ontario (computeontario.ca) and the Digital Research Alliance of Canada (alliancecan.ca).
This work is partially supported by the Cheriton School of Computer Science at the University of Waterloo under start-up grant and URF program.

\bibliography{bib}

\clearpage\newpage
\appendix

\section{Code and Data}

We have included \ijacocoTool (in \java), our experiment scripts (in \python), and data
(mostly in CSV format) in this package.  Please refer to README.md for how to use the
code and where to locate the data.

We will open-source this replication package upon the acceptance of this paper.

\section{Detailed Evaluation Results}

We include the plots showing detailed evaluation results in this document.

\begin{figure*}
\centering
\includegraphics[width=0.9\textwidth]{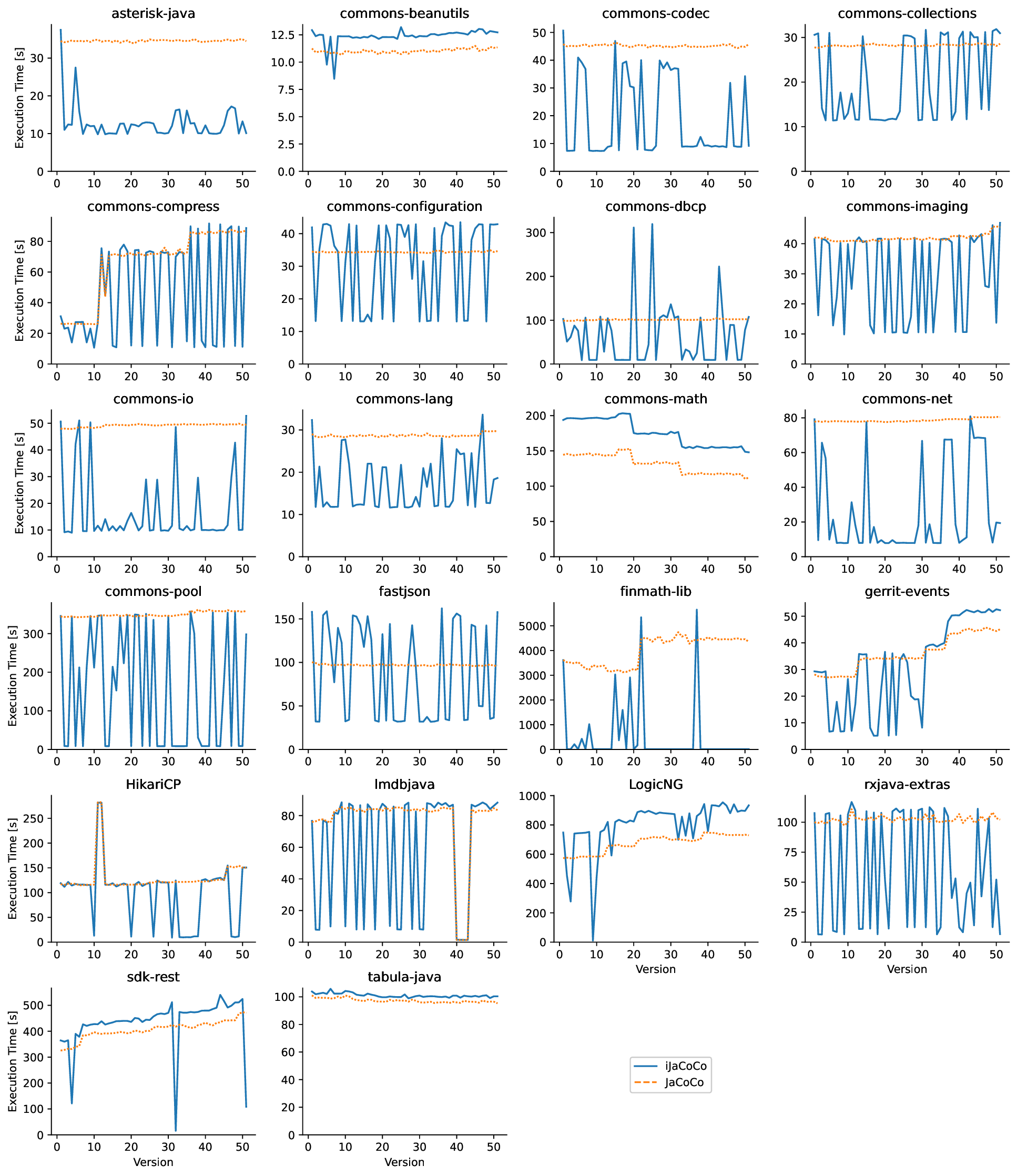}
\caption{\ijacocoTool vs. \jacoco end-to-end time. \label{fig:execution-time}}
\end{figure*}

Figure~\ref{fig:execution-time} compares the end-to-end time of \ijacocoTool and \jacoco.

\begin{figure*}
\centering
\includegraphics[width=0.9\textwidth]{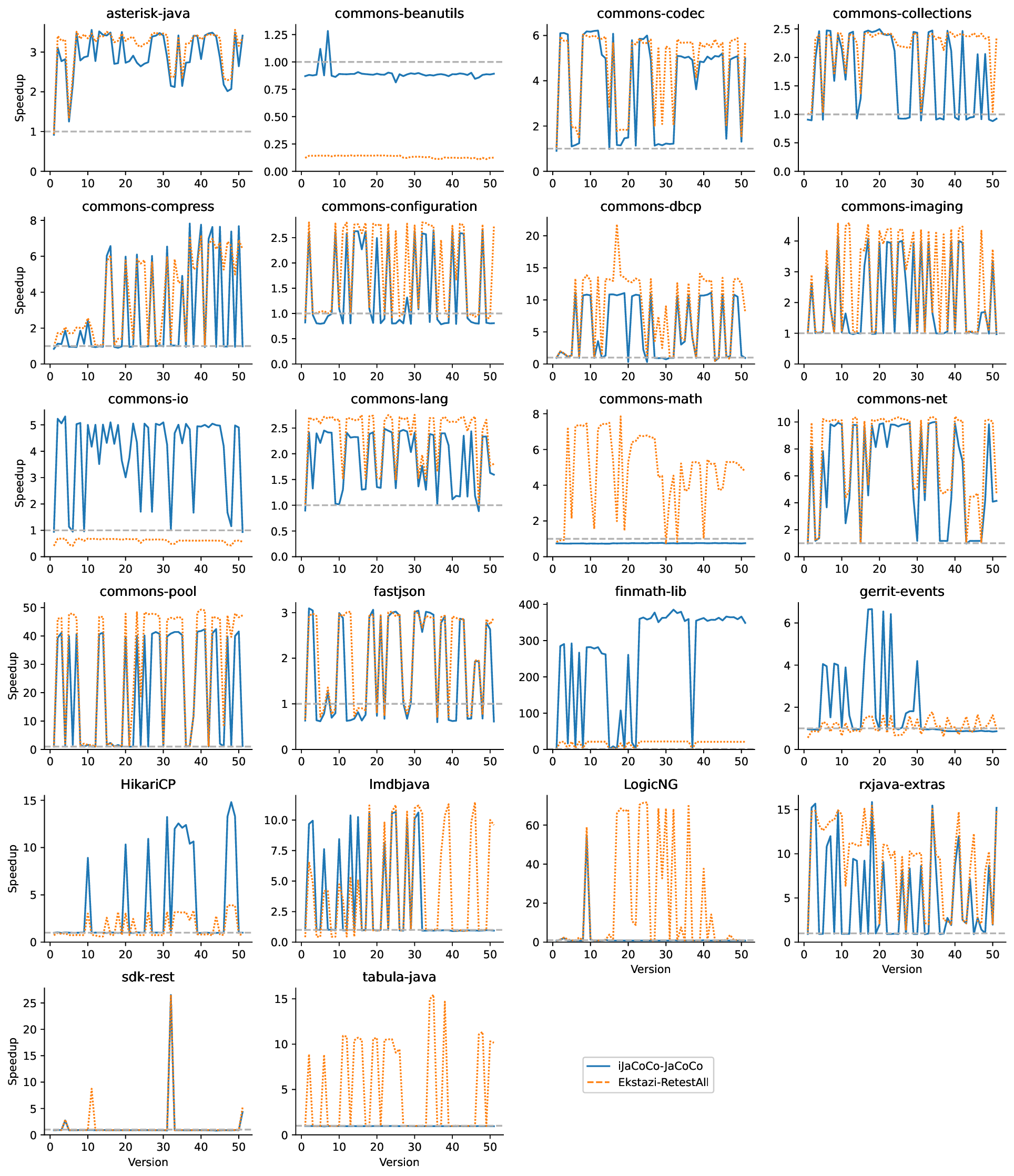}
\caption{Speedup of \ijacocoTool (w.r.t. \jacoco) vs. speedup of \ekstazi (w.r.t. \retestall). \label{fig:speedup-compare}}
\end{figure*}

Figure~\ref{fig:speedup-compare} shows the speedup of \ijacocoTool (w.r.t. \jacoco) and
compares it with the speedup of \ekstazi (w.r.t. \retestall).  The two speedup values are
mostly consistent.

\begin{figure*}
\centering
\includegraphics[width=0.9\textwidth]{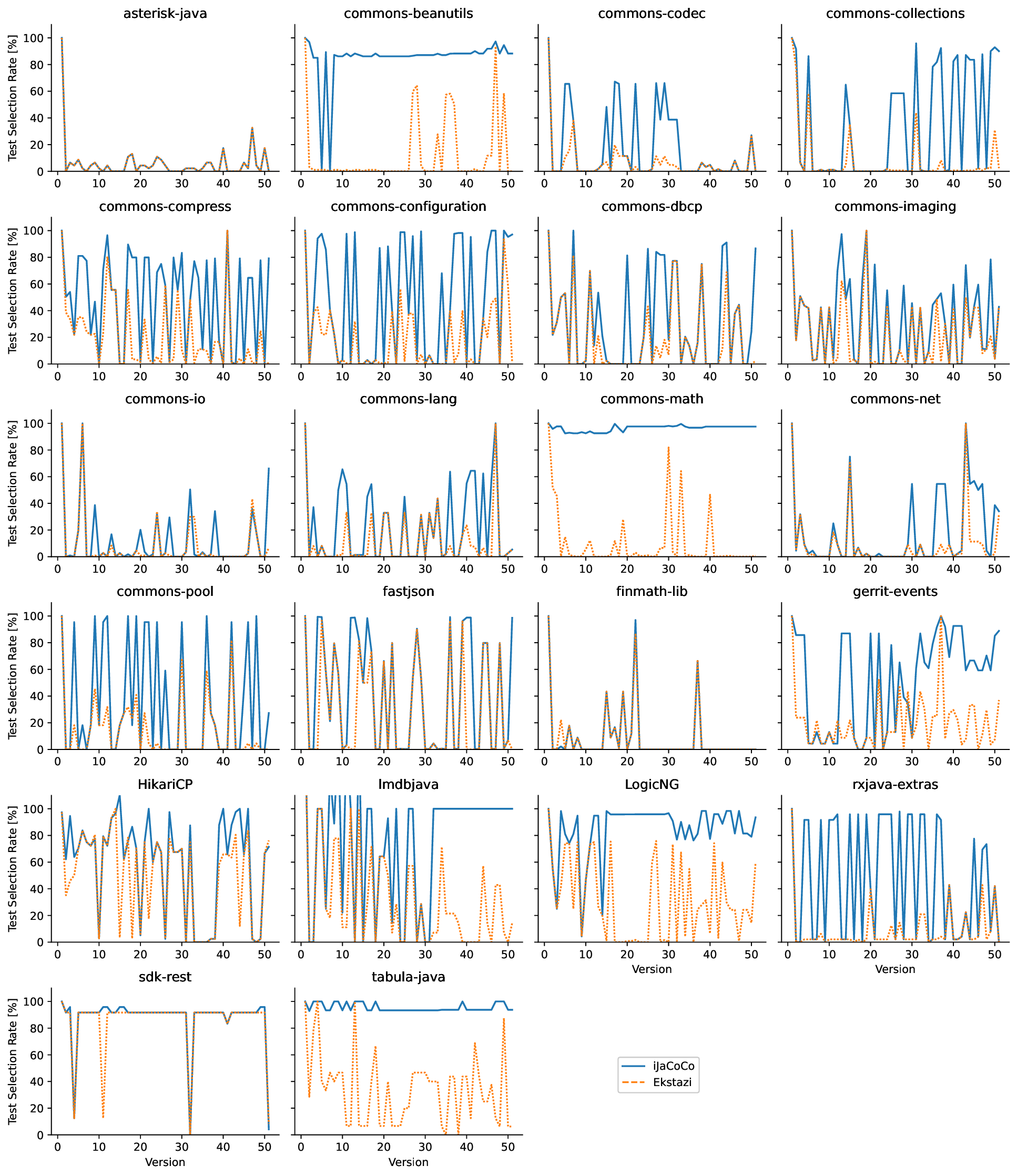}
\caption{\ijacocoTool vs. \ekstazi test selection rate. \label{fig:selection-rate}}
\end{figure*}

Figure~\ref{fig:selection-rate} compares the test selection rate of \ijacocoTool and \ekstazi.

\begin{figure*}
\centering
\includegraphics[width=0.9\textwidth]{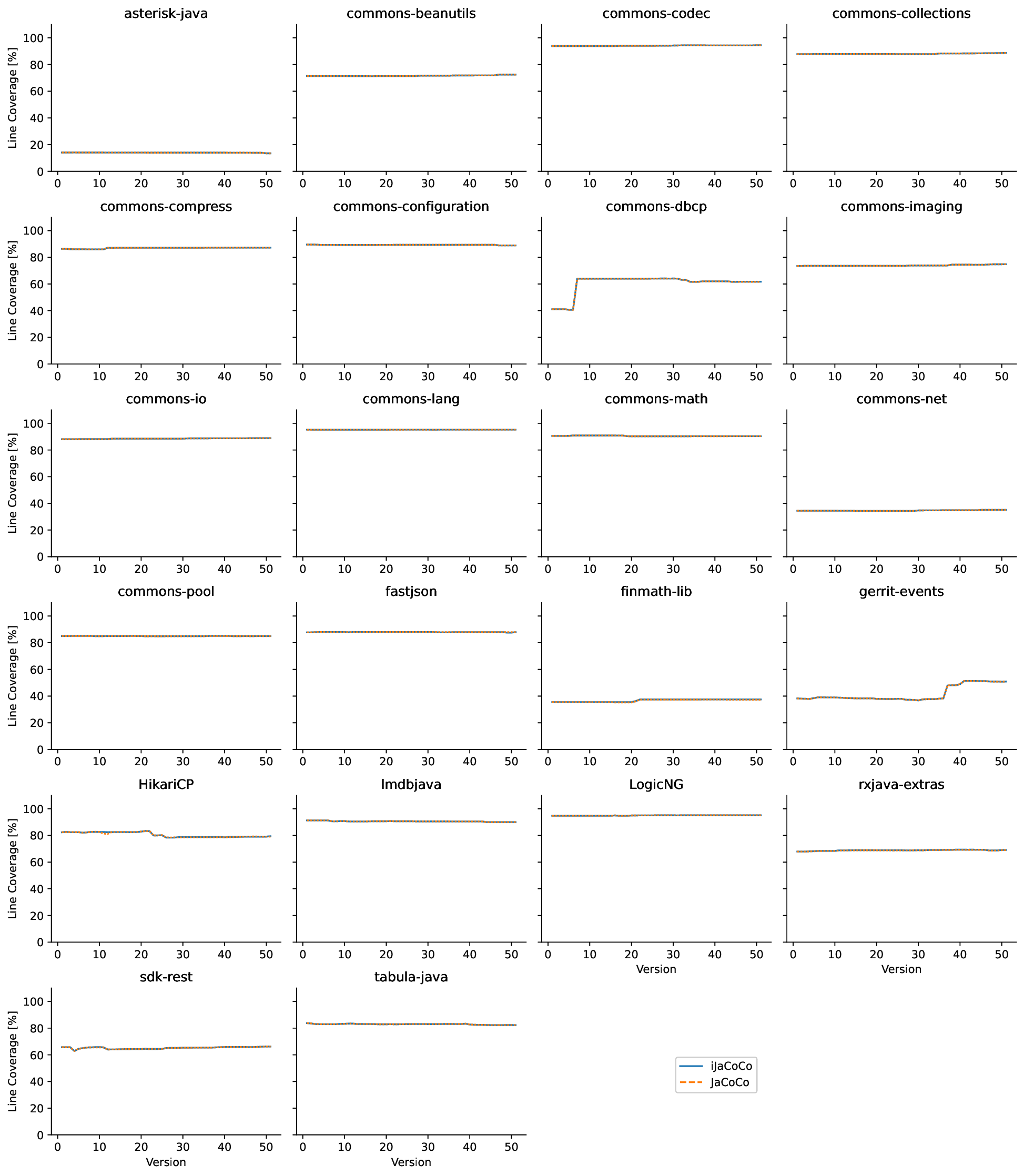}
\caption{\ijacocoTool vs. \bjacoco coverage scores. \label{fig:coverage}}
\end{figure*}

Figure~\ref{fig:coverage} compares the coverage scores of \ijacocoTool and \jacoco,
confirming that \ijacocoTool produces the correct \codecov results.

\end{document}